\documentclass[aps,prb,reprint,groupedaddress]{revtex4-2}

\usepackage{graphicx}

\begin{document}



\title{Superfluid flow in disordered superconductors with Dynes pair-breaking scattering: 
depairing current, kinetic inductance, and superheating field}


\author{Takayuki Kubo}
\email[]{kubotaka@post.kek.jp}
\affiliation{High Energy Accelerator Research Organization (KEK), Tsukuba, Ibaraki 305-0801, Japan}
\affiliation{The Graduate University for Advanced Studies (Sokendai), Hayama, Kanagawa 240-0193, Japan}



\begin{abstract}
We investigate the effects of Dynes pair-breaking scattering rate $\Gamma$ on the superfluid flow in a narrow thin-film  superconductor and a semi-infinite superconductor by self-consistently solving the coupled Maxwell and Usadel equations for the BCS theory in the diffusive limit for all temperature $T$, all $\Gamma$, and all superfluid momentum. 
We obtain the depairing current density $j_d(\Gamma, T)$ and the current-dependent nonlinear kinetic inductance $L_k(j_s, \Gamma, T)$ in a narrow thin-film and the superheating field $H_{sh}(\Gamma, T)$ and the current distribution in a semi-infinite superconductor, taking the nonlinear Meissner effect into account. 
The analytical expressions for $j_d(\Gamma,T)|_{T=0}$, $L_k(j_s, \Gamma, T)|_{T=0}$, and $H_{sh}(\Gamma, T)|_{T=0}$ are also derived.  
The theory suggests $j_d$ and $H_{sh}$ can be ameliorated by reducing $\Gamma$, 
and $L_k$ can be tuned by a combination of the bias current and $\Gamma$.
Tunneling spectroscopy can test the theory and also give insight into how to engineer $\Gamma$ via materials processing. 
Implications of the theory would be useful to improve performances of various superconducting quantum devices. 
\end{abstract}

\maketitle


\section{Introduction}

The physics of the superfluid flow in $s$-wave superconductors is closely tied with the operating principles and performances of various superconducting quantum devices such as superconducting nanowire single-photon detectors (SNSPDs)~\cite{2012_Natarajan, 2015_Engel}, 
resonators for microwave kinetic inductance detectors (MKIDs)~\cite{2012_Zmuidzinas, 2018_Mauskopf} and quantum computers~\cite{2013_Devoret, 2017_Wendin, 2019_Q_report}, 
and superconducting radio-frequency (SRF) resonant cavities for particle accelerators~\cite{2017_Padamsee, 2017_Gurevich_SUST, 2017_Liarte_SUST, 2017_Kubo_SUST}. 
The supercurrent density $j_s$ is proportional to the superfluid momentum $\hbar q$ and the superfluid density $n_s$.  
When $|q|$ is such a small value that $n_s$ is not significantly suppressed, 
$j_s$ linearly increases with $|q|$. 
However, as $|q|$ increases, the reduction of $n_s$ becomes significant, and $j_s$ ceases to increase~\cite{1969_Maki_Parks, 1963_Maki_I, 1963_Maki_II}. 
The maximum value of $j_s$ is called the depairing current density $j_d$ and determines the stability limit of the superfluid flow, 
above which finite electrical resistance necessarily appears.  
In SNSPDs, 
a superconducting nanowire is biased with a dc current close to $j_d$. 
An incident photon absorbed by the strip heats electrons and reduces the critical current below the bias current, resulting in measurable finite electrical resistance. 
The reset time after a detection event is often limited by the kinetic inductance~\cite{2006_Kerman}. 
In MKIDs, the kinetic inductance plays an essential role in its operating mechanism. 
Incoming photons with a frequency higher than the superconducting gap break Cooper pairs and lead to an increase of the kinetic inductivity $L_k \propto 1/n_s$. 
A resultant shift of resonant frequency $\delta f  \propto -\delta L_k$ can be detected. 
Besides, the bias current also reduces $n_s$ and increase $L_k$ and can be utilized to tune a resonator frequency~\cite{2015_Vissers} and to observe the nonlinear Meissner Effect~\cite{2010_Groll}. 
In SRF resonant cavities, 
charged particles are accelerated by the electric component of the microwave, 
which is proportional to the rf magnetic field at the surface. 
Vortex-free cavities~\cite{2014_Romanenko, 2016_Huang, 2016_Posen} exhibit huge quality factor $Q \sim 10^{10}$-$10^{12}$ at $T< 2\,{\rm K}$~\cite{2017_Romanenko, 2020_Romanenko, 2020_Posen} even under the strong rf magnetic field~\cite{2007_Geng, 2014_Kubo_IPAC, 2017_Grassellino, 2018_Dhakal} such that the nonlinear Meissner effect manifest itself. 
Here the achievable rf field is limited by the induced screening current at the surface, 
which cannot exceed $j_d$. 
The surface magnetic field that induces $j_d$ is coincident with the superheating field $H_{sh}$, 
which is the stability limit of the Meissner state. 
$H_{sh}$ is thought to define the upper limit of the accelerating field~\cite{2017_Gurevich_SUST, 2017_Liarte_SUST, 2017_Kubo_SUST} and is one of the main interests in fundamental SRF studies~\cite{2015_Posen_PRL, 2019_Keckert}.

Microscopic calculations of physical quantities relevant to these devices would provide us with a deeper understanding of experimental results and clues to improving device performances. 
Those for disordered materials are especially important because these devices are often made from high-resistance films or impurity-doped bulk materials~\cite{2013_Grassellino, 2013_Dhakal, 2017_Maniscalco, 2018_Yang, 2019_Gonnella}. 
Some 60 years ago, Maki calculated $j_d$ at the temperature $T\to 0$~\cite{1963_Maki_I, 1963_Maki_II}. 
Kupriyanov and Lukichev obtained $j_d$ for an arbitrary $T$~\cite{1980_Kupriyanov}.  
By using the Maki's results, $L_k$ for the current-carrying state was calculated afterwards~\cite{2010_Annunziata}. 
Then those for all $T$ and all $j_s$ up to $j_d$ were investigated~\cite{2012_Clem_Kogan}. 
Calculations of $H_{sh}$ also have a long history, starting from those for a clean-limit superconductor~\cite{1966_Galaiko, 2008_Catelani}. 
Effects of homogeneous~\cite{2012_Lin_Gurevich} and inhomogeneous~\cite{2019_Sauls} impurities on $H_{sh}$ were recently investigated.
Yet, theories including realistic materials features which can limit device performances have been studied lesser extent. 
Such theories would be useful to pin down causes of performance limitations, 
e. g., critical current below the ideal $j_d$ in nanowires and quenches below the ideal $H_{sh}$ in SRF cavities, etc.

\begin{figure}[tb]
   \begin{center}
   \includegraphics[width=0.75\linewidth]{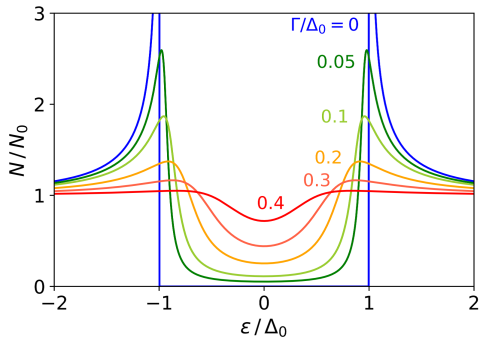}
   \end{center}\vspace{0 cm}
   \caption{
Quasiparticle DOS calculated for $T/T_{c0}=0.2$ and $\Gamma/\Delta_0 =0, 0.05, 0.1, 0.2, 0.3, 0.4$. 
Here $\Delta_0$ is the BCS pair potential at $T=0$ and $T_{c0}$ is the BCS critical temperature for $\Gamma=0$.  
   }\label{fig1}
\end{figure}

One of the common features among various superconducting materials is the broadening of the density of states (DOS), 
which has been observed in a numerous number of tunneling experiments~\cite{2003_Zasa} (see e.g., Refs.~\cite{2013_Dhakal, 2015_Becker, 2019_Groll} for SRF materials). 
Such a broadened DOS has been described by the Dynes formula~\cite{1978_Dynes, 1984_Dynes}, 
which is given by $N(\epsilon)/N_0 = {\rm Re} [(\epsilon + i\Gamma)/\sqrt{ (\epsilon + i\Gamma)^2 - \Delta^2}]$ (see Fig.~\ref{fig1}). 
Here $\Gamma$ is the Dynes pair-breaking scattering rate, 
resulting in a finite density of subgap states in the vicinity of the Fermi level $\Gamma/\Delta$ and a constant quasiparticle lifetime $\hbar/\Gamma$. 
A microscopic derivation of the Dynes formula has been investigated~\cite{2016_Herman, 2017_Herman, 2018_Herman}. 
Besides, independent of microscopic models of Dynes formula, 
it is possible to formulate the quasiclassical theory of the BCS model which incorporates $\Gamma$ ~\cite{2017_Gurevich_Kubo, 2019_Kubo_Gurevich, 2020_Kubo_jd}. 
Interestingly, it has been shown that the pair-breaking $\Gamma$ parameter and other pair breakers (e.g, current~\cite{1964_Maki_current, 1965_Fulde_current, 2003_Anthore}, magnetic impurities~\cite{1961_AG, 1966_Fulde_Maki_mag} and proximity-coupled normal layer~\cite{1996_Belzig, 1999_Belzig}) can reduce the rf dissipation in the weak- and the strong-rf regimes via a modification of the quasiparticle spectrum~\cite{2017_Gurevich_Kubo, 2019_Kubo_Gurevich, 2020_Kubo_jd, 2014_Gurevich_PRL}.

\begin{figure}[tb]
   \begin{center}
   \includegraphics[width=0.95\linewidth]{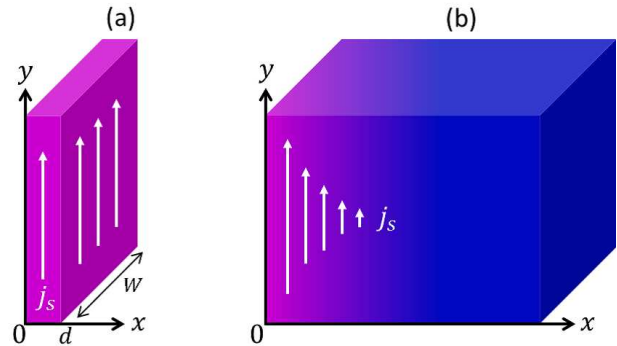}
   \end{center}\vspace{0 cm}
   \caption{
(a) Thin and narrow superconducting film carrying a uniform current $j_s$. 
We assume a thickness $d \ll \lambda$ and a width $W \ll \lambda^2/d$. 
(b) Semi-infinite superconductor carrying the Meissner current $j_s(x)$. 
We assume $\lambda$ is much larger than the coherence length. 
   }\label{fig2}
\end{figure}

In this work, 
we focus on effects of $\Gamma$ on the superfluid flow in disordered superconductors. 
We consider the geometries shown in Fig.~\ref{fig2}:
a thin and narrow superconducting film (relevant to, e.g., SNSPD, MKID) and a semi-infinite superconductor (relevant to, e.g., SRF cavities made from bulk materials or thick film). 
We evaluate the depairing current density $j_d(\Gamma, T)$, the current-dependent nonlinear kinetic inductance $L_k(j_s, \Gamma, T)$, and the superheating field $H_{sh}(\Gamma, T)$ for all $T$, all $\Gamma$, and all current. 
The results of this work would provide with clues to finding out causes of performance limitations and those used to improve performances of superconducting quantum devices.

The paper is organized as follows. 
In Section II, we briefly review the Eilenberger-Usadel formalism~\cite{1968_Eilenberger, 1969_LO, 1970_Usadel, Kopnin} of the BCS theory and express physical quantities with the Matsubara Green's functions. 
In Sec. III, we solve the Usadel equation for all $T$, all $\Gamma$, and all $q$. 
Some useful formulas of $\Delta$, $n_s$, $\lambda$, and $j_s$ at $T = 0$ and $\simeq T_c$ are also obtained. 
In Sec. IV, we consider a thin and narrow superconducting film [Fig.~\ref{fig2} (a)].  
We evaluate the depairing current density $j_d(\Gamma,T)$ and the current-dependent nonlinear kinetic inductance $L_k(j_s, \Gamma, T)$. 
In addition to numerical results, we present the analytical formulas for $j_d(\Gamma,T)$ and $L_k(j_s, \Gamma, T)$ at $T = 0$ and $\simeq T_c$. 
In Sec. V, we consider a semi-infinite superconductor [Fig.~\ref{fig2} (b)].  
We calculate the current distribution taking the nonlinear Meissner effect into account and evaluate the superheating field $H_{sh}(\Gamma, T)$. 
Analytical formula for $T\to 0$ and $T\simeq T_c$ are also derived.  
In Sec. VI, we discuss the implications of our results.

\section{Theory} \label{sec_theory}

We apply the well-established Eilenberger-Usadel formalism~\cite{1968_Eilenberger, 1969_LO, 1970_Usadel, Kopnin} of the BCS theory in the diffusive limit to the geometries shown in Fig.~\ref{fig2}. 
The normal and anomalous quasiclassical Matsubara Green's functions $G_{\omega_n}=\cos\theta$ and $F_{\omega_n}=\sin\theta$ obey the Usadel equation: 
\begin{eqnarray}
\frac{\hbar D}{2}\theta'' = s \sin\theta \cos \theta + (\hbar \omega_n + \Gamma) \sin \theta - \Delta \cos\theta . \label{thermodynamic_Usadel_0}
\end{eqnarray}
Here $D$ is the electron diffusivity, 
the prime denotes differentiation with respect to $x$, 
$s =  (q/q_{\xi})^2 \Delta_0$ is the superfluid flow parameter, 
$\Delta_0 = \Delta(s, \Gamma, T)|_{s=\Gamma=T=0}$ is the BCS pair potential at $T=0$, 
$\hbar q = 2 m v_s$ is the superfluid momentum, 
$v_s$ is the superfluid velocity, $m$ is the electron mass, 
$q_{\xi} = \sqrt{2\Delta_0/\hbar D}$ is the inverse of the coherence length, 
and $\hbar \omega_n = 2\pi k_B T(n + 1/2)$ is the Matsubara frequency. 
In Figs.~\ref{fig2} (a) and \ref{fig2} (b), the current distributes uniformly and varies slowly over the coherence length, respectively. 
In either cases, the $\theta''$ term is negligible and Eq.~(\ref{thermodynamic_Usadel_0}) reduces to
\begin{eqnarray}
\biggl( \Delta - \frac{s}{\sqrt{1+ \cot^2 \theta}} \biggr) \cot\theta = \hbar \omega_n + \Gamma .
\label{thermodynamic_Usadel}
\end{eqnarray}
Note that, in Fig.~\ref{fig2} (b), $\theta$ and $\Delta$ depend on $x$ via $s=s(x)$.  
The pair potential $\Delta(s,\Gamma,T)$ satisfies the self-consistency equation
\begin{eqnarray}
\ln \frac{T_{c0}}{T} = 2 \pi k_B T \sum_{\omega_n >0} \biggl( \frac{1}{\hbar \omega_n} - \frac{\sin\theta}{\Delta} \biggr) ,
\label{self-consistency}
\end{eqnarray}
where $k_B T_{c0}= \Delta_0 \exp(\gamma_E)/\pi  \simeq \Delta_0/1.76$ is the BCS critical temperature, 
and $\gamma_E=0.577$ is the Euler constant. 
The thermodynamic critical field $H_c$ is given by~\cite{2018_Herman, 2020_Kubo_jd} 
\begin{eqnarray}
&& \hspace{-0.5cm} H_c(\Gamma, T) = \sqrt{ -\frac{2}{\mu_0} \Omega (0, \Gamma, T) } , \label{Hc} \\
&& \hspace{-0.5cm} \Omega (0, \Gamma, T) = - 2\pi T N_0 \Delta \nonumber \\
&&\times  \sum_{\omega_n >0}  \biggl[ \frac{2(\hbar \omega_n + \Gamma)}{\Delta} ( \cos \theta -1) + \sin\theta \biggr] , \label{Omega}
\end{eqnarray}
where $\theta$ and $\Delta$ in Eq.~(\ref{Omega}) are evaluated for $s=0$. 
The superfluid density $n_s$, penetration depth $\lambda$ and supercurrent density $j_s$ are given by
\begin{eqnarray}
&&\frac{n_s(s, \Gamma, T)}{n_{s0}} = \frac{\lambda_0^2}{\lambda^2(s, \Gamma, T)}= \frac{4k_B T}{\Delta_0} \sum_{\omega_n > 0} \sin ^2 \theta 
, \label{superfluid_density} \\
&&\frac{j_s (s, \Gamma, T)}{j_{s0}} 
= \sqrt{\frac{\pi s}{\Delta_0}} \frac{n_s(s, \Gamma, T) }{n_{s0}}  . \label{supercurrent} 
\end{eqnarray}
Here $n_{s0} = n_s(0, 0, 0) = 2\pi m N_0 D \Delta_{0}/\hbar$ is the BCS superfluid density at $T=0$, 
$\lambda_0 = \lambda(0,0,0) = \sqrt{\hbar/\pi \mu_0 \Delta_0 \sigma_n}$ is the BCS penetration depth at $T=0$, 
$j_{s0} = H_{c0}/\lambda_0 = \sqrt{\pi} |e| N_0 D \Delta_0 q_{\xi}$, 
and $H_{c0}=H_c(0,0)=\sqrt{N_0/\mu_0} \Delta_0$ is the BCS thermodynamic critical field at $T=0$.

In the geometry shown in Fig.~\ref{fig2} (b), 
$n_s$, $\lambda$, and $j_s$ depend on $x$ through $s(x) = [q(x)/q_{\xi}]^2 \Delta_0$. 
Here $j_s(x)$, $H(x)$, and $q(x)$ obey the Maxwell equation, 
$j_s = -\partial_x H$ and $\mu_0 H=(\hbar/2|e|) \partial_x q$, namely,
\begin{eqnarray}
&&\frac{\partial^2 q}{\partial x^2} = \frac{q}{\lambda^2 (s,\Gamma, T)} , \label{Usadel_London} \\
&& \frac{H}{H_{c0}} = \sqrt{\pi}  \frac{\partial (q/q_{\xi})}{\partial (x/\lambda_0)} . \label{H}
\end{eqnarray}
By using the magnetic field at the surface $H_0$, 
the boundary conditions can be written as
\begin{eqnarray}
H(0) =H_0, \hspace{1cm}
\lim_{x\to \infty} q(x) \to 0. \label{boundary_conditions}
\end{eqnarray}
Self-consistent calculations of Eqs.~(\ref{thermodynamic_Usadel}), (\ref{self-consistency}), and (\ref{superfluid_density})-(\ref{boundary_conditions}) yield the distributions of $\theta(x)$, $\Delta(x)$, $\lambda(x)$, $j_s(x)$, $q(x)$, and $H(x)$ for a given set of $H_0$ and $T$.

In the Ginzburg-Landau (GL) regime, $T\simeq T_c$, 
we can expand the Matsubara Green's functions in powers of $\delta = \Delta/2\pi k_B T \ll 1$~\cite{2020_Kubo_jd}:
\begin{eqnarray}
&&\sin\theta = a \delta - \frac{1}{2} \biggl( a^3 - \frac{s}{2\pi k_B T} a^4 \biggr) \delta^3 , \label{sinGL} \\
&&\cos\theta = 1 -\frac{a^2}{2} \delta^2 - \frac{1}{8} \biggl( \frac{2s}{\pi k_B T}  a^5 - 3 a^4 \biggr) \delta^4 . \label{cosGL}
\end{eqnarray}
Here $a=(n+1/2+ s/2\pi k_B T +\Gamma/2\pi k_B T)^{-1}$. 
Substituting Eqs.~(\ref{sinGL}) and (\ref{cosGL}) into Eq.~(\ref{self-consistency}), 
we obtain the GL equation~\cite{2020_Kubo_jd}, 
\begin{eqnarray}
1 - \frac{T}{T_{c}} = \frac{\pi s}{4 k_B T_{c}} + \frac{7\zeta(3)}{8\pi^2 k_B^2 T_{c}^2} \Delta^2 , \label{GL}
\end{eqnarray}
for $\Delta, \, s, \, \Gamma \ll 2\pi k_B T$ and $T \simeq T_{c}$. 
Note here $T_c$ depends on $\Gamma$.

In the following, we use $\Delta_{0}$ as a unit of energy and use dimensionless quantities 
$\tilde{s} = s/\Delta_{0}$, 
$\tilde{\omega}_n= \hbar \omega_n/\Delta_{0}$, 
$\tilde{\Gamma} = \Gamma/\Delta_{0}$, $\tilde{\Delta}= \Delta/\Delta_{0}$, 
$\tilde{T}= k_B T/\Delta_{0}$, etc. 
For brevity, we omit all these tildes.

\section{Solutions} \label{sec_solutions}

\subsection{Zero-current state ($s=0$)}\label{sec_zero_current}

\begin{figure}[tb]
   \begin{center}
   \includegraphics[height=0.43\linewidth]{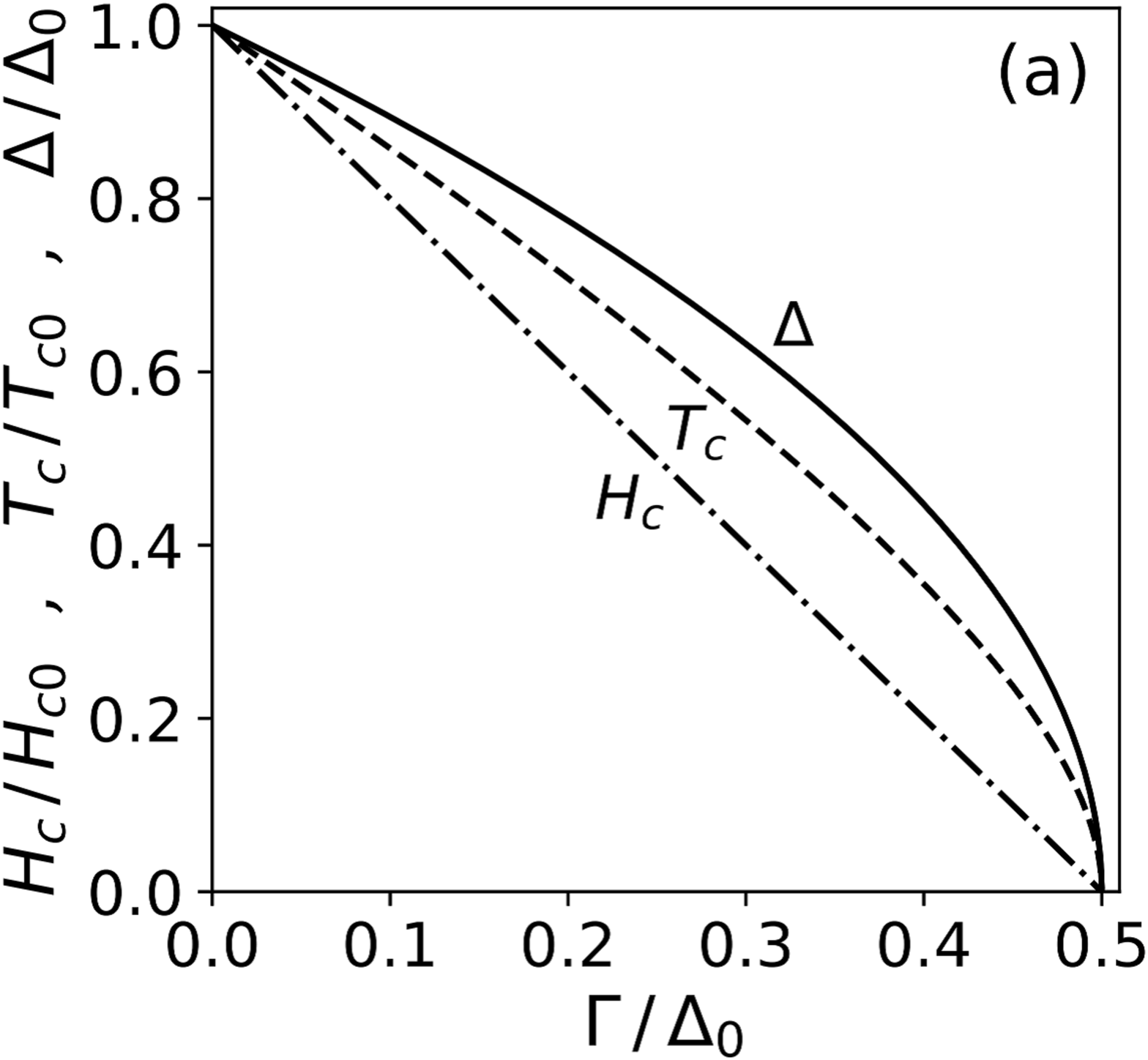}
   \includegraphics[height=0.43\linewidth]{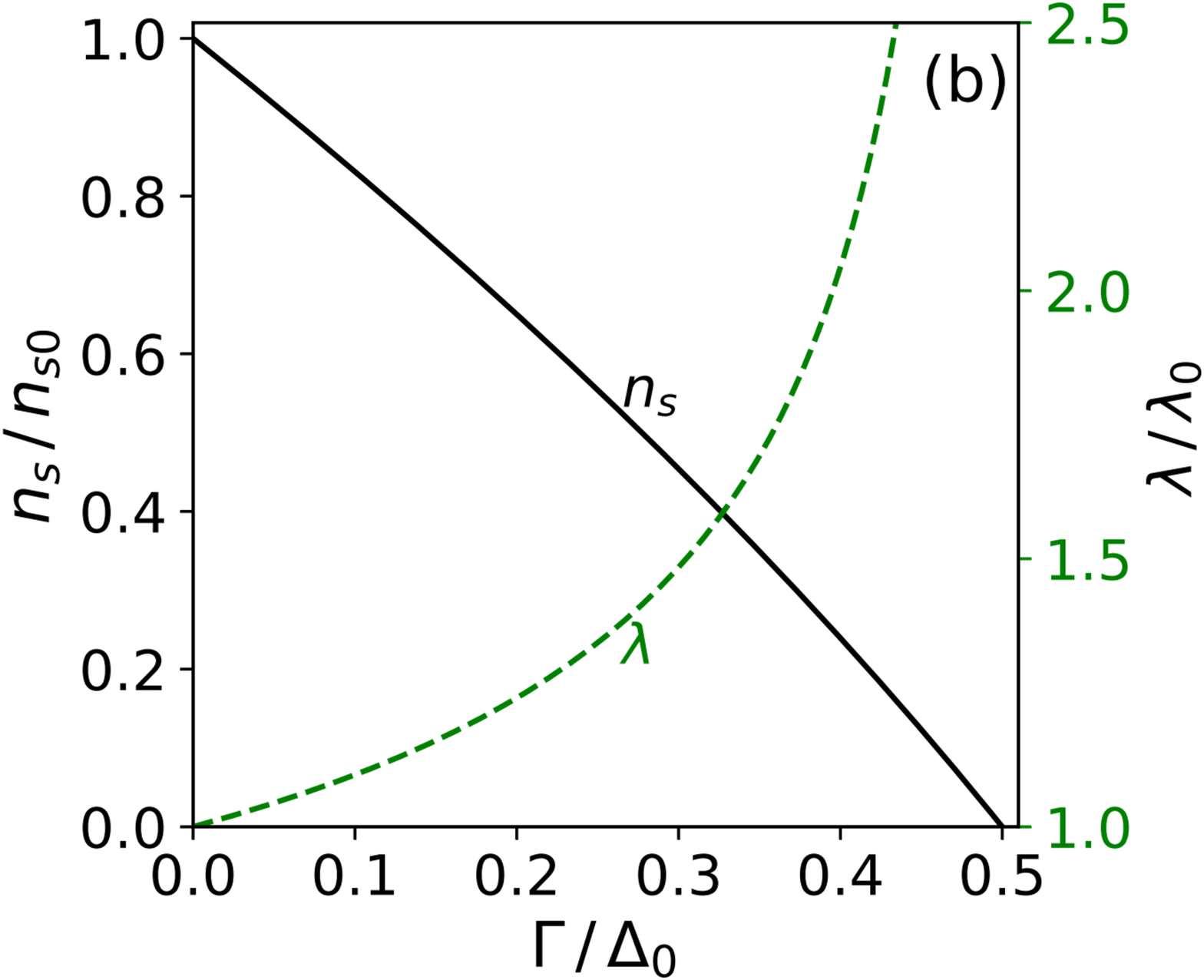} 
\end{center}\vspace{0 cm}
   \caption{
(a) $\Delta(0, \Gamma, T)|_{T\to 0}$, $T_c(\Gamma)$, 
$H_c(\Gamma,T)|_{T\to 0}$,
(b) $n_s(0, \Gamma, T)|_{T\to 0}$, and $\lambda(0, \Gamma, T)|_{T\to 0}$ as functions of $\Gamma$. 
   }\label{fig3}
\end{figure}

First, consider the zero-current state ($s=0$) at the low-temperature limit ($T=0$). 
The pair potential $\Delta$ can be calculated from Eqs.~(\ref{thermodynamic_Usadel}) and (\ref{self-consistency}). 
The solution of Eq.~(\ref{thermodynamic_Usadel}) is given by $\cot\theta = (\hbar \omega_n  + \Gamma)/\Delta$. 
Substituting $\theta$ into Eq.~(\ref{self-consistency}), 
we find (see Appendix~\ref{a1})
\begin{eqnarray}
\Delta(0, \Gamma, 0)  = \exp\biggl[ -\sinh^{-1} \frac{\Gamma}{\Delta(0, \Gamma, 0)} \biggr] ,  \label{Delta_0_Gamma_0}
\end{eqnarray}
which reduces to $\Delta \simeq 1- \Gamma$ for $\Gamma \ll 1$. 
The equation for $T_c$ is obtained by solving Eqs.~(\ref{thermodynamic_Usadel}) and (\ref{self-consistency}) for $(\theta, \Delta) \ll 1$ and $T \simeq T_c$. 
Then, we have~\cite{2017_Gurevich_Kubo, 2020_Kubo_jd}, 
\begin{eqnarray}
\ln \frac{T_{c}}{T_{c0}} = \psi \biggl(\frac{1}{2} \biggr) - \psi\biggl( \frac{1}{2} + \frac{\Gamma}{2\pi T_{c}} \biggr) , \label{Tc} 
\end{eqnarray}
which reduces to $T_c \simeq T_{c0} -\pi\Gamma/4$ for $\Gamma \ll 1$. 
The thermodynamic critical field $H_c$ can be calculated from Eqs.~(\ref{Hc}) and (\ref{Omega}). 
At $T = 0$, we find (see Appendix~\ref{a1})
\begin{eqnarray}
\frac{H_c(\Gamma, 0)}{H_{c0}} = \Delta \sqrt{ 1 + \frac{2\Gamma^2}{\Delta^2} - \frac{2\Gamma}{\Delta} \sqrt{1 + \frac{\Gamma^2}{\Delta^2} } } \label{Hc_Gamma_0} ,
\end{eqnarray}
which reduces to $H_c/H_{c0} \simeq 1-2\Gamma$ for $\Gamma \ll 1$.  
The superfluid density $n_s$ and the penetration depth $\lambda$ can be calculated from Eq.~(\ref{superfluid_density}). 
At $T = 0$, we have~\cite{2017_Gurevich_Kubo, 2020_Kubo_jd}: 
\begin{eqnarray}
&&\frac{n_s(0, \Gamma, 0)}{n_{s0}} = \frac{\lambda_0^2}{\lambda^2(0, \Gamma, 0)} \nonumber \\
&&= \Delta(0, \Gamma, 0) \biggl[ 1-\frac{2}{\pi} \tan^{-1} \frac{\Gamma}{\Delta(0, \Gamma, 0)} \biggr] , \label{ns_0_Gamma_0}
\end{eqnarray}
which reduces to $n_s/n_{s0}\simeq 1- (1+2/\pi)\Gamma$ for $\Gamma \ll 1$. 
Shown in Fig.~\ref{fig3} are $\Delta(0,\Gamma,T)|_{T= 0}$, $T_c(\Gamma)$, $H_c(\Gamma,T)|_{T= 0}$, and $n_s(0,\Gamma,T)|_{T= 0}$ as functions of $\Gamma$, 
which monotonically decrease as $\Gamma$ increases and vanish at $\Gamma=1/2$.

For $T\simeq T_c$, the GL regime, 
we can calculate $\Delta$ from Eq.~(\ref{GL}). 
Then, we have~\cite{2020_Kubo_jd}
\begin{eqnarray}
&&\Delta (0, \Gamma, T) 
= \sqrt{ \frac{8\pi^2 T_{c}^2}{7\zeta(3)} \biggl(1- \frac{T}{T_{c}} \biggr)   }, \label{GL_Delta_zero_current}  \\
&&H_c (\Gamma, T) 
= \sqrt{ \frac{8\pi^2 T_{c}^2 N_0}{7\zeta(3) \mu_0}} \biggl( 1- \frac{T}{T_{c}} \biggr) , \label{GL_Hc_zero_current} \\
&&\frac{n_s (0, \Gamma, T)}{n_{s0}} 
= \frac{\lambda_0^{2}}{\lambda^{2} (0, \Gamma, T)}
= \frac{4\pi^2 T_{c}}{7\zeta(3)} \biggl(1- \frac{T}{T_{c}} \biggr) 
,  \label{GL_ns_zero_current} 
\end{eqnarray}
for $T \simeq T_c(\Gamma)$. 
Here Eqs.~(\ref{Hc})-(\ref{superfluid_density}) are used. 
Eqs.~(\ref{GL_Delta_zero_current})-(\ref{GL_ns_zero_current}) are coincident with the usual GL results for the zero-current state. 
The only difference is that $T_c$ depends on $\Gamma$ through Eq.~(\ref{Tc}).

\begin{figure}[tb]
   \begin{center}
   \includegraphics[height=0.43\linewidth]{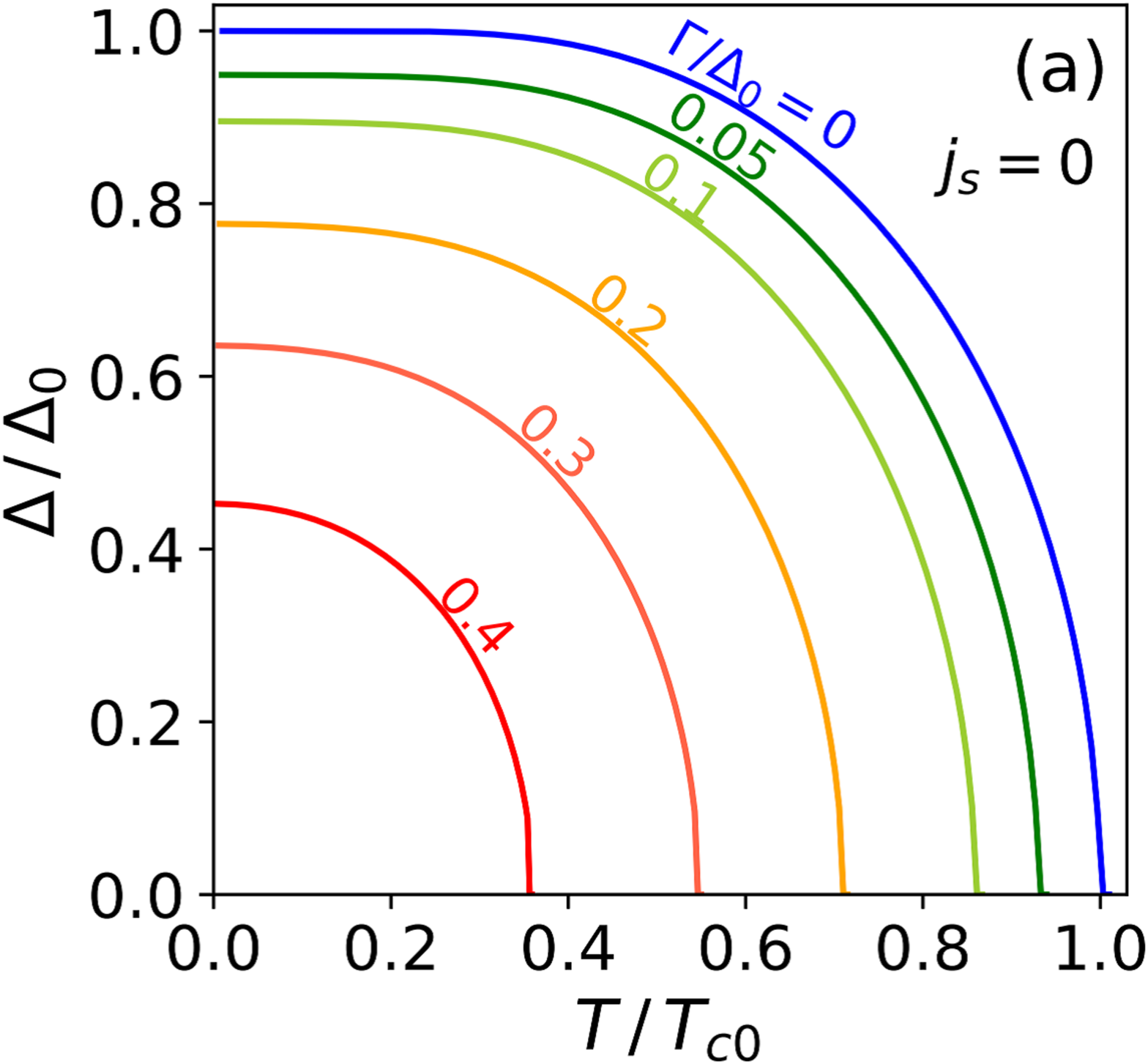}
   \includegraphics[height=0.43\linewidth]{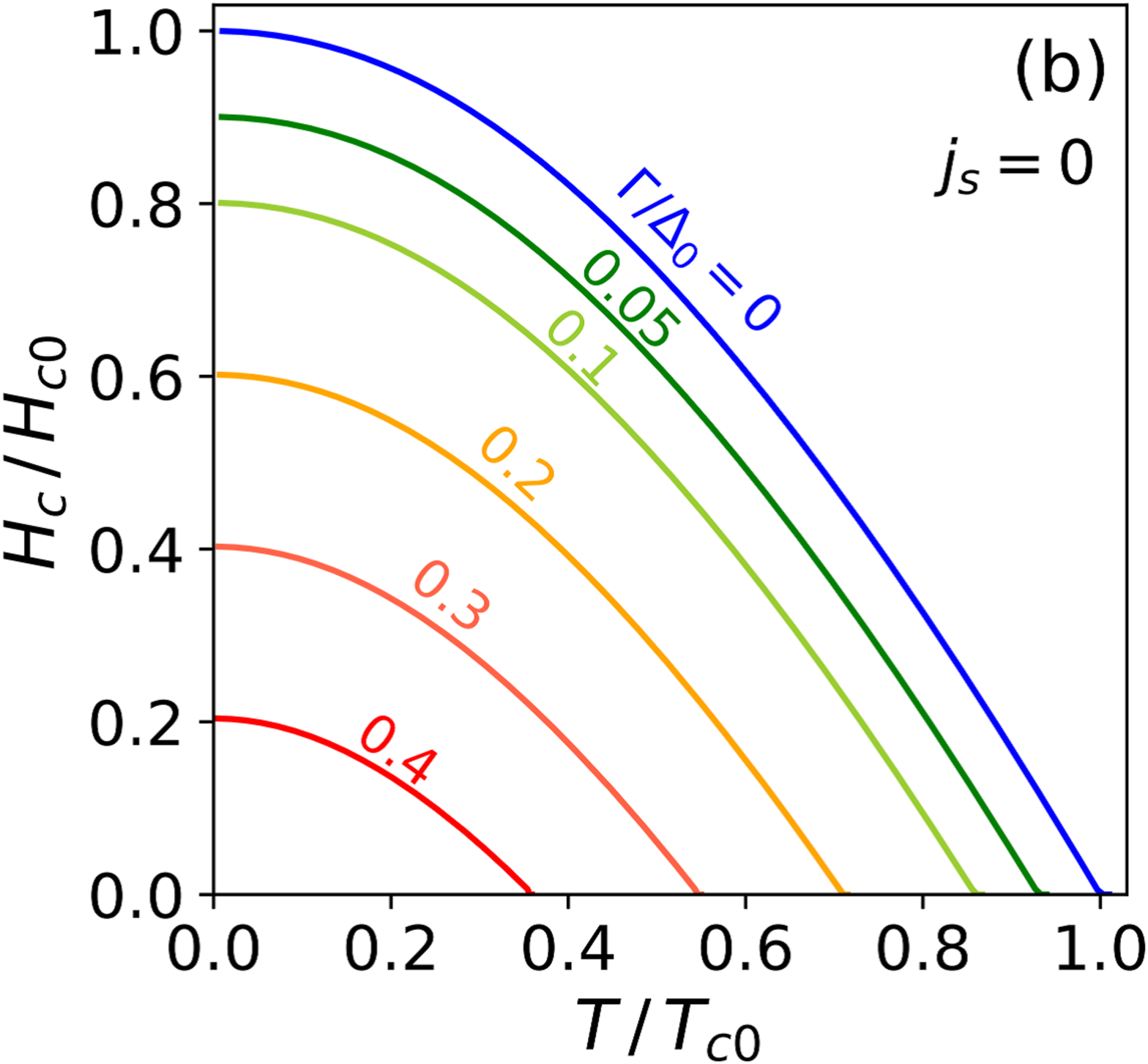}
   \includegraphics[height=0.43\linewidth]{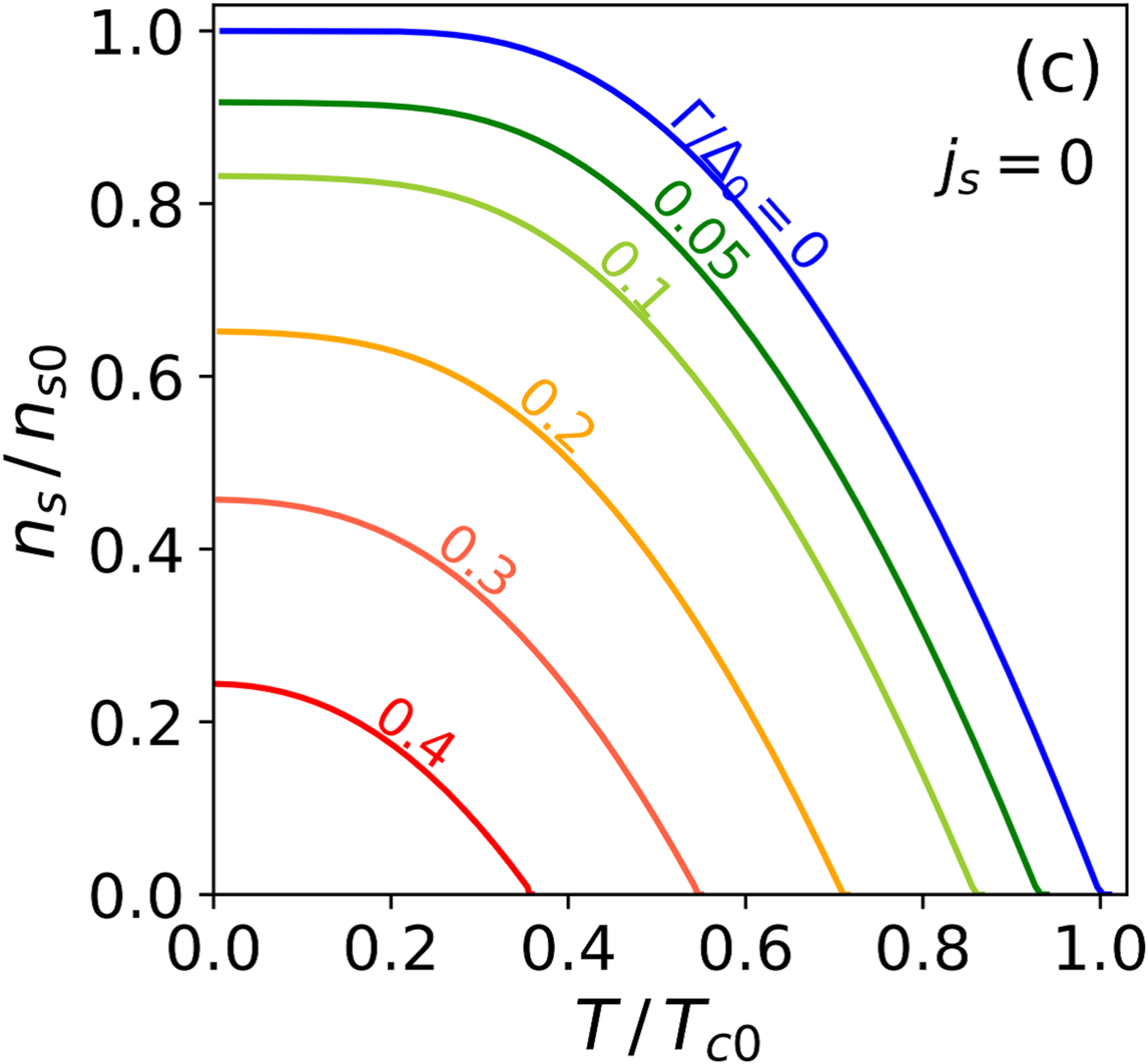}
   \includegraphics[height=0.43\linewidth]{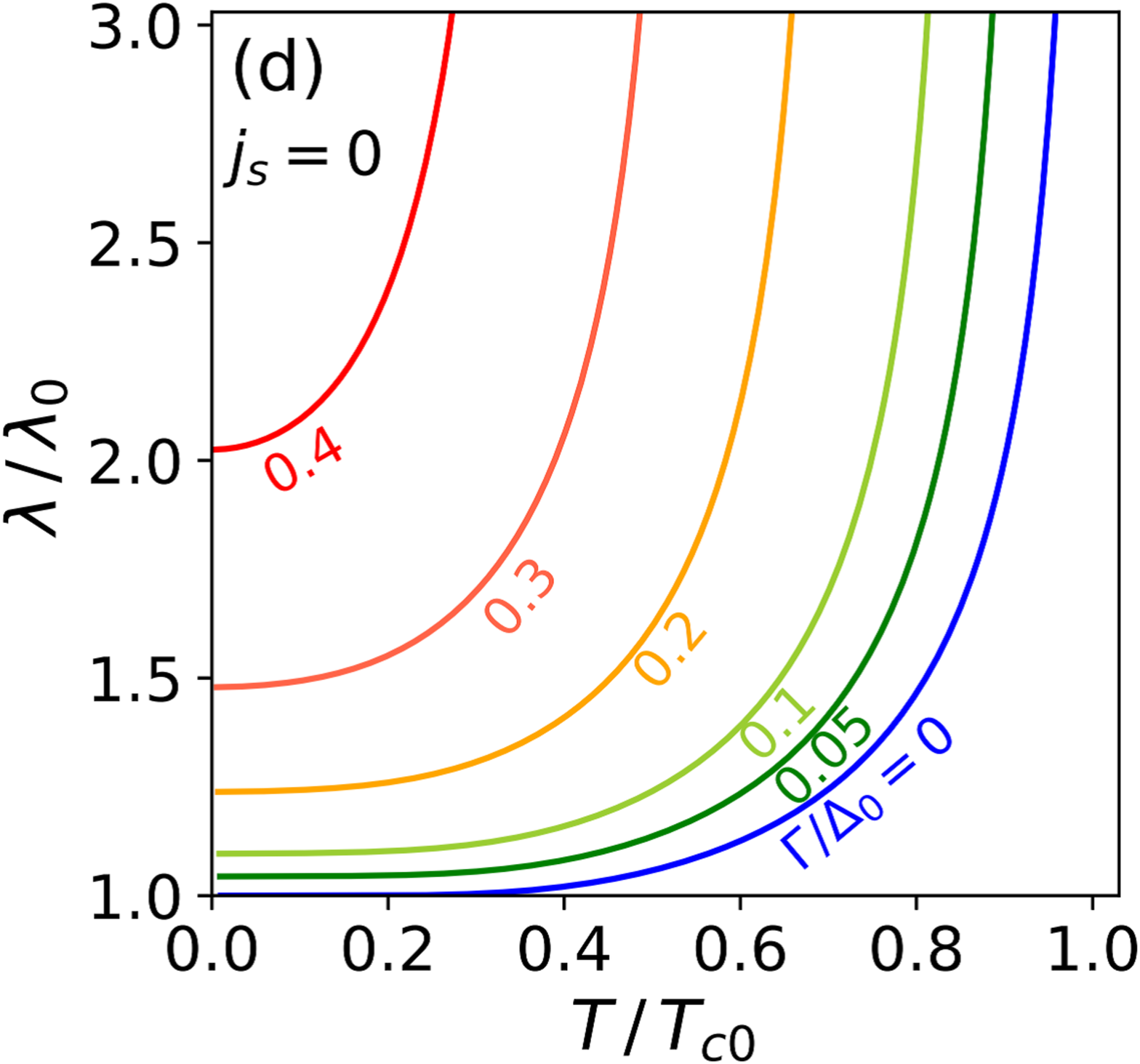} 
\end{center}\vspace{0 cm}
   \caption{
(a) $\Delta(0,\Gamma,T)$, (b) $H_c(\Gamma,T)$, (c) $n_s(0,\Gamma,T)$, and (d) $\lambda(0,\Gamma,T)$ as functions of $T$ calculated for different $\Gamma$. 
   }\label{fig4}
\end{figure}

For $0< T < T_c$, 
we numerically solve Eqs.~(\ref{thermodynamic_Usadel})-(\ref{superfluid_density}). 
Shown in Fig.~\ref{fig4} are $\Delta(0,\Gamma,T)$, $H_c(\Gamma,T)$, $n_s(0,\Gamma,T)$, and $\lambda(0,\Gamma,T)$  as functions of $T$. 
Then, $\Delta$, $H_c$, and $n_s \propto \lambda^{-2}$ are monotonically decreasing functions of $T$ and $\Gamma$~\cite{2017_Gurevich_Kubo, 2018_Herman, 2020_Kubo_jd}.

\subsection{Current-carrying state ($s> 0$)}\label{sec_current_carrying}

Now consider the current-carrying state ($s>0$). 
In addition to the Dynes pair-breaking scattering rate $\Gamma$, 
the superfluid momentum $q$ give rise to pair-breaking effects, 
suppressing $\Delta$ and $n_s$ and resulting in nonlinearity of $j_s$ respect with $q$.

For $T = 0$, the self-consistency equation results in (see Appendix~\ref{a2})
\begin{eqnarray}
&&\Delta(s,\Gamma,0) = \exp\biggl[ -\sinh^{-1}u_0  \nonumber \\
&&-\frac{s}{2\Delta(s,\Gamma,0)} \biggl( \frac{\pi}{2} -\tan^{-1} u_0 - \frac{u_0}{1+u_0^2}  \biggr) \biggr] ,  \label{self-consistency_Tzero_1} 
\end{eqnarray}
where $u_0(s,\Gamma)$ is defined by $[\Delta(s,\Gamma,0) - s/\sqrt{1+u_0^2} ] u_0 = \Gamma$. 
The superfluid density $n_s(s,\Gamma,0)$ can be calculated from Eq.~(\ref{superfluid_density}). 
Hence (see Appendix~\ref{a2})
\begin{eqnarray}
\frac{n_s(s,\Gamma,0)}{n_{s0}} 
= \frac{\lambda_0^2}{\lambda^2(s,\Gamma,0)} 
&=& \Delta(s,\Gamma,0) \biggl[ 1- \frac{2}{\pi} \tan^{-1} \! u_0 \biggr] \nonumber \\
&-& \frac{4s}{3\pi} \biggl\{ 1 - \frac{u_0 (3+2u_0^2)}{2(1+u_0^2)^{\frac{3}{2}}} \biggr\}   
\label{superfluid_density_Tzero_1}.
\end{eqnarray}
Note that Eqs.~(\ref{self-consistency_Tzero_1}) and (\ref{superfluid_density_Tzero_1}) reproduce Eqs.~(\ref{Delta_0_Gamma_0}) and (\ref{ns_0_Gamma_0}) at the zero-current limit ($s \to 0$).  
The supercurrent density $j_s$ is calculated from Eq.~(\ref{supercurrent}): 
\begin{eqnarray}
j_s(s, \Gamma, 0) = \sqrt{\pi s} \frac{n_s(s,\Gamma,0)}{n_{s0}} \frac{H_{c0}}{\lambda_0}. \label{js_Tzero_1}
\end{eqnarray}
The set of Eqs.~(\ref{self-consistency_Tzero_1})-(\ref{js_Tzero_1}) are the general formulas for $\Delta(s,\Gamma,T)|_{T=0}$, $n_s(s,\Gamma,T)|_{T=0}$, $\lambda(s,\Gamma,T)|_{T=0}$, and $j_s(s,\Gamma,T)|_{T=0}$ valid for all current and all $\Gamma$. 
When $\Gamma=0$, the well-known Maki's formulas~\cite{1963_Maki_I, 1963_Maki_II, 2012_Clem_Kogan} are reproduced.

For $\Gamma \ll 1$ such that $u_0=\Gamma/(\Delta-s) \ll 1$,  
Eqs.~(\ref{self-consistency_Tzero_1})-(\ref{js_Tzero_1}) reduce to the approximate formulas: 
\begin{eqnarray}
&& \Delta (s, \Gamma, 0) = \exp\biggl[ -\frac{\pi s}{4\Delta (s, \Gamma, 0)} - \frac{\Gamma}{\Delta(s, \Gamma, 0)} \biggr], \label{self-consistency_Tzero_2} \\
&& \frac{n_s(s, \Gamma, 0)}{n_{s0}} = \frac{\lambda_0^2}{\lambda^2(s, \Gamma, 0)}
= \Delta(s, \Gamma, 0) -\frac{4s}{3\pi} - \frac{2\Gamma}{\pi} , \label{superfluid_density_Tzero_2} \\
&& j_s (s,\Gamma,0) = \sqrt{\pi s} \biggl[ \Delta (s,\Gamma,0) - \frac{4s}{3\pi} -\frac{2\Gamma}{\pi} \biggr] \frac{H_{c0}}{\lambda_0}.  \label{js_Tzero_2}
\end{eqnarray}
For $(s,\Gamma) \ll 1$, Eqs.~(\ref{self-consistency_Tzero_2})-(\ref{js_Tzero_2}) reduce to even simpler formulas:
\begin{eqnarray}
\Delta (s, \Gamma, 0) &=& 1 -\frac{\pi s}{4} - \Gamma, \label{self-consistency_Tzero_3} \\
\frac{n_s(s, \Gamma, 0)}{n_{s0}} &=& \frac{\lambda_0^2}{\lambda^2(s, \Gamma, 0)} \nonumber \\
&=& 1 -\biggl( \frac{\pi}{4} + \frac{4}{3\pi} \biggr) s - \biggl( 1+ \frac{2}{\pi} \biggr) \Gamma , \label{superfluid_density_Tzero_3}  \\
j_s (s,\Gamma,0) &=& \sqrt{\pi s} \biggl[  1 -\biggl( \frac{\pi}{4} + \frac{4}{3\pi} \biggr) s - \biggl( 1+ \frac{2}{\pi} \biggr) \Gamma \biggr] \frac{H_{c0}}{\lambda_0} . \nonumber \\  \label{js_Tzero_3}
\end{eqnarray}
Eqs.~(\ref{self-consistency_Tzero_2})-(\ref{js_Tzero_2}) are valid under an arbitrary current density as long as $\Gamma$ satisfies $\Gamma/(\Delta-s) \ll 1$. 
Eqs.~(\ref{self-consistency_Tzero_3})-(\ref{js_Tzero_3}) are useful when we consider the small-current regime.

\begin{figure}[tb]
   \begin{center}
   \includegraphics[height=0.44\linewidth]{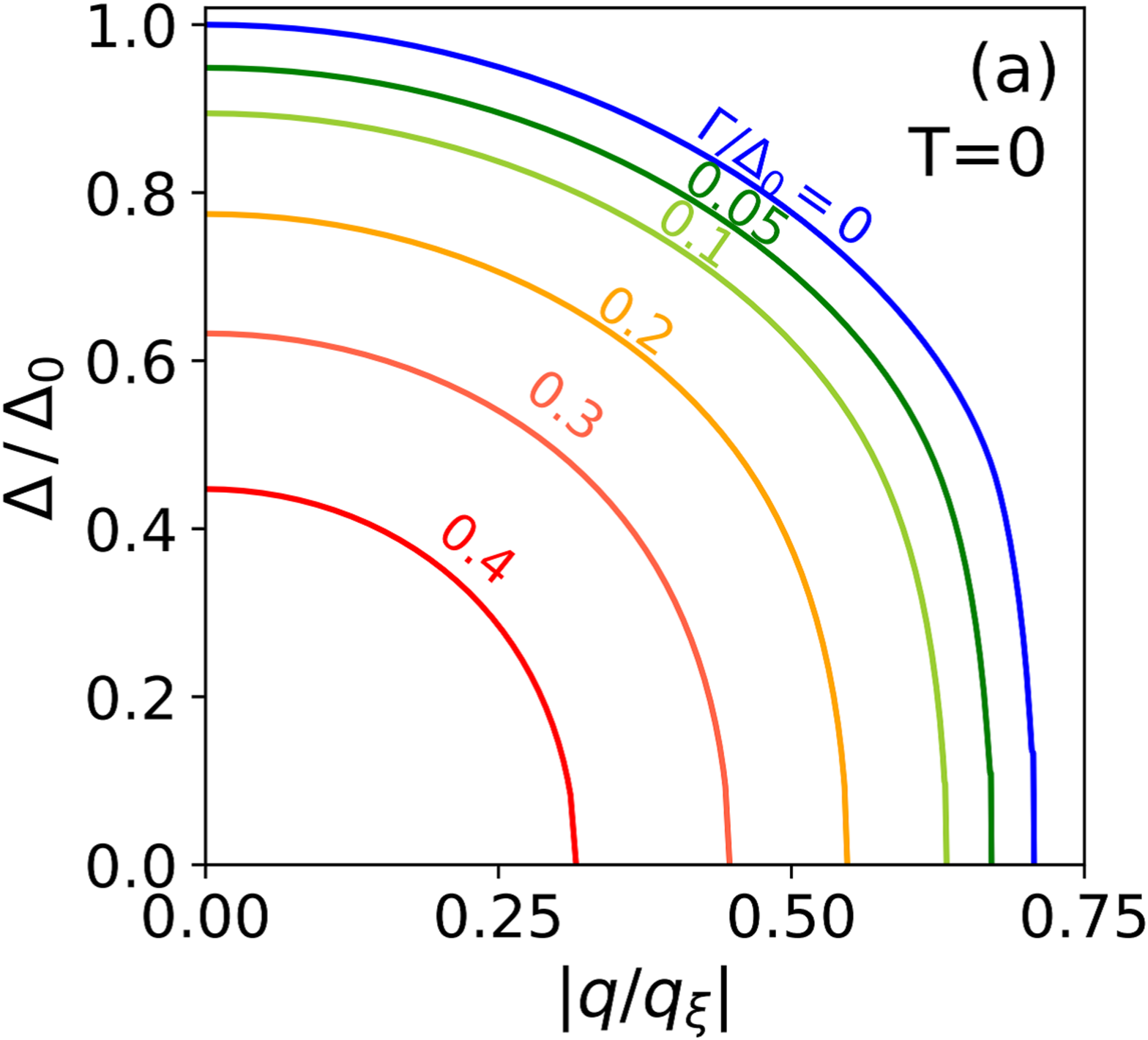}  
   \includegraphics[height=0.44\linewidth]{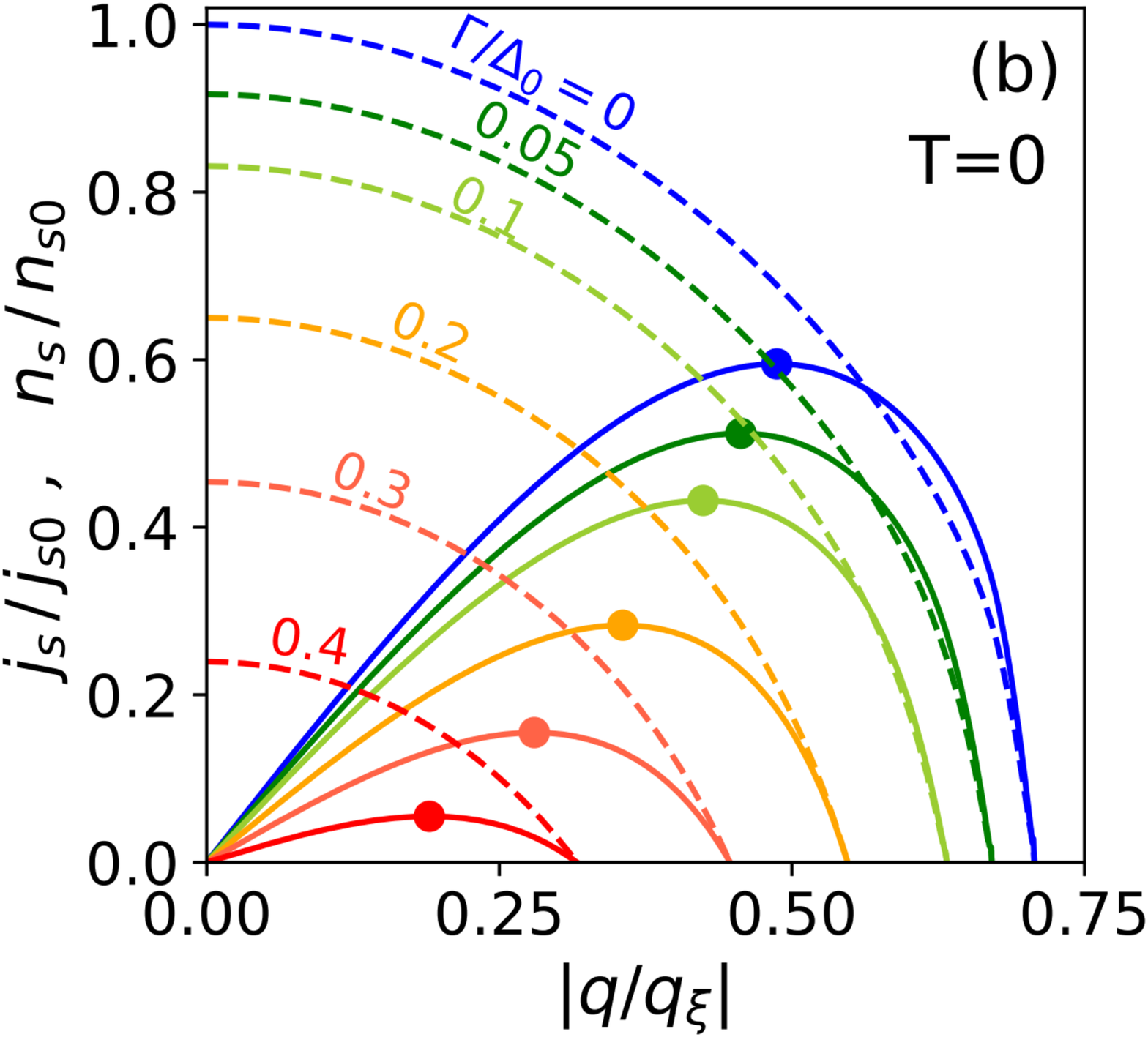}
\end{center}\vspace{0 cm}
   \caption{
$\Delta(s,\Gamma,T)|_{T= 0}$, $n_s(s,\Gamma,T)|_{T= 0}$, and $j_s(s,\Gamma,T)|_{T= 0}$ calculated from the formulas given by Eqs.~(\ref{self-consistency_Tzero_1})-(\ref{js_Tzero_1}). 
(a) $\Delta$, (b) $j_s$ (solid curves) and $n_s$ (dashed curves) as functions of $|q/q_{\xi}| (= \sqrt{s})$ for different $\Gamma$. 
Each colored blob represents the maximum current density, namely, the deparing current density. 
   }\label{fig5}
\end{figure}

Shown in Fig.~\ref{fig5} are $\Delta$, $n_s$, $j_s$ at $T=0$ as functions of the superfluid momentum $|q/q_{\xi}|=\sqrt{s}$ calculated from the exact formulas given by Eqs.~(\ref{self-consistency_Tzero_1})-(\ref{js_Tzero_1}). 
While $\Delta$ and $n_s (\propto \lambda^{-2})$ are monotonically decreasing functions, 
$j_s$ exhibits non-monotonic behaviors. 
At smaller $|q|$ regions, $j_s$ is proportional to $|q|$. 
As $|q|$ increases, $j_s$ becomes dominated by a rapid reduction of $n_s$ and ceases to increase. 
At a threshold value $q_d$ or $s_d = (q_d/q_{\xi})^2$, $j_s$ reaches the maximum value (colored blob), 
which is the depairing current density $j_d$.

\begin{figure}[tb]
   \begin{center}
   \includegraphics[height=0.44\linewidth]{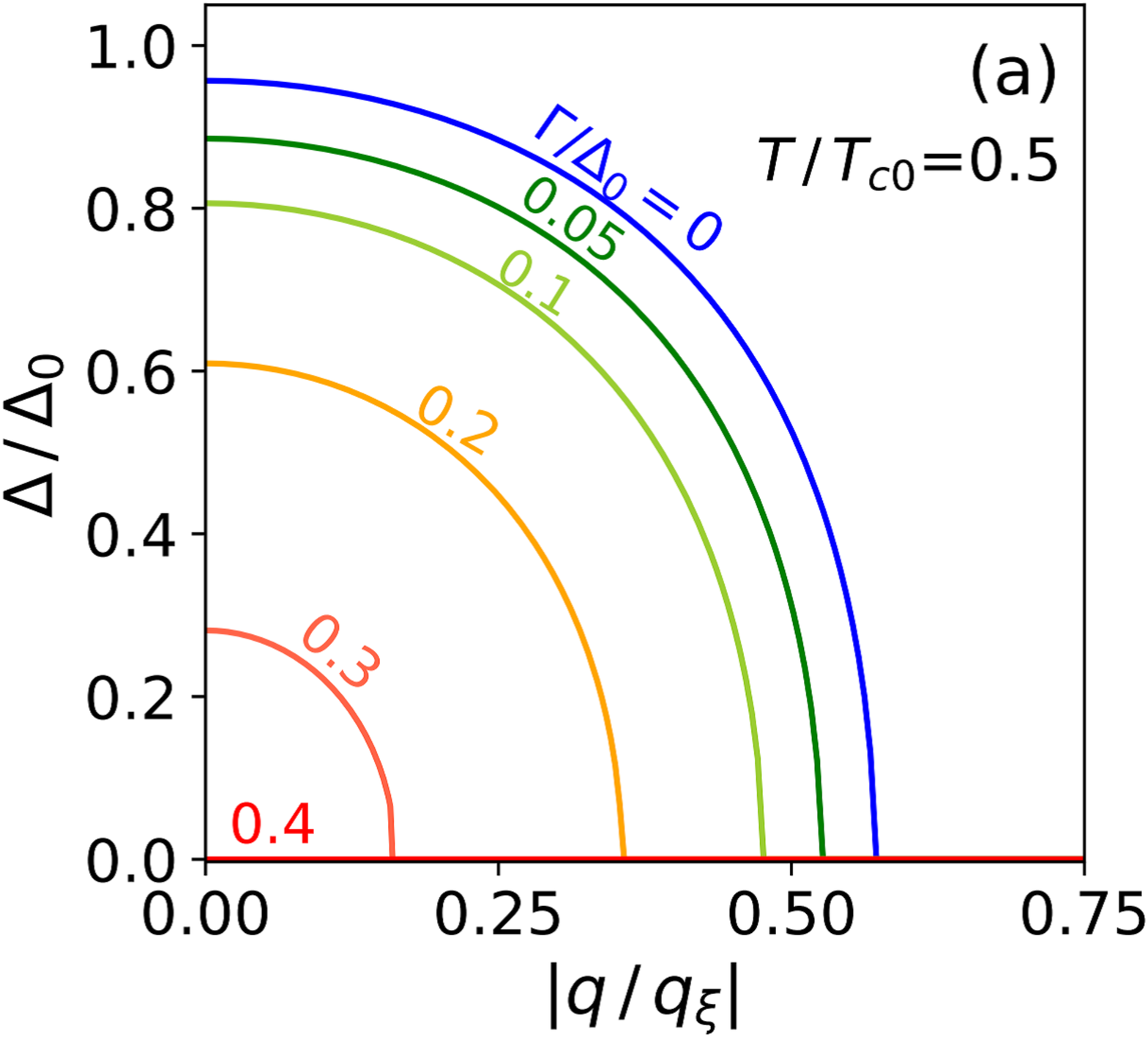}  
   \includegraphics[height=0.44\linewidth]{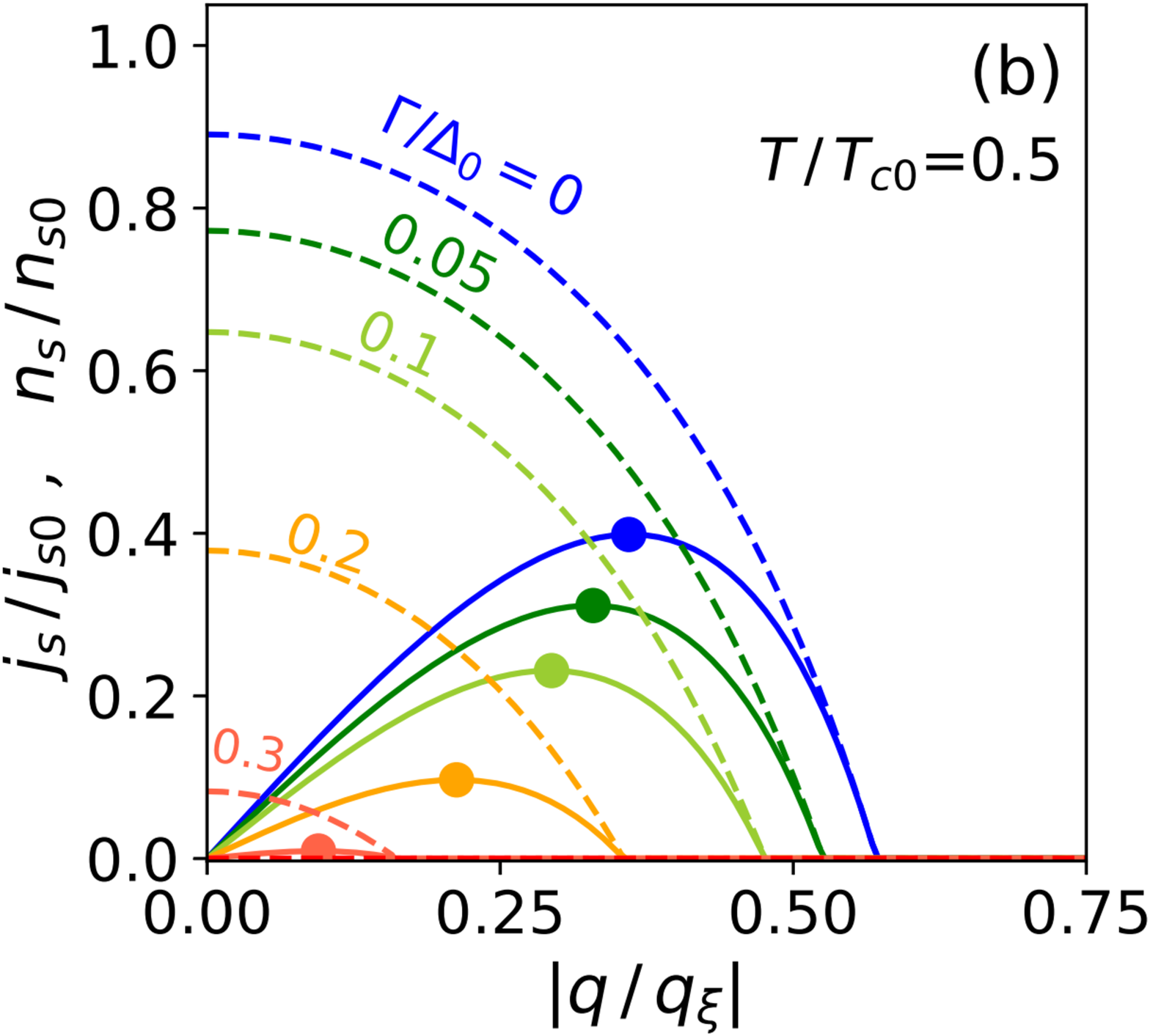}
\end{center}\vspace{0 cm}
   \caption{
$\Delta (s, \Gamma, T)$, $n_s (s, \Gamma, T)$, and $j_s (s, \Gamma, T)$ at a finite temperature obtained by numerically solving Eqs.~(\ref{thermodynamic_Usadel})-(\ref{supercurrent}).  
(a) $\Delta$, (b) $j_s$ (solid curves) and $n_s$ (dashed curves). 
Note that a superconductor with $\Gamma=0.4$ (red) has $T_c=0.36 T_{c0} < T$, being normal for all $q$. 
   }\label{fig6}
\end{figure}

For $T\simeq T_c$, the GL regime, 
$\Delta$ in the current-carrying state is calculated from Eq.~(\ref{GL}). 
Then, we obtain~\cite{2020_Kubo_jd}
\begin{eqnarray}
&&\Delta (s, \Gamma, T) 
= \sqrt{ \frac{8\pi^2 T_{c}^2}{7\zeta(3)} 
\biggl(1- \frac{T}{T_{c}} \biggr)  \biggl( 1-\frac{s}{s_m} \biggr) }, \label{GL_Delta}  \\
&&\frac{n_s (s, \Gamma, T)}{n_{s0}} 
= \frac{\lambda_0^{2}}{\lambda^{2} (s, \Gamma, T)}
=\frac{\Delta^2 (s, \Gamma, T)}{2T_{c}} ,  \label{GL_ns} \\
&&j_s (s, \Gamma, T) 
= \sqrt{ \frac{\pi}{2(T_{c}-T)} } \sqrt{s} \biggl( 1- \frac{s}{s_m} \biggr) \frac{H_c(\Gamma, T)}{\lambda(0, \Gamma, T)} . \label{GL_js}
\end{eqnarray}
for $T \simeq T_c(\Gamma)$. 
Here $s_m(\Gamma,T) = (4T_{c}/\pi) (1-T/T_{c})$.

For $0< T < T_c$, 
we need to numerically solve Eqs.~(\ref{thermodynamic_Usadel})-(\ref{supercurrent}) for a finite $s$. 
Shown in Figs.~\ref{fig6} (a) and \ref{fig6} (b) are $\Delta (s, \Gamma, T)$, $n_s (s, \Gamma, T)$, and $j_s (s, \Gamma, T)$ as functions of $|q/q_{\xi}|=\sqrt{s}$ at $T/T_{c0}=0.5$ for different $\Gamma$. 
The similar calculations for other $T$ are straightforward (see also Ref.~\cite{2020_Kubo_jd}).  
The $q$ dependences of $\Delta$, $n_s$, and $j_s$ resemble those for $T=0$ (see Fig.~\ref{fig5}). 
As $T$ increases, all these values monotonically decrease,  
and the depairing current densities (colored blobs) also shift to lower values.

\section{Thin and narrow film}\label{sec_film}

In this section, we consider the geometry shown in Fig.~\ref{fig2} (a): a narrow thin-film. 
We calculate the depairing current density and the kinetic inductance using the results obtained in Sec.~\ref{sec_solutions}. 

\subsection{Depairing current density}\label{sec_film_jd}

\begin{figure}[tb]
   \begin{center}
   \includegraphics[width=0.494\linewidth]{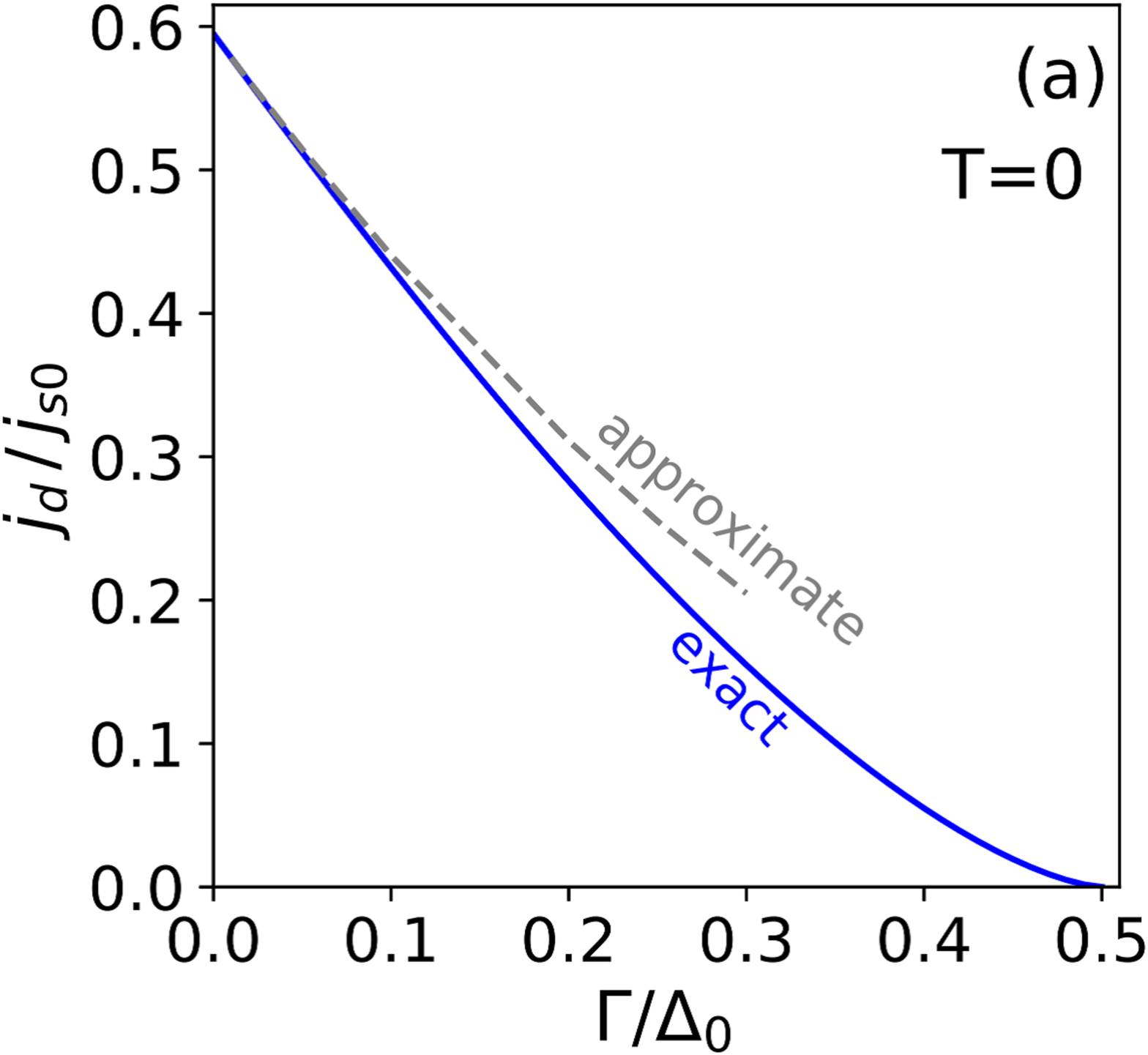}  
   \includegraphics[width=0.494\linewidth]{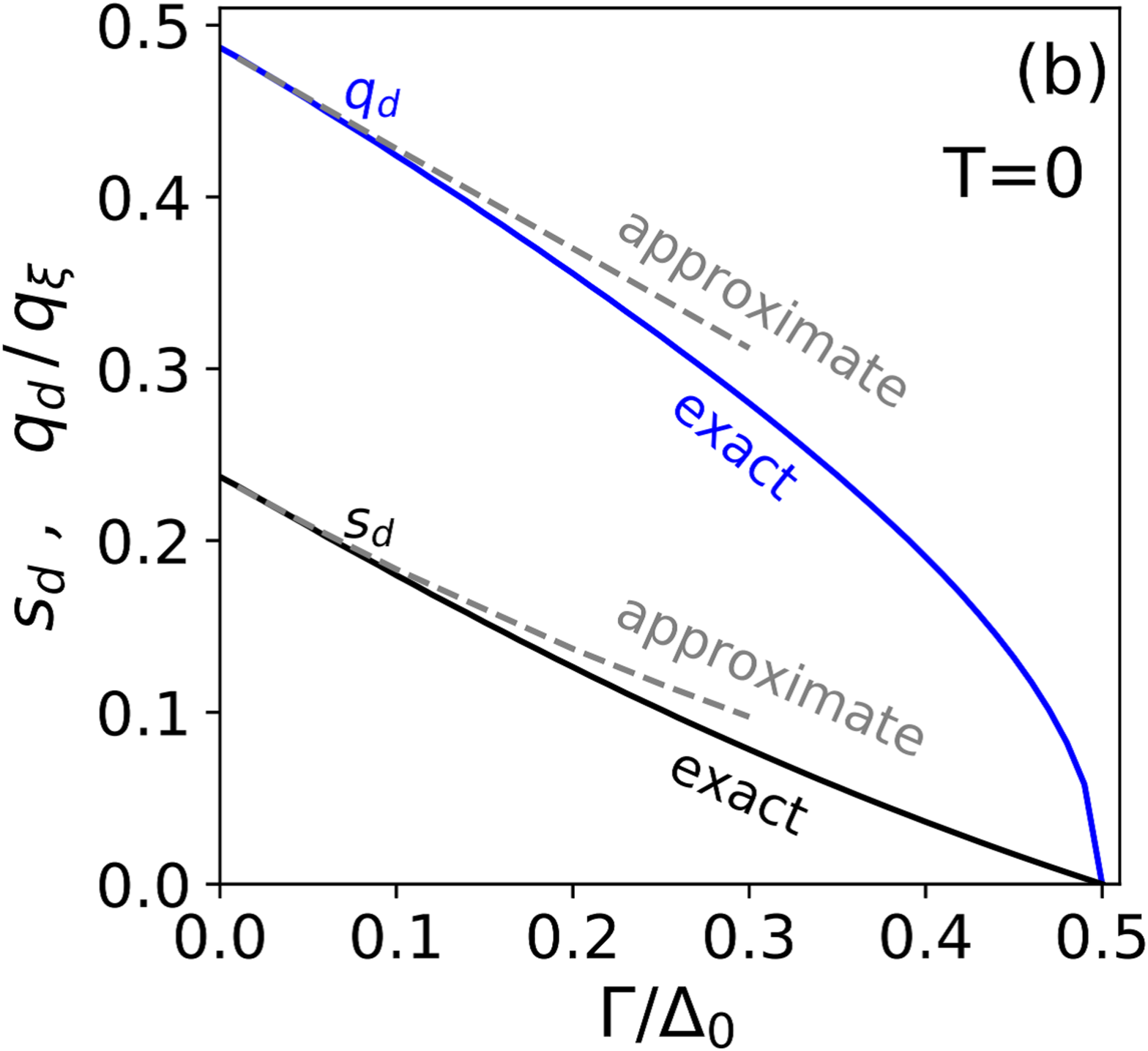}
\end{center}\vspace{0 cm}
   \caption{
(a) Depairing current density $j_d$ at $T=0$ as a function of $\Gamma$. 
(b) Depairing momentum $q_d$ and $s_d = (q_d/q_{\xi})^2$ at $T=0$ as functions of $\Gamma$. 
The solid curves represent the exact results calculated from Eqs.~(\ref{self-consistency_Tzero_1})-(\ref{js_Tzero_1}). 
The dashed curves are calculated from the approximate formulas given by Eqs.~(\ref{jd_approx})-(\ref{alpha}). 
   }\label{fig7}
\end{figure}

For $T=0$, the depairing current density $j_d$ is already obtained for some $\Gamma$ values (colored blobs in Fig.~\ref{fig5}). 
That for an arbitrary $\Gamma$ can also be calculated from the formulas, Eqs.~(\ref{self-consistency_Tzero_1})-(\ref{js_Tzero_1}), by finding the maximum values of $j_s$.  
Shown as the solid curves in Fig.~\ref{fig7} are $j_d$, $q_d$ and $s_d= (q_d/q_{\xi})^2$ as functions of $\Gamma$. 
An increase of $\Gamma$ leads to decreases of $n_s$ and $j_s$, 
resulting in a monotonic decrease of $j_d$.

For $\Gamma \ll 1$, we can derive an analytical formula for $j_d$. 
Using Eqs.~(\ref{self-consistency_Tzero_2})-(\ref{js_Tzero_2}) and the condition $\partial_s j_s =0$, 
we find (see Appendix~\ref{a3})
\begin{eqnarray}
&&j_d(\Gamma, 0) = \sqrt{\pi s_d} \biggl[ \Delta_d - \frac{4s_d}{3\pi} -\frac{2\Gamma}{\pi} \biggr] \frac{H_{c0}}{\lambda_0} , \label{jd_approx} \\
&& \Delta_{d} = \Delta_{d0} \exp \biggl[ \biggl( \frac{\pi \alpha}{4} -1 \biggr)\frac{\Gamma}{\Delta_{d0}} \biggr] ,  \label{Delta_d} \\
&& s_{d} = (s_{d0} -\alpha \Gamma) \exp \biggl[ \biggl( \frac{\pi \alpha}{4} -1 \biggr)\frac{\Gamma}{\Delta_{d0}} \biggr]  , \label{sd} \\
&&\zeta_{d0}= \frac{2}{\pi} + \frac{3\pi}{8} - \sqrt{\biggl( \frac{2}{\pi} + \frac{3\pi}{8} \biggr)^2 -1}  = 0.300 ,  
\\
&& \Delta_{d0} = e^{-\frac{\pi}{4} \zeta_{d0}} = 0.790, \label{Delta_d0} \\
&& s_{d0} = \Delta_{d0} \zeta_{d0} = 0.237 , \label{s_d0} \\
&& \alpha = \frac{1+ \pi/2 -2 (1+8/\pi) \zeta_{d0}}{2+ 3\pi^2/8 - \pi \zeta_{d0}} = 0.365 .\label{alpha}
\end{eqnarray}
Shown as the dashed curves in Fig.~\ref{fig7} are calculated from Eqs.~(\ref{jd_approx})-(\ref{alpha}), 
which agree well with the exact results (solid curves) at $\Gamma \ll 1$.  
For $\Gamma=0$, we have $\Delta_d = \Delta_{d0} = 0.790$ and $s_d=s_{d0} = 0.237$, 
reproducing the well-known result, $j_d (0, 0)=0.595 H_{c0}/\lambda_0$~\cite{1963_Maki_I, 1963_Maki_II, 2012_Clem_Kogan}.

For $T\simeq T_c$, we can use Eqs.~(\ref{GL_Delta})-(\ref{GL_js}). 
The condition $\partial_s j_s =0$ yields $s_d = s_m/3 = (4T_{c}/3\pi) (1-T/T_{c})$, 
and we have~\cite{2020_Kubo_jd}
\begin{eqnarray}
j_d (\Gamma, T) 
= \biggl( \frac{2}{3} \biggr) ^{\frac{3}{2}}\frac{H_c(\Gamma, T)}{\lambda(0, \Gamma, T)} 
= 0.544 \frac{H_c(\Gamma, T)}{\lambda(0, \Gamma, T)}  . \label{jd_GL_1}
\end{eqnarray}
Eq.~(\ref{jd_GL_1}) can be written as
\begin{eqnarray}
j_d (\Gamma, T) = j_{d0}^{\rm GL} \biggl( 1 - \frac{T}{T_c} \biggr)^{\frac{3}{2}} \label{jd_GL_2} , 
\end{eqnarray}
where
\begin{eqnarray}
j_{d0}^{\rm GL} =\frac{16 j_{s0}}{21 \zeta(3)} \sqrt{\frac{\pi}{3}} 
\biggl( \frac{e^{\gamma_E}T_c}{T_{c0}} \biggr)^{\frac{3}{2}} 
= \frac{8\pi^2 \sqrt{2\pi}}{21 \zeta(3) e} \sqrt{\frac{(k_B T_c)^3}{\hbar v_F \rho (\rho/\tau)}}
 ,\nonumber \\ 
\label{jd0_GL}
\end{eqnarray}
Eqs.~(\ref{jd_GL_1})-(\ref{jd0_GL}) have the same form as the well-known GL depairing current density except $T_c$ depends on $\Gamma$ through Eq.~(\ref{Tc}).

\begin{figure}[tb]
   \begin{center}
   \includegraphics[width=0.494\linewidth]{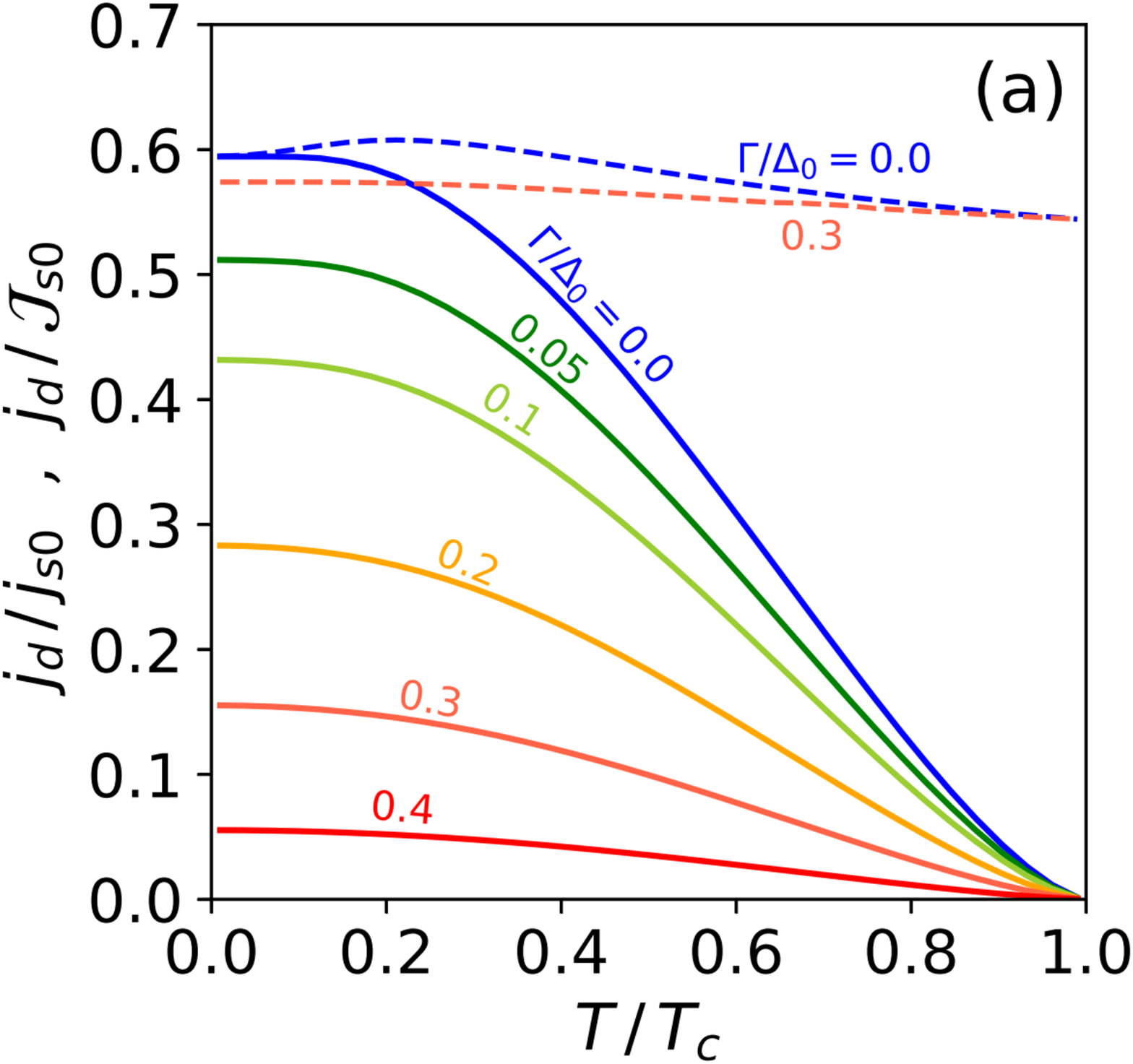}  
   \includegraphics[width=0.494\linewidth]{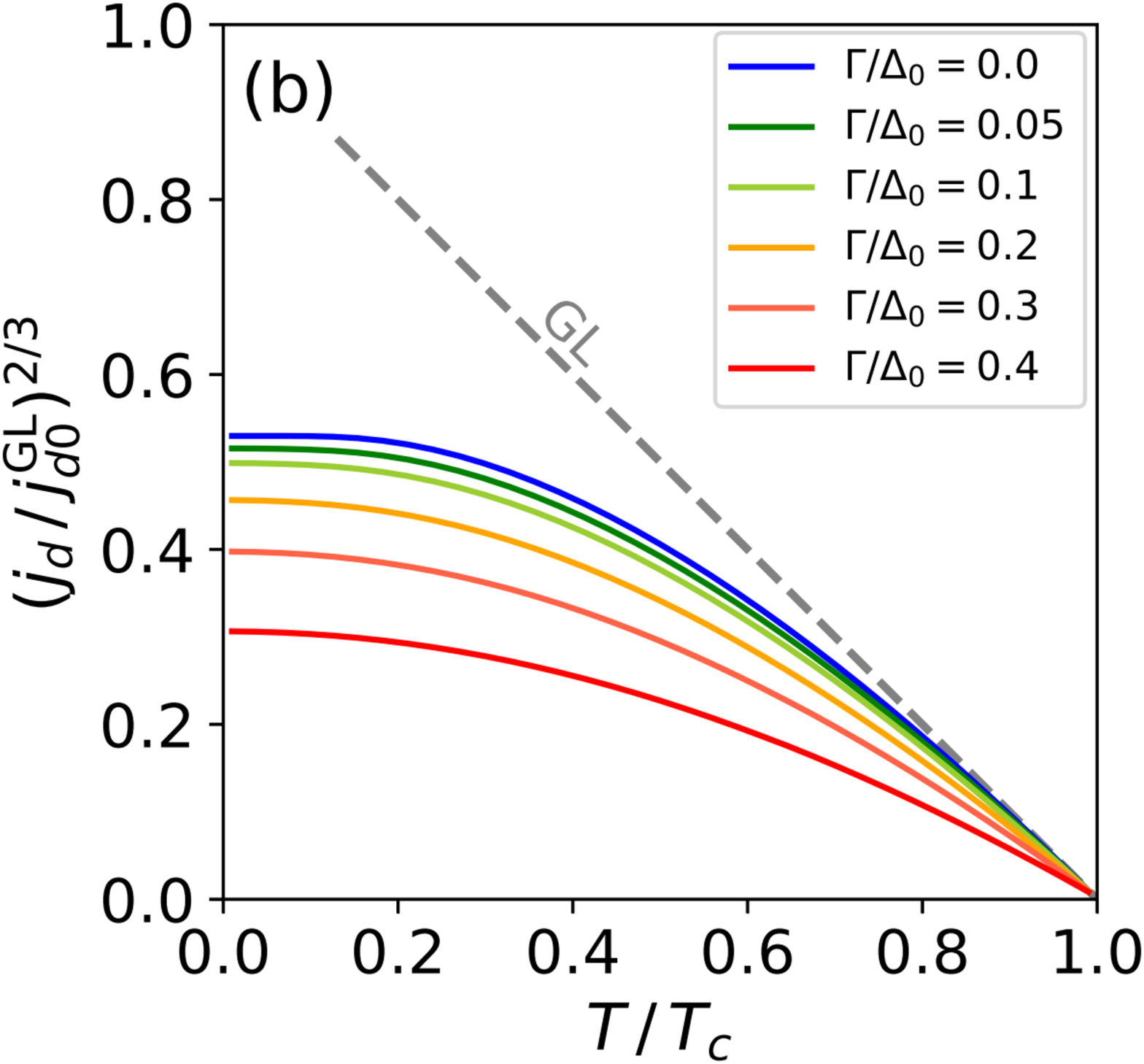}
\end{center}\vspace{0 cm}
   \caption{
(a) Depairing current density $j_d(\Gamma,T)/j_{s0}$ and $j_d(\Gamma,T)/\mathcal{J}_{s0}(\Gamma,T)$ as functions of $T/T_c$ for different $\Gamma$. 
Here $\mathcal{J}_{s0}(\Gamma,T)=H_{c}(\Gamma,T)/\lambda(0,\Gamma,T)$ and $j_{s0}=\mathcal{J}_{s0}(0,0)=H_{c0}/\lambda_0$. 
(b) Depairing current density $(j_d/j_{d0}^{\rm GL})^{2/3}$ as functions of $T/T_c$. 
Here $j_{d0}^{\rm GL}$ is defined by Eq.~(\ref{jd0_GL}). 
The dashed gray line is the GL result [Eq.~(\ref{jd_GL_2})] extrapolated to lower $T$ regions. 
   }\label{fig8}
\end{figure}

For $0 < T < T_c$, we can obtain $j_d(\Gamma, T)$ by numerically solving Eqs.~(\ref{thermodynamic_Usadel})-(\ref{supercurrent}) for different $\Gamma$ and $T$ and finding the maximum value of $j_s$. 
Shown as solid curves in Fig.~\ref{fig8} (a) are $j_d(\Gamma,T)$ as functions of $T$ for different $\Gamma$. 
The Kupriyanov-Lukichev-Maki's result~\cite{1963_Maki_I, 1963_Maki_II, 1980_Kupriyanov} is reproduced for the ideal BCS superconductor with $\Gamma=0$ (blue curve), 
while a finite $\Gamma$ reduces $j_d$ for all $T$. 
The dashed curves represent $j_d(\Gamma,T)/\mathcal{J}_{s0}(\Gamma,T)$, 
where $\mathcal{J}_{s0}(\Gamma,T)=H_{c}(\Gamma,T)/\lambda(0,\Gamma,T)$.  
The curves merge at $T \to T_c$ and take 0.544, consistent with Eq.~(\ref{jd_GL_1}). 
It is sometimes useful to plot $(j_d/j_{d0}^{\rm GL})^{2/3}$ as functions of $T/T_c$ (see, e.g., Refs.~\cite{1982_Romijn, 2004_Rusanov, 2019_Sun}). 
Shown as solid curves in Fig.~\ref{fig8} (b) are $(j_d/j_{d0}^{\rm GL})^{2/3}$ obtained by numerically solving Eqs.~(\ref{thermodynamic_Usadel})-(\ref{supercurrent}). 
The dashed line represents the GL result given by Eq.~(\ref{jd_GL_2}), which is valid only at $T\simeq T_c$.

\subsection{Kinetic inductance}\label{sec_film_Lk}

The kinetic inductance of a narrow thin-film is given by $L_{\rm film} = (\ell /Wd) L_k$ for the length $\ell$, width $W$, and thickness $d$. 
The kinetic inductivity $L_k$ is defined by $L_k \dot{j}_s = -\dot{A}$~\cite{1989_Anlage, 2012_Clem_Kogan}, 
the dot denotes differentiation with respect to the time $t$,  
and $A=\hbar q/2|e|$ is the vector potential. 
Hence, 
\begin{eqnarray}
L_k (s, \Gamma, T)
= \frac{\hbar (-\dot{q}/\dot{j_s})}{2|e|}  
= \mu_0\lambda_0^2 \frac{\dot{q}}{ \dot{q} \frac{n_s}{n_{s0}} + q \frac{\dot{n}_s}{n_{s0}} } .
\label{Lk_formula0}
\end{eqnarray}
Here $\hbar q_{\xi}/2\sqrt{\pi}|e|j_{s0} = \mu_0 \lambda_0^2$ is used. 
In the following, we investigate $L_k$ for the zero-current limit and the current-carrying state in the fast- and the slow-measurement regimes~\cite{1989_Anlage, 2012_Clem_Kogan} for all $T$, all $\Gamma$, and all $j_s$ up to $j_d(\Gamma)$.

\subsubsection{Zero-current limit ($s \to 0$)} \label{sec_film_Lk_Zero}

\begin{figure}[tb]
   \begin{center}
   \includegraphics[height=0.435\linewidth]{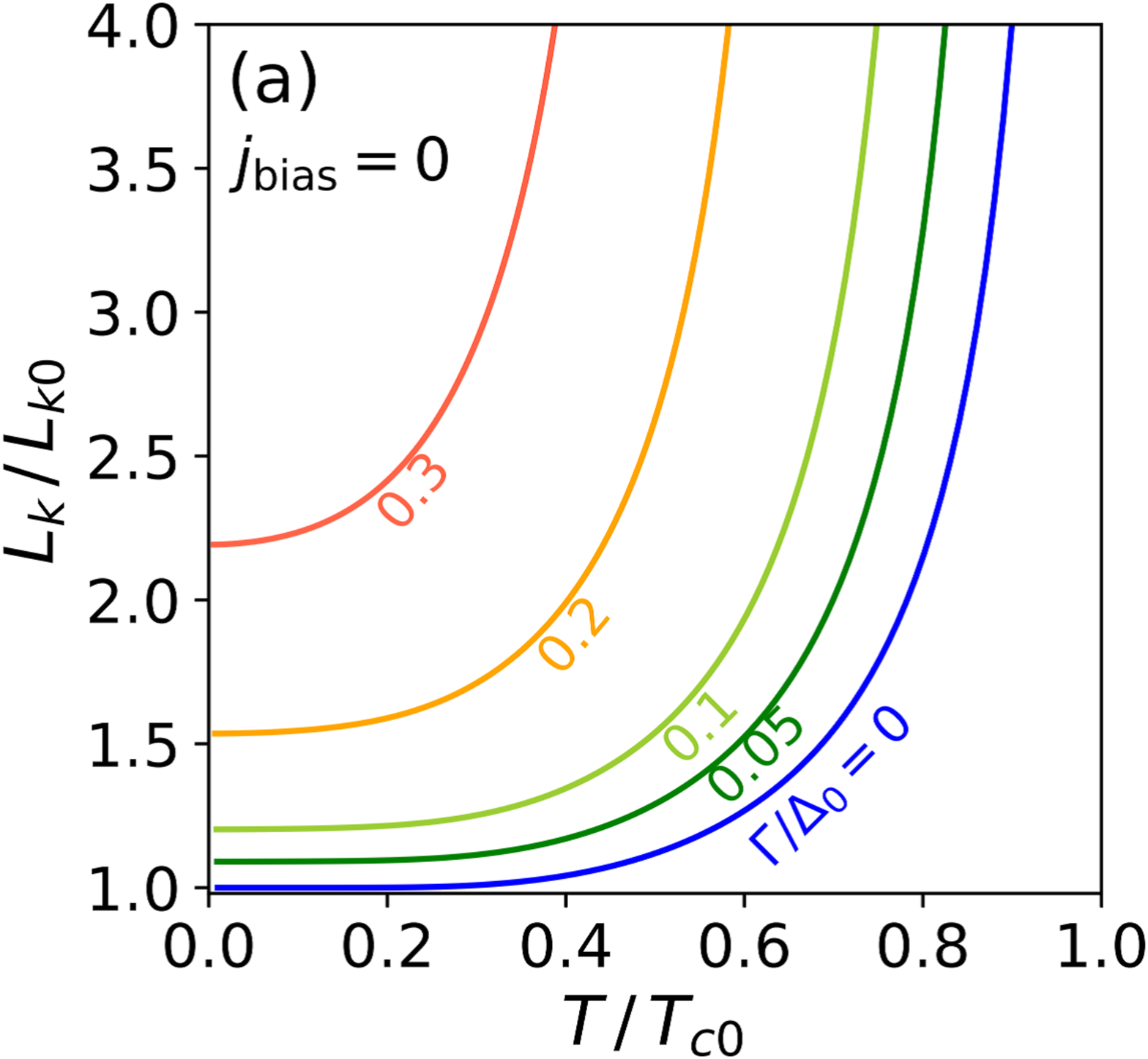}
   \includegraphics[height=0.435\linewidth]{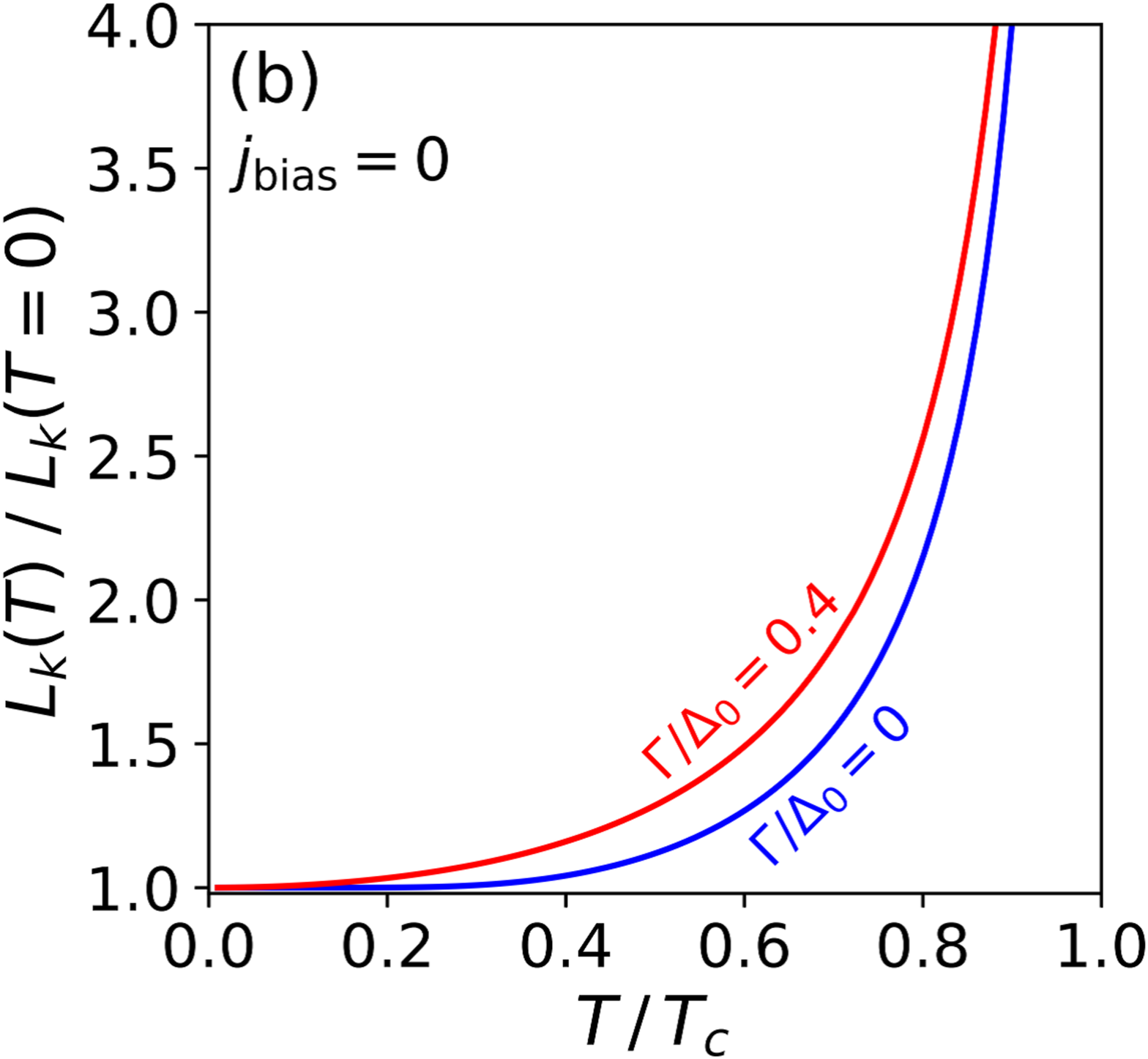}
\end{center}\vspace{0 cm}
   \caption{
Zero-current kinetic inductance $L_k (0, \Gamma, T)$ calculated from the formula given by Eq.~(\ref{Lk_0_Gamma_T}). 
(a) $L_k(0, \Gamma, T)$ as functions of $T$. Here, $L_{k0}=L_k(0,0,0)=\mu_0 \lambda_0^2$. 
(b) $L_k(0,\Gamma,T)/L_k(0,\Gamma,0)$ as functions of the normalized temperature $T/T_c$. 
   }\label{fig9}
\end{figure}

We start from the zero-current limit. 
Taking $q \to 0$, Eq.~(\ref{Lk_formula0}) reduces to $L_k(0,\Gamma,T)=\mu_0\lambda_0^2 n_{s0}/n_s(0,\Gamma,T) =\mu_0\lambda^2 (0,\Gamma,T)$. 
Using the analytical expression for $n_s(0,\Gamma,T)$~\cite{2017_Gurevich_Kubo, 2020_Kubo_jd}, 
we find the general formula for the zero-current kinetic inductance, 
\begin{eqnarray}
L_k(0, \Gamma, T) 
= \frac{\mu_0 \lambda_0^2}{\frac{2\Delta(0, \Gamma, T)}{\pi} {\rm Im} \psi \bigl[\frac{1}{2}+  \frac{\Gamma+i\Delta(0, \Gamma, T)}{2\pi T} \bigr] } ,  \label{Lk_0_Gamma_T}
\end{eqnarray}
where $\Delta(0,\Gamma,T)$ can be calculated from Eq.~(\ref{self-consistency}) (see also Fig.~\ref{fig4}). 
For the ideal BCS superconductor ($\Gamma=0$), 
Eq.~(\ref{Lk_0_Gamma_T}) reproduces the well-known result~\cite{2010_Annunziata, 2012_Clem_Kogan, 2016_Santavicca}: 
\begin{eqnarray}
L_k(0, 0, T) 
= \mu_0 \lambda_0^2 \frac{1}{\tanh \frac{\Delta(0,0,T)}{2T}}
 .  \label{Lk_0_0_T}
\end{eqnarray}
For $T=0$ and $\Gamma \ge 0$, Eq.~(\ref{Lk_0_Gamma_T}) reduces to
\begin{eqnarray}
L_k(0, \Gamma, 0) = \frac{\mu_0 \lambda_0^2}{\Delta(0,\Gamma,0) \bigl[ 1- \frac{2}{\pi} \tan^{-1} \frac{\Gamma}{\Delta(0,\Gamma,0)} \bigr]} ,  \label{Lk_0_Gamma_0}
\end{eqnarray}
where $\Delta(0,\Gamma,0)$ can be calculated from Eq.~(\ref{Delta_0_Gamma_0}). 
At $T=\Gamma=0$, Eqs.~(\ref{Lk_0_0_T}) and (\ref{Lk_0_Gamma_0}) yield 
\begin{eqnarray}
L_{k0} = L_k(0, 0, 0) = \mu_0 \lambda_0^2 .
\end{eqnarray}
In the GL regime, we can use Eq.~(\ref{GL_ns_zero_current}): 
\begin{eqnarray}
L_k(0, \Gamma, T)
= \frac{7\zeta(3) \mu_0 \lambda_0^2}{4\pi^2 T_c} \biggl( 1-\frac{T}{T_c}\biggr)^{-1} .
\label{Lk_0_Gamma_T_GL}
\end{eqnarray}
for $T\simeq T_c$.

Shown in Fig.~\ref{fig9} (a) are $L_k$ in the zero-current state as functions of $T$ for different $\Gamma$ calculated from Eq.~(\ref{Lk_0_Gamma_T}). 
As $T$ or $\Gamma$ increase, the superfluid density $n_s$ decreases, 
and $L_k$ increases and diverges at $T=T_c$ or $\Gamma=1/2$. 
Shown in Fig.~\ref{fig9} (b) are $L_k(0,\Gamma,T)$ normalized with $L_k(0,\Gamma,T)|_{T=0}$ as functions of $T/T_c(\Gamma)$. 
The blue curve, which represents the ideal BCS superconductor with $\Gamma=0$, 
obeys Eq.~(\ref{Lk_0_0_T}). 
The red curve ($\Gamma=0.4$) represents a superconductor with large densities of subgap states (see also Fig.~\ref{fig1}), 
which grows up with $T/T_c$ faster the ideal one. 

\subsubsection{Current carrying state ($s>0$): Fast measurement} \label{sec_film_Lk_Fast}

\begin{figure}[tb]
   \begin{center}
   \includegraphics[height=0.435\linewidth]{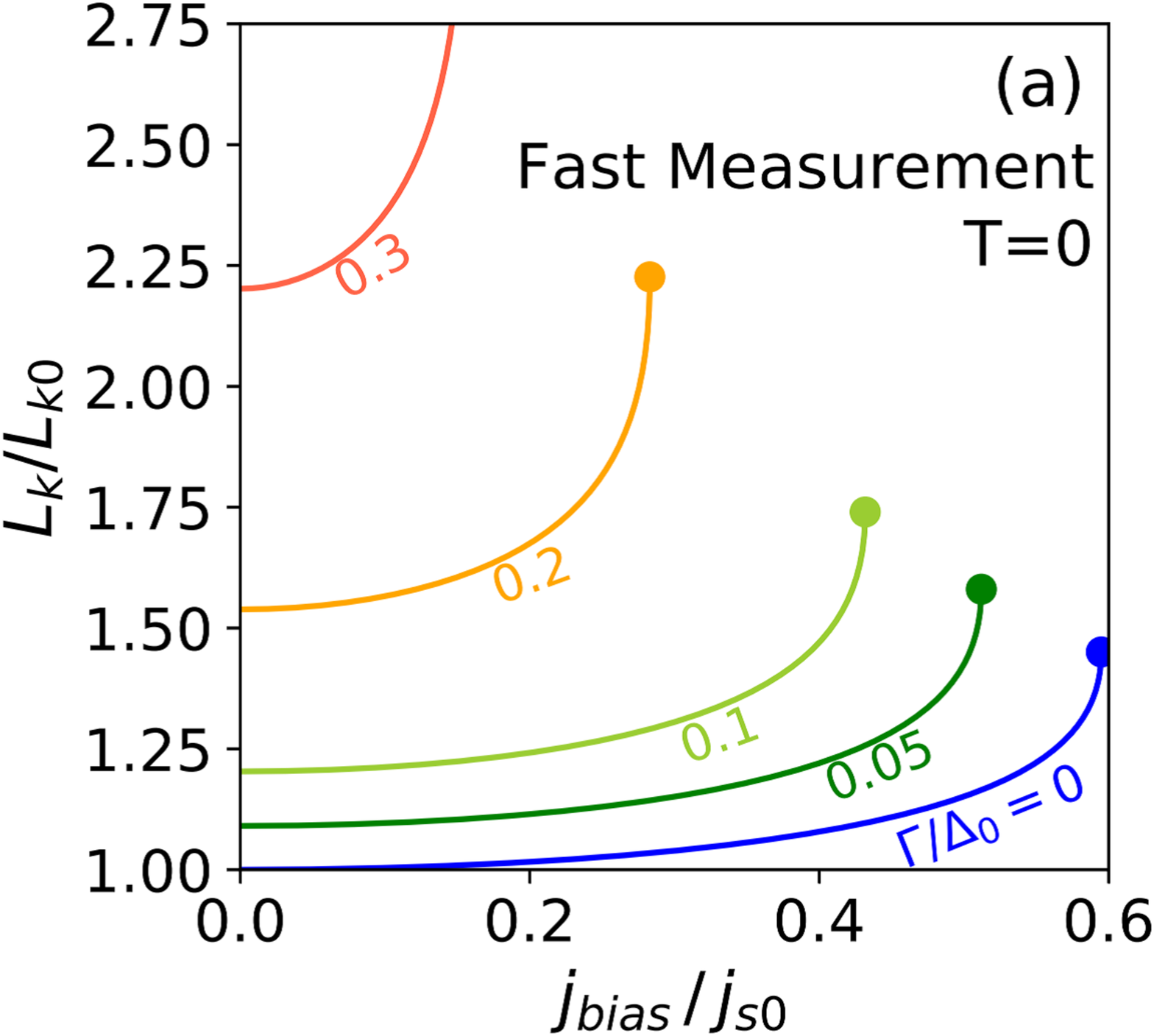}  
   \includegraphics[height=0.435\linewidth]{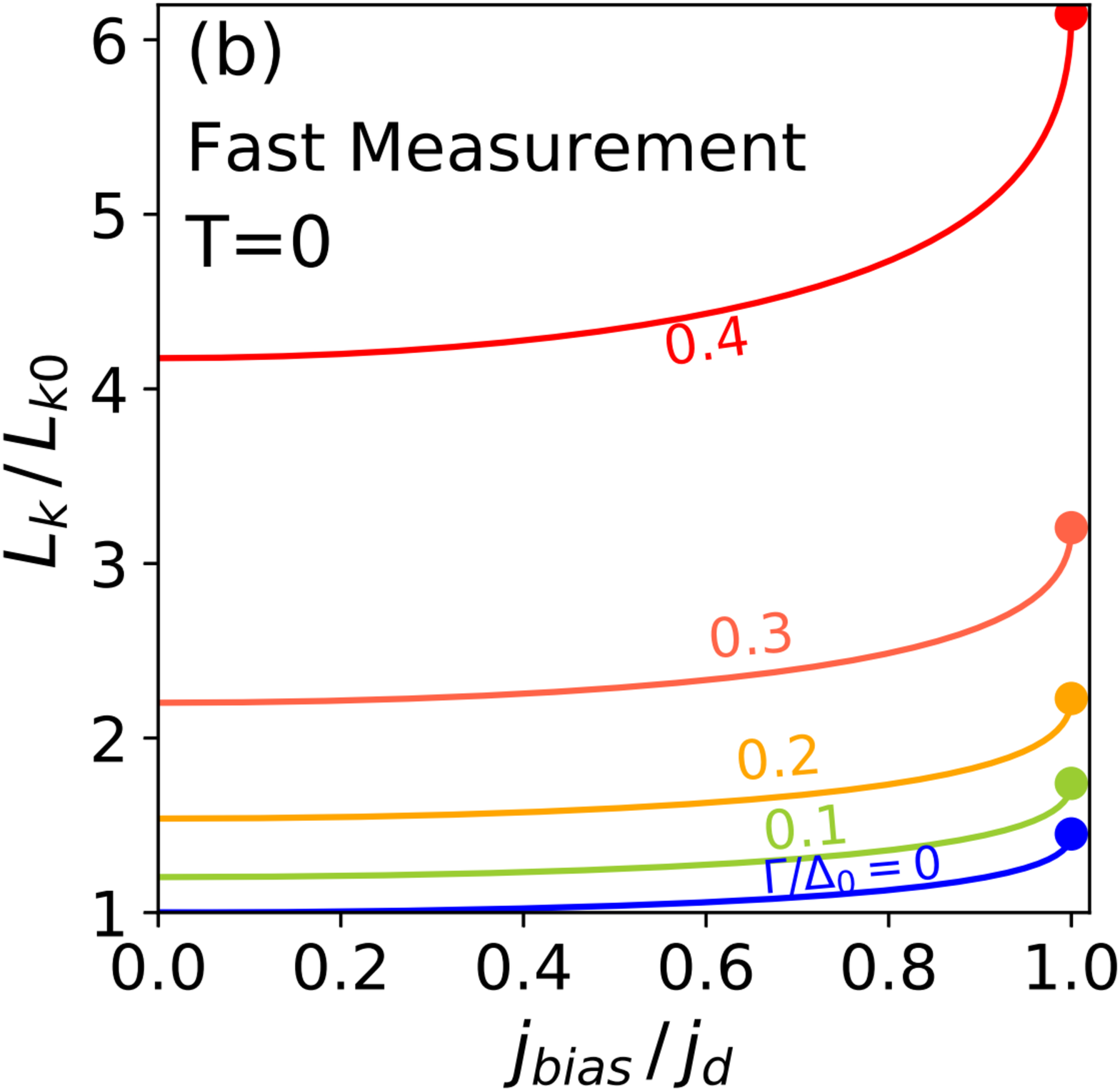}
   \includegraphics[height=0.435\linewidth]{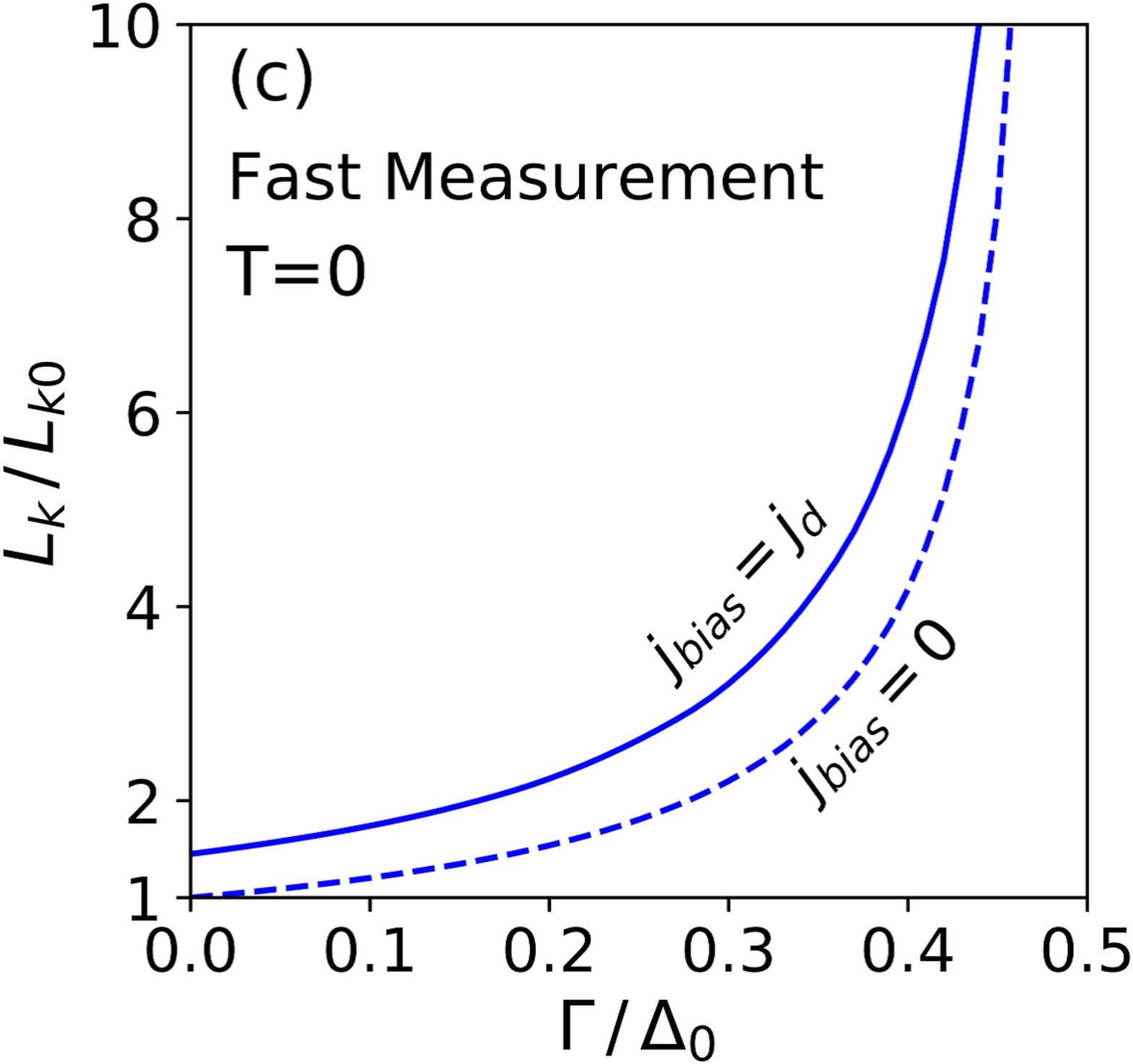}
   \includegraphics[height=0.435\linewidth]{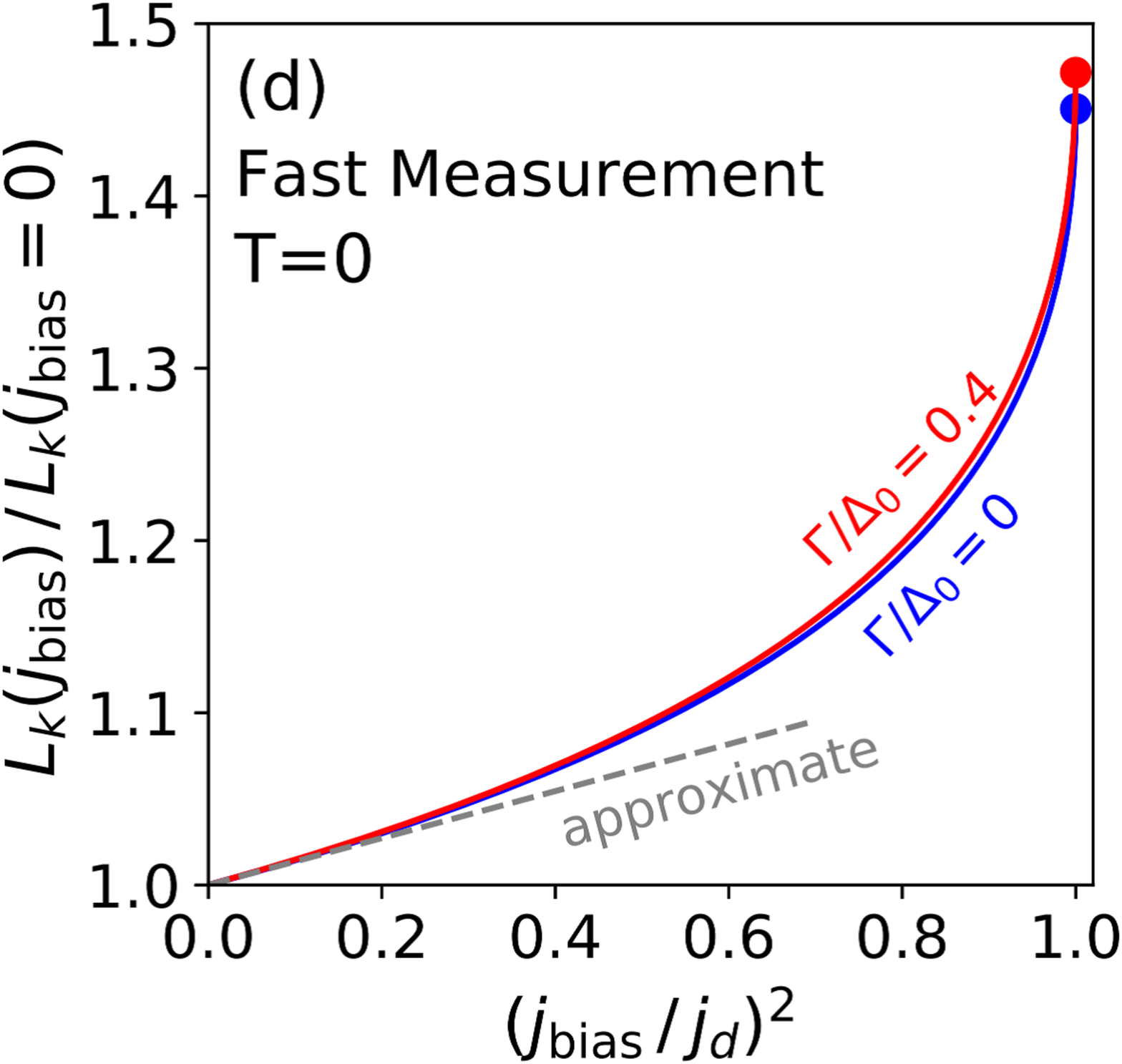}
\end{center}\vspace{0 cm}
   \caption{
Fast-measurement kinetic inductance $L_k (j_{\rm bias}, \Gamma, T)|_{T=0}$ calculated from the formulas given by Eqs.~(\ref{self-consistency_Tzero_1})-(\ref{js_Tzero_1}) and (\ref{Lk_s_Gamma_T_fast}). 
(a) $L_k(j_{\rm bias}, \Gamma, T)|_{T=0}$ as functions of $j_{\rm bias}$. 
(b) $L_k(j_{\rm bias}, \Gamma, T)|_{T=0}$ as functions of $j_{\rm bias}/j_d(\Gamma)$. 
(c) $L_k(j_{\rm bias}, \Gamma, T)|_{T=0}$ as functions of $\Gamma$ for $j_{\rm bias} = 0$ (dashed curve) and $j_{\rm bias} = j_d$ (solid curve). 
(d) $L_k(j_{\rm bias}, \Gamma, 0)/L_k(0, \Gamma, 0)$ as functions of $(j_{\rm bias}/j_d)^2$. 
The dashed gray line is calculated from the approximate formulas given by Eq.~(\ref{C_fast_measure_1}). 
   }\label{fig10}
\end{figure}

Now consider a current-carrying superconductor in the fast measurement regime~\cite{1989_Anlage, 2012_Clem_Kogan}, 
in which the time-dependent current changes rapidly about its time average on a time scale much shorter than the relaxation time of $n_s$. 
We assume $n_s$ cannot follow the time dependence of the current and take $\dot{n}_s \to 0$. 
Then, Eq.~(\ref{Lk_formula0}) reduces to
\begin{eqnarray}
\!\!\!\!\!\!\!L_k(s, \Gamma, T) 
= \mu_0 \lambda_0^2 \frac{n_{s0}}{ n_s(\langle s \rangle, \Gamma, T) \rangle } 
= \mu_0  \lambda^2(\langle s \rangle , \Gamma,T) , \label{Lk_s_Gamma_T_fast}
\end{eqnarray}
where $\langle s \rangle$ is the time average of $s(t)$. 
Experiments in this regime is found in e.g., Ref.~\cite{2016_Santavicca}.

We consider a superconductor under a dc biased rf current: 
$q(t) = q_{\rm bias} + q_{\rm rf}(t)$, 
$s(t)  = [q(t)/q_{\xi}]^2 = s_{\rm bias} + s_{\rm rf}(t)$, 
and $j_s = j_s (s_{\rm bias} + s_{\rm rf}(t))=j_{\rm bias} + j_{\rm rf}(t)$.
In this case, we have $\langle s \rangle = s_{\rm bias} = (q_{\rm bias}/q_{\xi})^2$ and $j_{\rm bias}=j_s(s_{\rm bias})$. 
Note that, when $s_{\rm bias}=0$, Eq.~(\ref{Lk_s_Gamma_T_fast}) reduces to the zero-current kinetic inductance, Eq.~(\ref{Lk_0_Gamma_T}).

For $T=0$, we can use Eqs.~(\ref{self-consistency_Tzero_1})-(\ref{js_Tzero_1}) to evaluate Eq.~(\ref{Lk_s_Gamma_T_fast}). 
Shown in Fig.~\ref{fig10} (a) are $L_k(j_{\rm bias}, \Gamma,T)|_{T=0}$ as functions of the dc bias current $j_{\rm bias}$ for different $\Gamma$. 
As $j_{\rm bias}$ increases, $L_k$ monotonically increases and reaches the maximum at the depairing current density $j_d$ (colored blob). 
Shown in Fig.~\ref{fig10} (b) are $L_k(j_{\rm bias}, \Gamma,0)$ as functions of the normalized current $j_{\rm bias}/j_d$. 
The effects of $\Gamma$ are significant rather than those of $j_{\rm bias}$. 
Shown in Fig.~\ref{fig10} (c) are $L_k(j_{\rm bias}, \Gamma,0)$ as functions of $\Gamma$ for $j_{\rm bias}=0$ (dashed curve) and $j_{\rm bias}=j_d$ (solid curve). 
While both the curves quickly increase with $\Gamma$ and diverge at $\Gamma=1/2$, 
the difference between the solid and dashed curves is always smaller than factor 1.5. 
Shown in Fig.~\ref{fig10} (d) are $L_k(j_{\rm bias}, \Gamma,0)/L_k(0, \Gamma,0)$ as functions of $(j_{\rm bias}/j_d)^2$ for different $\Gamma$. 
The blue ($\Gamma=0$) and red ($\Gamma=0.4$) curves almost overlap, 
and the effects of $j_{\rm bias}$ is less than 1.5 independent of $\Gamma$. 
It should be noted that the blue curves in Figs.~\ref{fig10} (a), \ref{fig10} (b), and \ref{fig10} (d), 
which represent the ideal BCS superconductor with $\Gamma=0$, 
are coincident with the results in the previous study~\cite{2012_Clem_Kogan}.

To understand the nonlinear $L_k$ for small current regions, 
we use the approximate formulas, Eqs.~(\ref{self-consistency_Tzero_3})-(\ref{js_Tzero_3}). 
After some calculations, we find (see Appendix~\ref{a4}): 
\begin{eqnarray}
&&L_k (j_{\rm bias}, \Gamma, 0) = L_k(0,\Gamma, 0) \biggl[ 1 + C_{\rm fm} \biggl( \frac{j_{\rm bias}}{j_d} \biggr)^2 \biggr] ,  \label{C_fast_measure_1} \\
&&C_{\rm fm} = \frac{(3\pi^2 +16) s_{d0}}{12\pi} \biggl( \Delta_{d0} - \frac{4 s_{d0}}{3\pi} \biggr)^2 =0.136 , 
\label{C_fast_measure_2}
\end{eqnarray}
for $T=0$ and $(j_{\rm bias}/j_d)^2 \ll 1$. 
Here $\Delta_{d0}$ and $s_{d0}$ are given by Eqs.~(\ref{Delta_d0}) and (\ref{s_d0}). 
Shown as the dashed gray line in Fig.~\ref{fig10} (d) is calculated from Eq.~(\ref{C_fast_measure_1}), 
which agrees well with the exact results (solid curves) at $(j_{\rm bias}/j_d)^2 \ll 1$.

For $T\simeq T_c$, 
Eq.~(\ref{Lk_s_Gamma_T_fast}) can be calculated from Eqs.~(\ref{GL_Delta})-(\ref{GL_js}). 
Hence, 
\begin{eqnarray}
L_k(s_{\rm bias}, \Gamma, T) 
&=& \frac{7\zeta(3) \mu_0 \lambda_0^2}{4\pi^2 T_c} \biggl( 1-\frac{T}{T_c}\biggr)^{-1} \biggl( 1-\frac{s_{\rm bias}}{3s_d} \biggr)^{-1} \nonumber \\
&=& L_k(0, \Gamma, T) \biggl( 1-\frac{s_{\rm bias}}{3s_d} \biggr)^{-1} ,  
\end{eqnarray}
for $T\simeq T_c$. 
Here $s_d = s_m/3 = (4T_{c}/3\pi) (1-T/T_{c})$ and Eq.~(\ref{Lk_0_Gamma_T_GL}) are used. 
When $s_{\rm bias} = s_d$, we have $L_k(s_d, \Gamma, T)/L_k(0, \Gamma, T)|_{T\simeq T_c} = 1.5$. 
The bias momentum parameter $s_{\rm bias}$ can be converted to $j_{\rm bias}/j_d$ by using Eqs.~(\ref{GL_js}) and (\ref{jd_GL_1}). 
For small current regions, 
we have $s_{\rm bias}/3s_d = (4/27) (j_{\rm bias}/j_d)^2$ and obtain 
\begin{eqnarray}
&& L_k(j_{\rm bias}, \Gamma, T) = L_k(0, \Gamma, T) \biggl[ 1+ C_{\rm fm}^{\rm GL} \biggl( \frac{j_{\rm bias}}{j_d} \biggr)^2 \biggr] ,  \label{C_fast_measure_GL_1} \\
&& C_{\rm fm}^{\rm GL} =  \frac{4}{27} = 0.148 , \label{C_fast_measure_GL_2}
\end{eqnarray}
for $T\simeq T_c$ and $(j_{\rm bias}/j_d)^2 \ll 1$. 
Eq.~(\ref{C_fast_measure_GL_1}) has the same form as that obtained in the previous study~\cite{1989_Anlage} except $T_c$ and $j_d$ depend on $\Gamma$. 
Note the value of $C_{\rm fm}^{\rm GL}$ differs from that of $C_{\rm fm}$ at $T=0$.

\begin{figure}[tb]
   \begin{center}
   \includegraphics[height=0.45\linewidth]{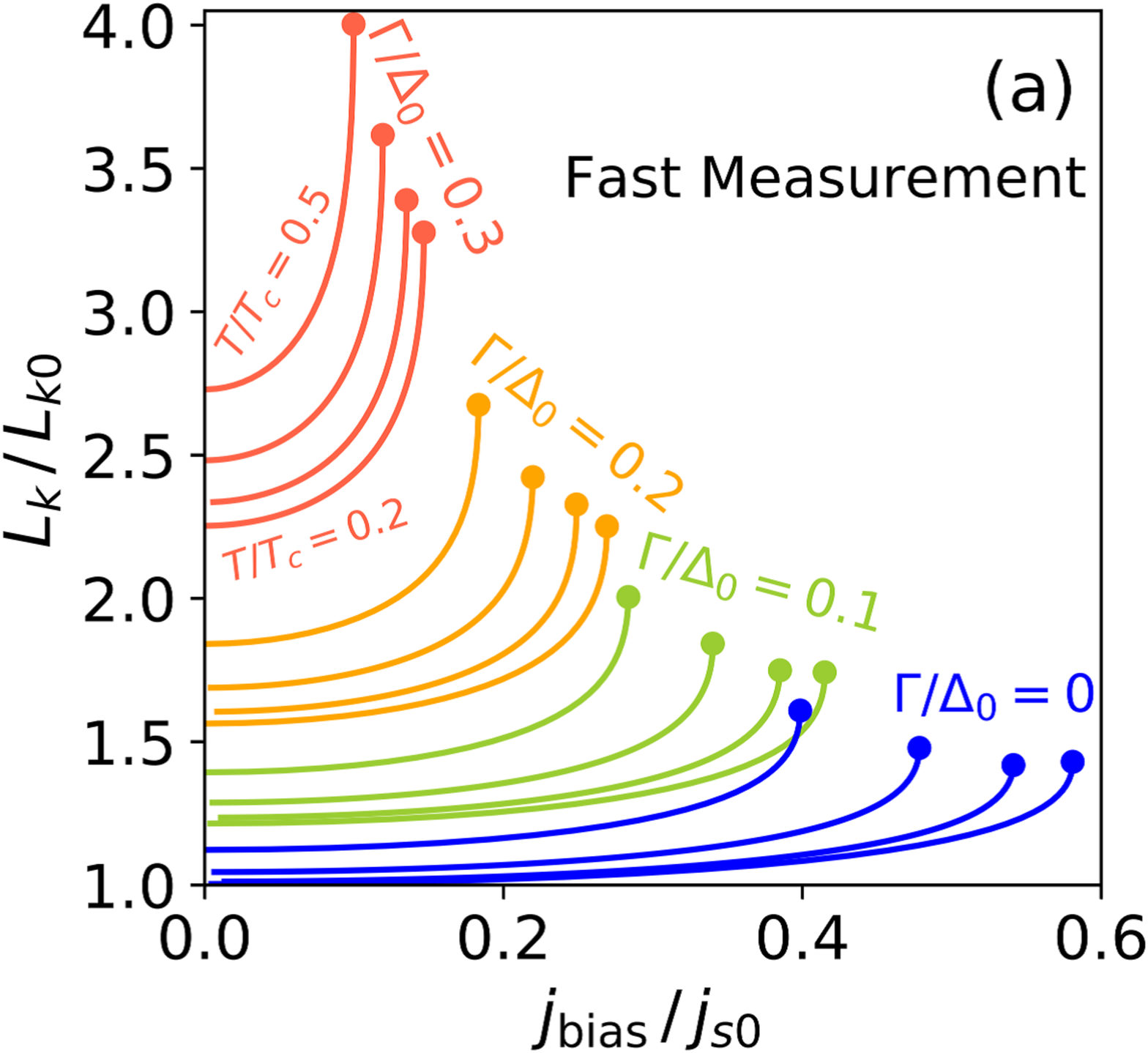}  
   \includegraphics[height=0.45\linewidth]{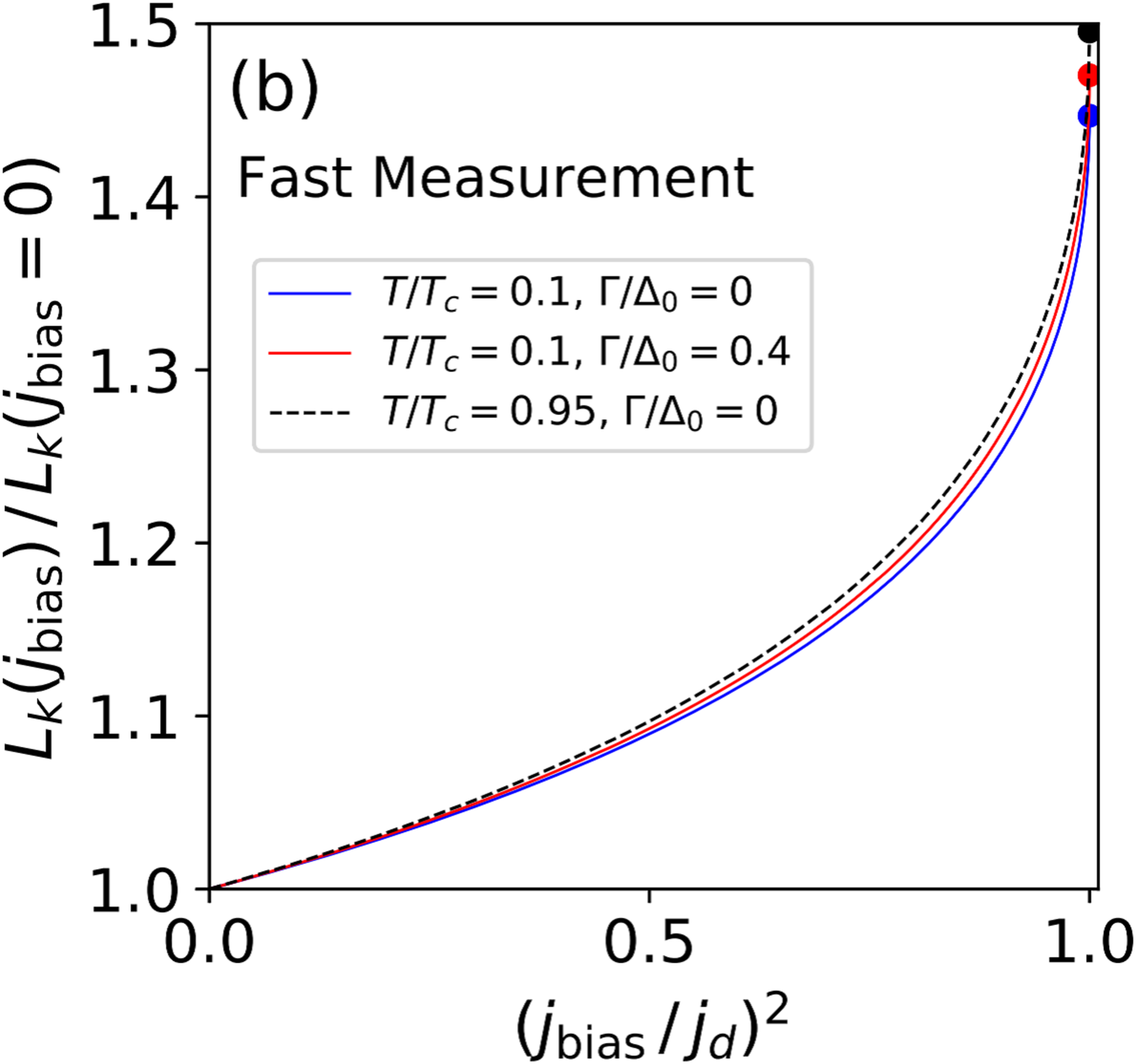}
\end{center}\vspace{0 cm}
   \caption{
Fast measurement kinetic inductance $L_k(j_{\rm bias}, \Gamma, T)$  for a finite $T$. 
(a) $L_k(j_{\rm bias}, \Gamma, T)$ as functions of $j_{\rm bias}$ up to $j_d$ calculated for $\Gamma=0, 0.1, 0.2, 0.3$ and $T/T_c=0.2, 0.3, 0.4, 0.5$. 
The colored blobs are the depairing points. 
(b) $L_k(j_{\rm bias},\Gamma,T)/L_k(0,\Gamma,T)$ as functions of $(j_{\rm bias}/j_d)^2$ for $T/T_c = 0, 0.95$ and $\Gamma = 0, 0.4$.
   }\label{fig11}
\end{figure}

For $0 < T < T_c$, we use the numerical solutions of Eqs.~(\ref{thermodynamic_Usadel})-(\ref{supercurrent}). 
Shown in Fig.~\ref{fig11} (a) are $L_k(j_{\rm bias}, \Gamma, T)$ as functions of $j_{\rm bias}$ for different $\Gamma$ and $T$. 
The blobs represent the depairing points. 
Shown in Fig.~\ref{fig11} (b) are $L_k(j_{\rm bias}, \Gamma, T)/L_k(0, \Gamma, T)$ for different $\Gamma$ and $T$, 
which are not sensitive neither to $\Gamma$ nor $T/T_c$ [see also Fig.~\ref{fig10} (d) for $T=0$].  
The similar curves for $\Gamma=0$ are found in Ref.~\cite{2012_Clem_Kogan}.

\subsubsection{Current carrying state ($s>0$): Slow measurement}\label{sec_film_Lk_Slow}

Consider the other limit, the slow measurement regime~\cite{1989_Anlage, 2012_Clem_Kogan}, 
in which the time-dependent current changes on a time scale much longer than the relaxation time of $n_s$. In this case, we can assume $n_s$ instantly follows the time dependence of the current: $n_s = n_s(q(t))$. 
Substituting $\dot{n}_s = \dot{q} \partial_q n_s $ into Eq.~(\ref{Lk_formula0}), 
we find 
\begin{eqnarray}
L_k (s, \Gamma, T)
&=& \mu_0\lambda_0^2 \biggl[ (1 +  q  \partial_q ) \frac{n_s(s,\Gamma,T)}{n_{s0}} \biggr]^{-1} 
\label{Lk_s_Gamma_T_slow_0} \\
&=& \mu_0\lambda_0^2 \sqrt{\pi} \biggl| \frac{\partial (j_s/j_{s0})}{\partial (q/q_{\xi})} \biggr|^{-1} , 
\label{Lk_s_Gamma_T_slow}
\end{eqnarray}
where Eq.~(\ref{Lk_s_Gamma_T_slow}) corresponds with the expression given in Ref.~\cite{2012_Clem_Kogan}.

\begin{figure}[tb]
   \begin{center}
   \includegraphics[height=0.44\linewidth]{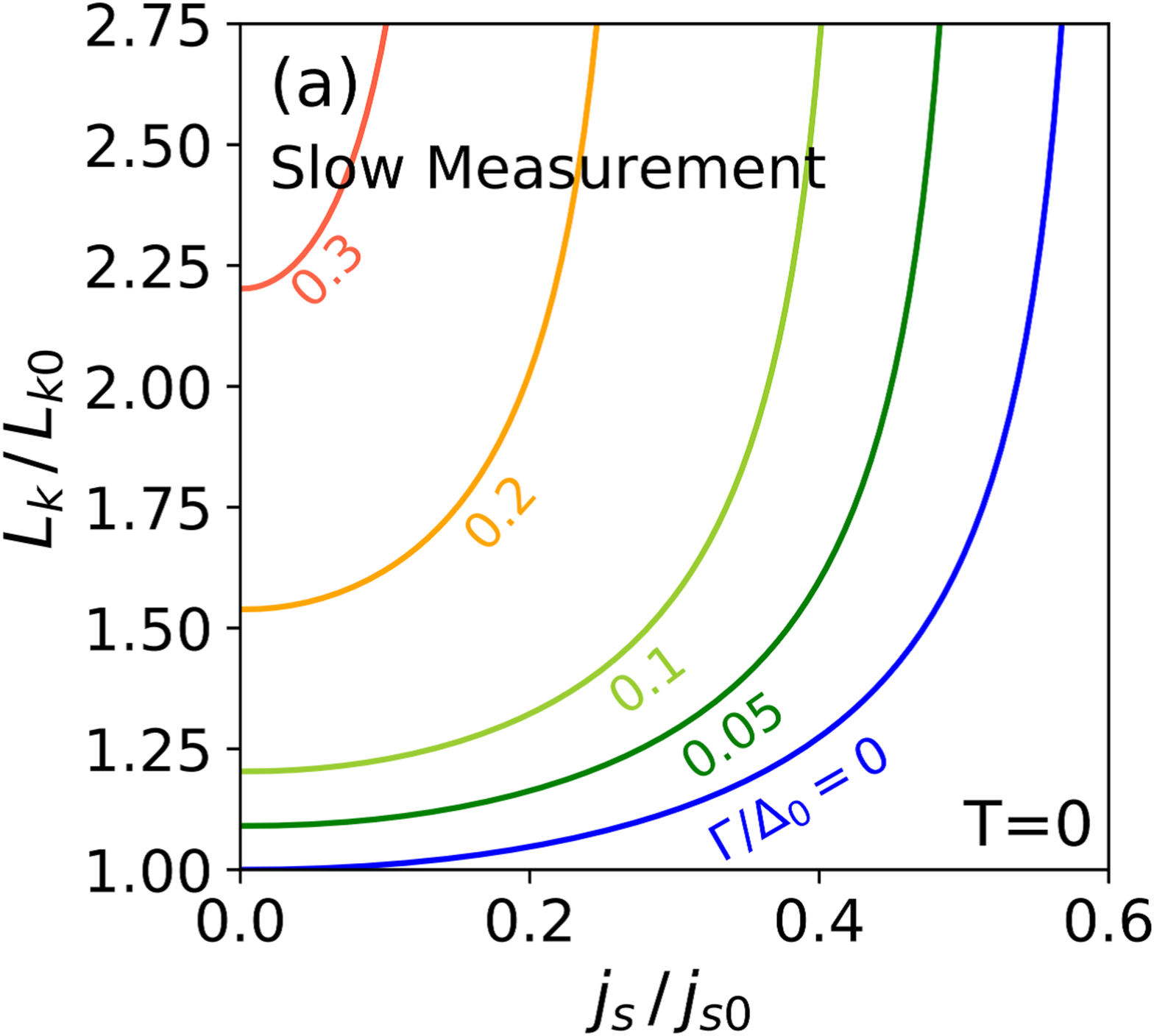}  
   \includegraphics[height=0.44\linewidth]{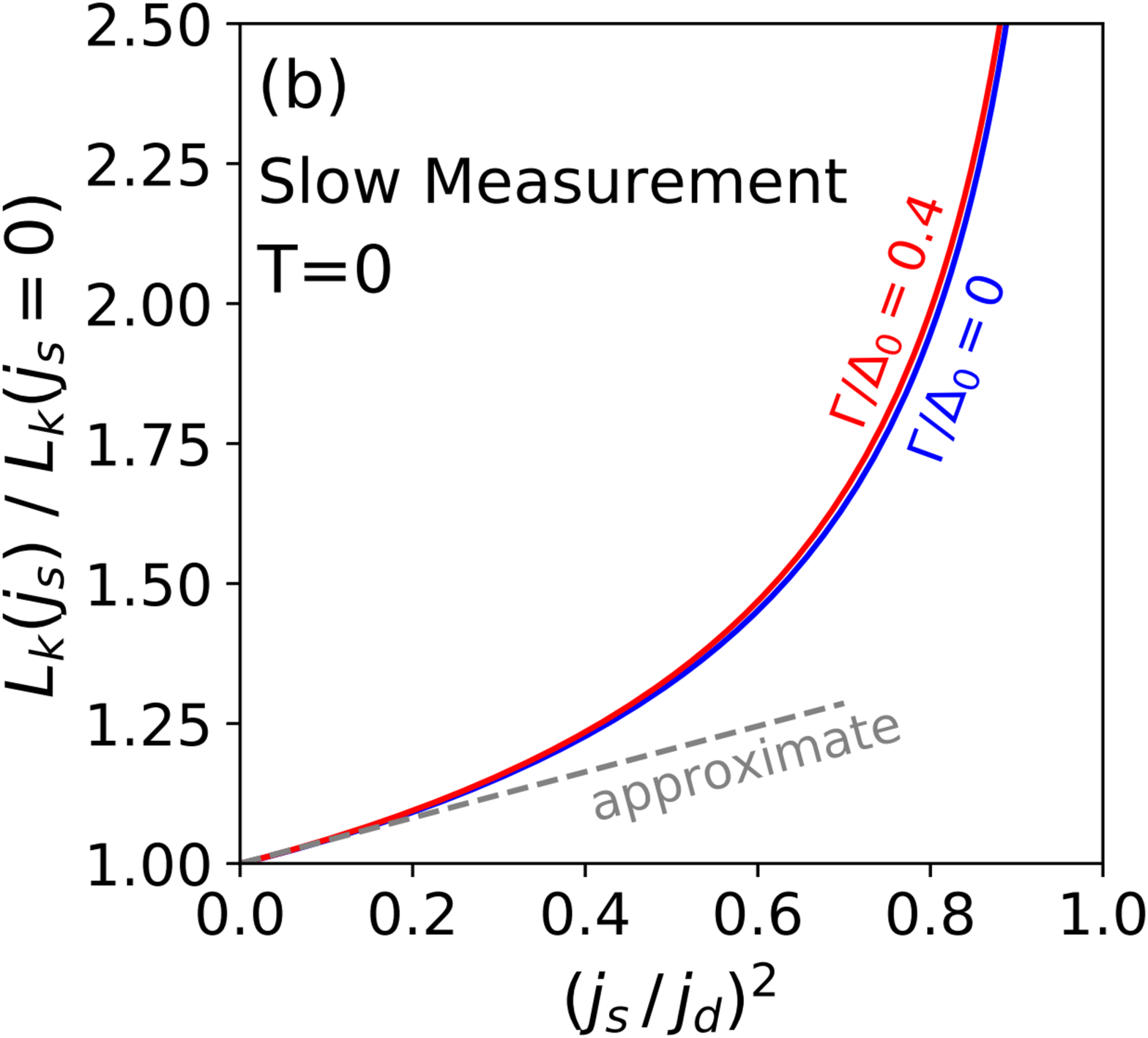}
\end{center}\vspace{0 cm}
   \caption{
Slow-measurement kinetic inductance $L_k (j_s, \Gamma, T)|_{T=0}$ calculated from the formulas given by Eqs.~(\ref{self-consistency_Tzero_1})-(\ref{js_Tzero_1}) and (\ref{Lk_s_Gamma_T_slow}). 
(a) $L_k$ as functions of $j_s$. 
(b) $L_k$ normalized with $L_k(j_s=0)$ as functions of $(j_s/j_d)^2$. 
The dashed gray line is calculated from the approximate formulas given by Eq.~(\ref{C_slow_measure_1}). 
   }\label{fig12}
\end{figure}

For $T = 0$, we can evaluate Eq.~(\ref{Lk_s_Gamma_T_slow}) using the solutions of Eqs.~(\ref{self-consistency_Tzero_1})-(\ref{js_Tzero_1}). 
Shown in Fig.~\ref{fig12} (a) are $L_k(j_s, \Gamma,T)|_{T=0}$ as functions of $j_s$, 
which diverge at $j_s=j_d(\Gamma)$. 
These divergences come from $\partial_q j_s =0$ at $j_s=j_d$ (see also Fig.~\ref{fig5}). 
Shown in Fig.~\ref{fig12} (b) are $L_k(j_s, \Gamma,0)/L_k(0,\Gamma,0)$ as functions of the normalized current $(j_s/j_d)^2$, 
which are not sensitive to $\Gamma$.

For small-current regions, we have an useful formula to calculate $L_k$ (see Appendix~\ref{a4}): 
\begin{eqnarray}
&&L_k (j_s, \Gamma, 0) = L_k (0, \Gamma, 0) \biggl[ 1 + C_{\rm sm} \biggl( \frac{j_s}{j_d} \biggr)^2 \biggr] , 
\label{C_slow_measure_1}\\
&& C_{\rm sm} =
\frac{(3\pi^2+16) s_{d0}}{4\pi} \biggl( \Delta_{d0} -\frac{4 s_{d0}}{3\pi}  \biggr)^2 = 0.544 ,
\label{C_slow_measure_2}
\end{eqnarray}
for $T=0$ and $(j_s/j_d)^2 \ll 1$. 
Hence, we have a relation $C_{\rm sm} = 4C_{\rm fm}$ [see Eqs.~(\ref{C_fast_measure_GL_1}) and (\ref{C_fast_measure_GL_2})]. 
Shown as the dashed gray line in Fig.~\ref{fig12} (b) is the normalized $L_k$ calculated from Eq.~(\ref{C_slow_measure_1}), 
which agrees well with the exact results (solid curves) at $(j_s/j_d)^2 \ll 1$.

For $T\simeq T_c$, we use the GL results. 
Substituting Eq.~(\ref{GL_js}) into Eq.~(\ref{Lk_s_Gamma_T_slow}), we obtain 
\begin{eqnarray}
L_k (s, \Gamma, T)
 = L_k (0, \Gamma, T) \biggl( 1 - \frac{s}{s_d} \biggr)^{-1} .
\label{Lk_GL_slow}
\end{eqnarray}
When $s\ll s_d$, we have $s/s_d = (4/9) (j_s/j_d)^2$ and obtain
\begin{eqnarray}
&&L_k (s, \Gamma, T)
 = L_k (0, \Gamma, T) \biggl[ 1 + C_{\rm sm}^{\rm GL} \biggl( \frac{j_s}{j_d} \biggr)^2 \biggr] , \label{Lk_GL_slow_small_current_0}
\\
&& C_{\rm sm}^{\rm GL}  = \frac{4}{9} = 0.444, 
\label{Lk_GL_slow_small_current}
\end{eqnarray}
for $T\simeq T_c$ and $(j_s/j_d)^2 \ll 1$. 
We find $C_{\rm sm}$ for $T\simeq T_c$ is smaller than that at $T=0$. 
Eq.~(\ref{Lk_GL_slow_small_current_0}) has the same form as the well-known result~\cite{1989_Anlage} except $T$ and $j_d$ depend on $\Gamma$.

\begin{figure}[tb]
   \begin{center}
   \includegraphics[height=0.442\linewidth]{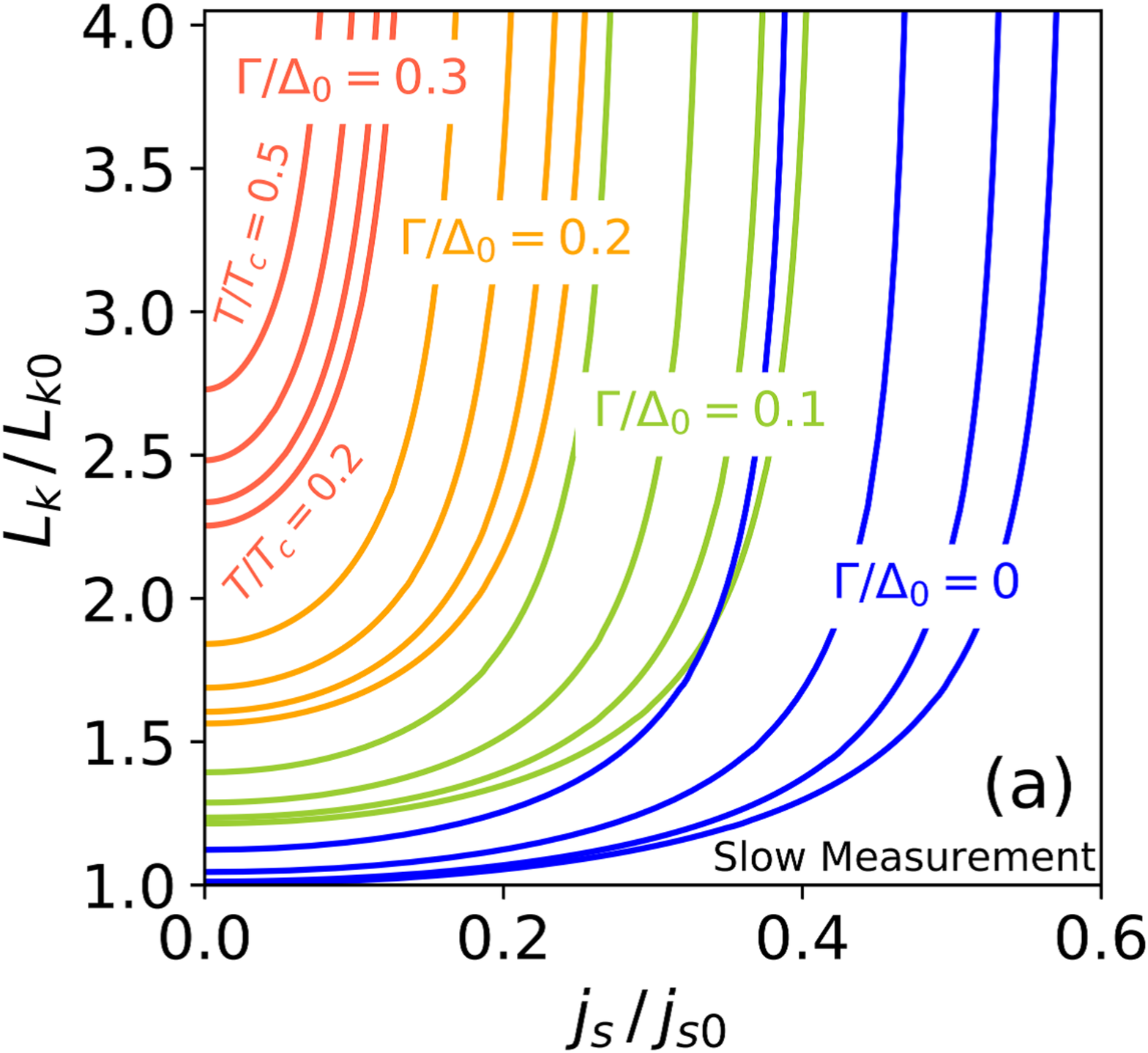}  
   \includegraphics[height=0.442\linewidth]{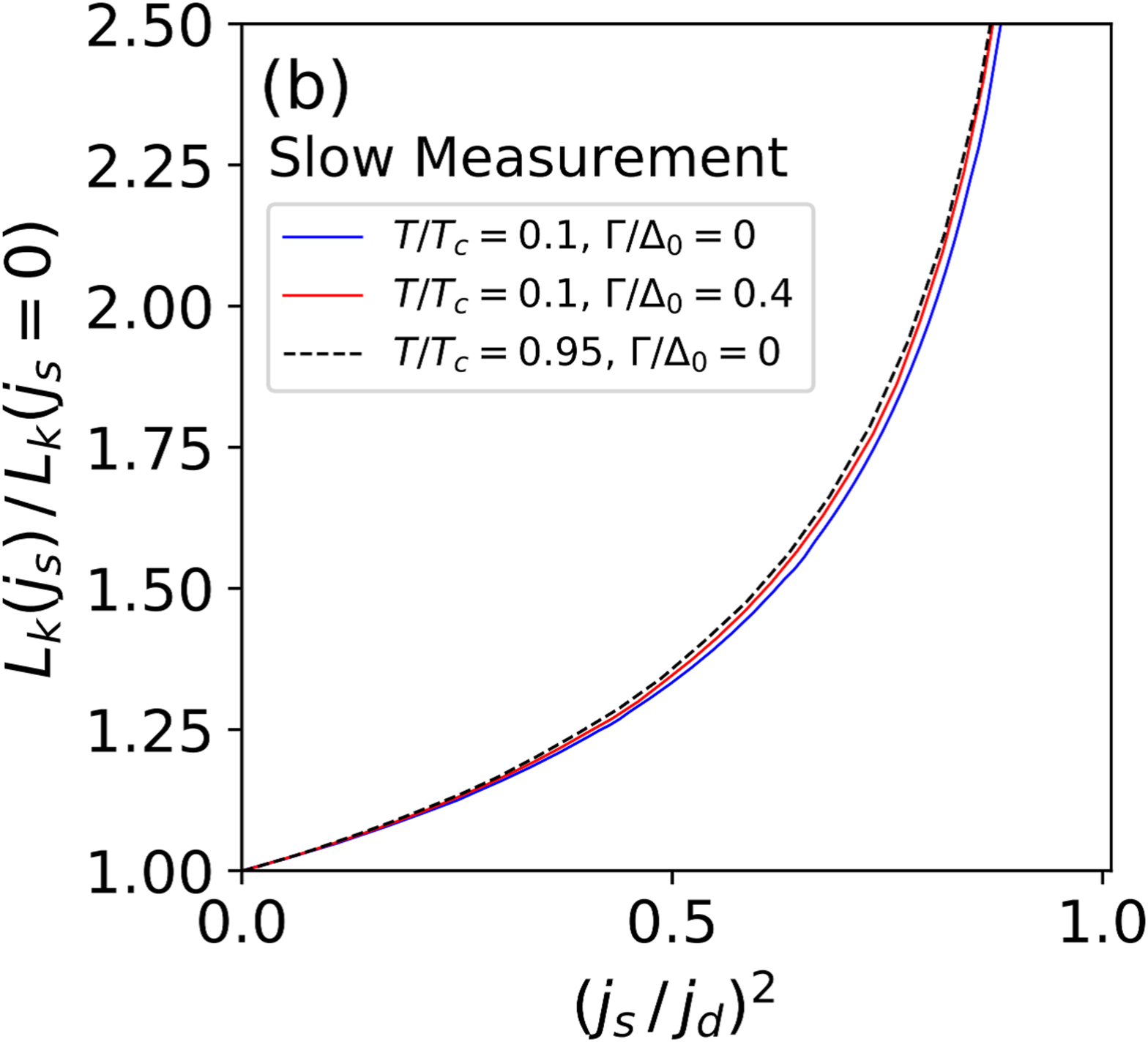}
\end{center}\vspace{0 cm}
   \caption{
(a) Slow-measurement kinetic inductance $L_k (j_s, \Gamma, T)$ at finite temperatures as functions of $j_s$ up to $j_d$ calculated for $\Gamma=0, 0.1, 0.2, 0.3$ and $T/T_c=0.2, 0.3, 0.4, 0.5$. 
(b) $L_k(j_s,\Gamma, T)/L_k(0,\Gamma,T)$ as functions of $(j_s/j_d)^2$ for $\Gamma=0, 0.4$ and $T/T_c=0.1, 0.95$. 
   }\label{fig13}
\end{figure}

For $0< T< T_c$, we use the numerical solutions of Eqs.~(\ref{thermodynamic_Usadel})-(\ref{supercurrent}). 
Shown in Fig.~\ref{fig13} (a) are $L_k(j_s, \Gamma, T)$ as functions of $j_s$ for different $\Gamma$ and $T$, 
which increase with $j_s$ and diverge at $j_s=j_d(\Gamma)$, $\Gamma=1/2$, and $T=T_c(\Gamma)$. 
Shown in Fig.~\ref{fig13} (b) are $L_k(j_s, \Gamma, T)/L_k(0, \Gamma, T)$ as functions of $(j_s/j_d)^2$ for different $\Gamma$ and $T$, 
sensitive neither to $\Gamma$ nor $T$. 
This insensitivity resembles that for the fast measurement case [see also Fig.~\ref{fig11} (b)].  
See also Ref.~\cite{2012_Clem_Kogan} for $\Gamma=0$.

\section{Semi-infinite superconductor} \label{sec_bulk}

In this section, we consider the geometry shown in Fig.~\ref{fig2} (b): 
a semi-infinite superconductor occupying $x \ge 0$. 
We calculate the current distribution and the superheating field.

\subsection{Current distribution}

In the Meissner state, 
the current distributes within the depth $\sim \lambda$ from the surface. 
When the superfluid flow is small ($|q| \ll q_d$), 
the pair-breaking effect due to a finite $q$ is negligible, 
and the distributions of the current $j_s(x)$ and the magnetic field $H(x)$ obey the London equation. 
As $|q|$ increases, the London equation ceases to be valid due to the current-induced pair-breaking effect (nonlinear Meissner effect~\cite{2010_Groll, 1995_Sauls}).  
To obtain the current and field distributions, 
we need the self-consistent solutions of the coupled Maxwell and Usadel equations,  Eqs.~(\ref{thermodynamic_Usadel})-(\ref{boundary_conditions}).

\begin{figure}[tb]
   \begin{center}
   \includegraphics[width=0.494\linewidth]{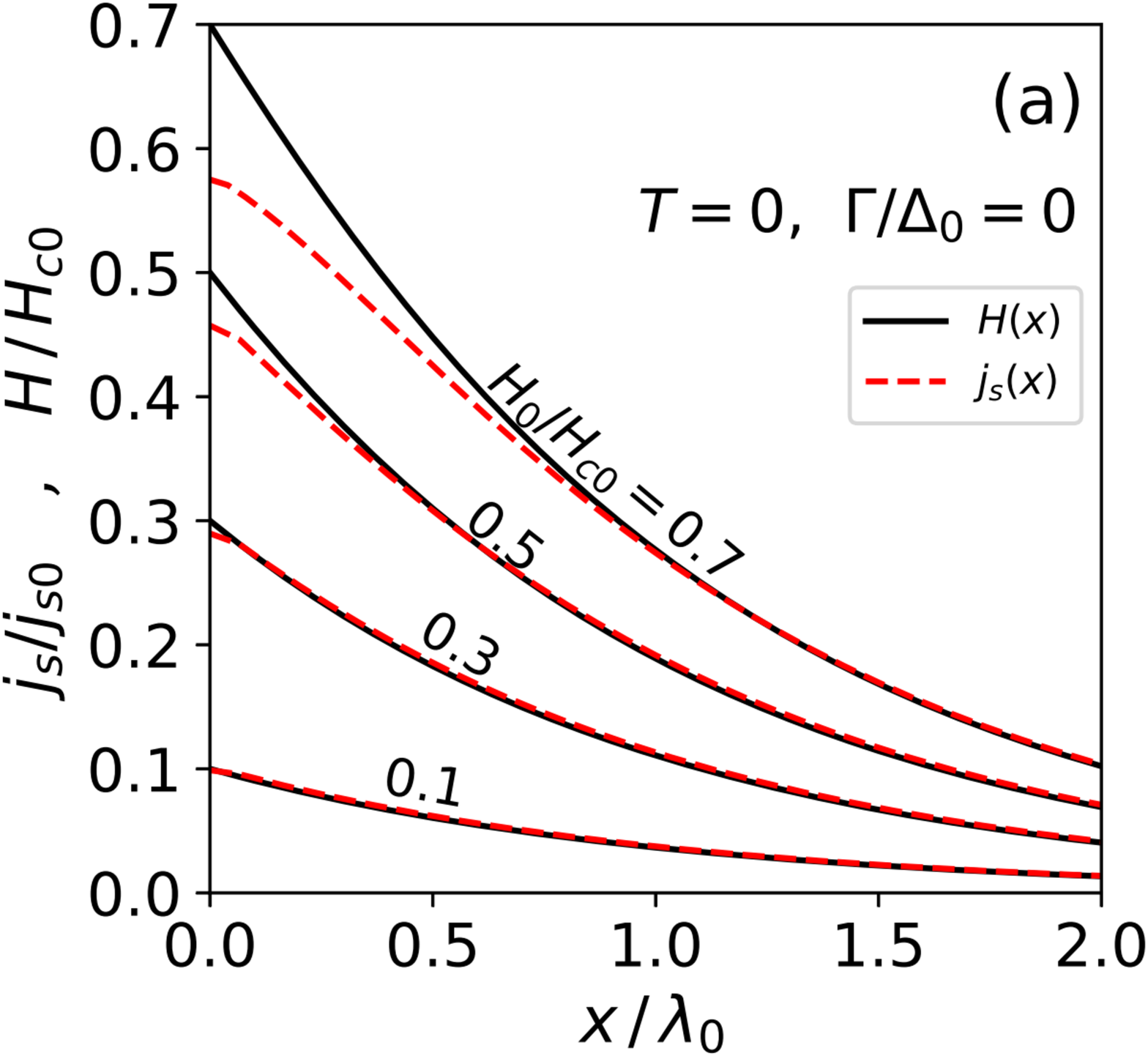}
   \includegraphics[width=0.494\linewidth]{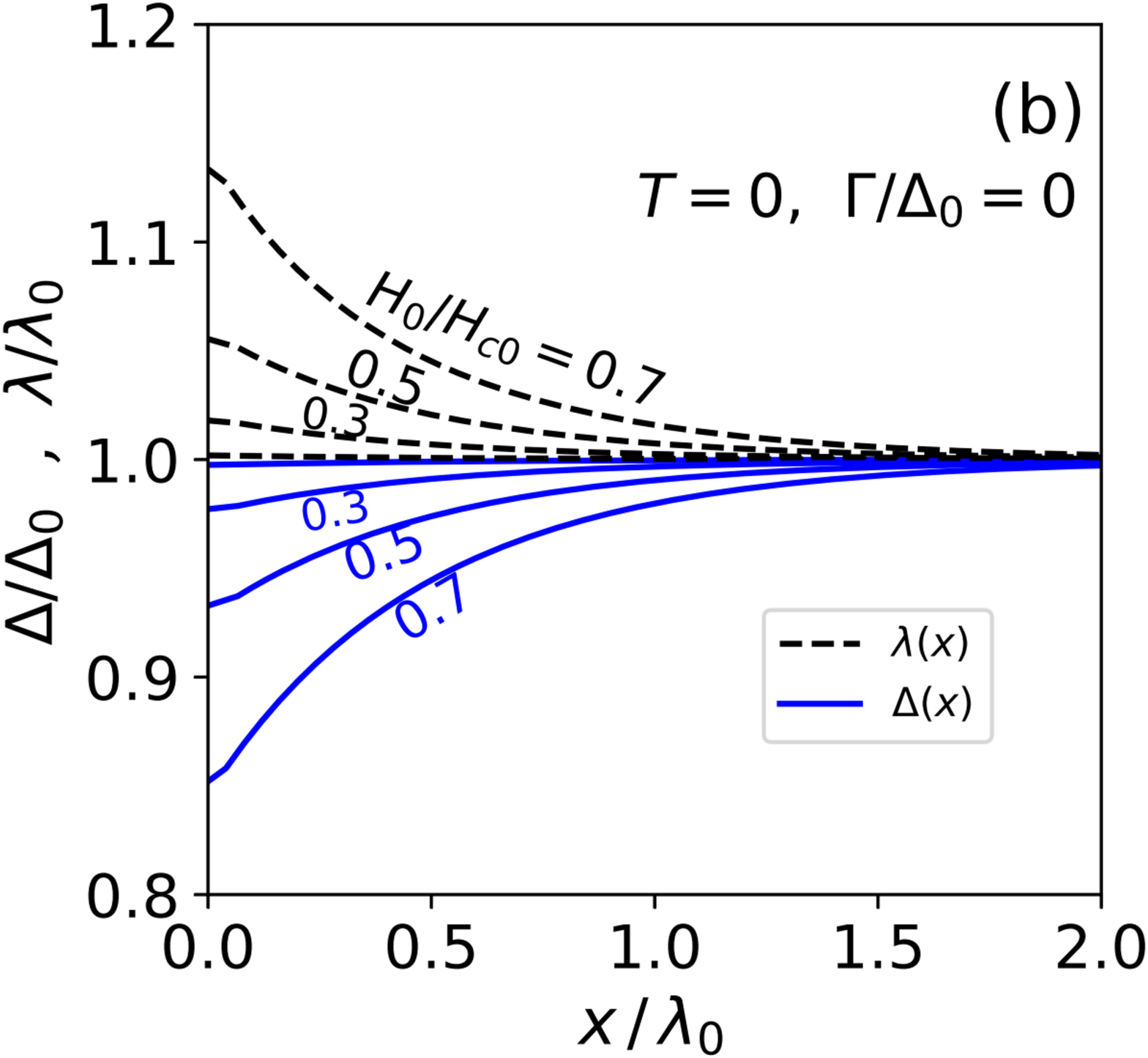}
   \end{center}\vspace{0 cm}
   \caption{
Distributions of (a) $H$, $j_s$, (b) $\Delta$, and $\lambda$ in the semi-infinite superconductor for the surface magnetic field $H_0/H_{c0} = 0.1, 0.3, 0.5, 0.7$. 
   }\label{fig14}
\end{figure}

Let us consider the simplest case: $T=0$ and $\Gamma=0$. 
For $H_0 \to 0$, we can use the London equation,  
which gives $H(x)/H_{c0}=j_s(x)/j_{s0} = H_0/H_{c0} \exp(-x/\lambda_0)$: 
the curves for $H(x)/H_{c0}$ and $j_s(x)/j_{s0}$ completely overlap. 
Shown in Fig.~\ref{fig14} (a) are $H(x)$ and $j_s(x)$ as functions of $x$ for different $H_0$ calculated from the self-consistent solutions of Eqs.~(\ref{thermodynamic_Usadel})-(\ref{boundary_conditions}). 
For small $H_0$ regions, in fact, $H(x)/H_{c0}$ (solid curves) and $j_s(x)/j_{s0}$ (dashed curves) almost overlap, expected from the London equation. 
However, as $H_0$ increases, the dashed curves deviate from the solid curves in the vicinity of the surface: 
the nonlinear Meissner effect manifests itself. 
Shown in Fig.~\ref{fig14} (b) are $\Delta(x)$ and $\lambda(x)$, 
which differ from the zero-current values ($\Delta_0$ and $\lambda_0$) at $x \lesssim \lambda$ 
but approach $\Delta_0$ and $\lambda_0$ as $x$ increases.

\subsection{Superheating field} \label{sec_bulk_Hsh}

The superheating field $H_{sh}$ is given by the value of $H_0$ which induces $j_s(x_0) = j_d$. 
Here $x_0$ is the depth at which the distribution $j_s(x)$ takes the maximum. 
Note here it is not necessarily the case that we have $x_0=0$.  
For instance, in the multilayer structure~\cite{2006_Gurevich, 2014_Kubo, 2015_Gurevich, 2017_Kubo_SUST, 2019_Kubo_Gurevich} or a superconductor including inhomogeneous impurities in the vicinity of the surface~\cite{2017_Kubo_SUST, 2019_Sauls, 2019_Kubo_Gurevich}, 
the surface current can be suppressed, 
and $j_s(x)$ takes the maximum at the inside ($x_0 > 0$).

In our semi-infinite superconductor,  
the current is a monotonically decreasing function of $x$ as shown in Fig.~\ref{fig14} (a). 
Hence, $x_0=0$. 
In this case, we can derive a simple formula of $H_{sh}$. 
Integrating both the sides of Eq.~(\ref{Usadel_London}) from $x=0$ to $\infty$, 
we obtain $q'(0)^2 = -2 \int_0^{\infty} q q' \lambda^{-2}(s,\Gamma,T) dx$. 
Then, Using Eqs.~(\ref{H}) and (\ref{boundary_conditions}), 
we find the relation between the applied magnetic field $H_0$ and the superfluid flow at the surface $s(0)$: 
\begin{eqnarray}
\frac{H_0^2}{H_{c0}^2}
= \int_0^{s(0)} \!\!\!\frac{\pi \lambda_0^2 ds}{\lambda^2(s,\Gamma,T)}  
=  \pi  \int_0^{s(0)} \!\!\!\!\!\!ds  \frac{n_s(s,\Gamma,T)}{n_{s0}} .
\end{eqnarray}
$H_{sh}$ can be calculated by substituting $s(0) =s_d$: 
\begin{eqnarray}
H_{sh} (\Gamma,T)
= H_{c0} \sqrt{ \pi  \int_0^{s_d(\Gamma,T)} \!\!\!\!ds  \frac{n_s(s,\Gamma,T)}{n_{s0}}   } ,
\label{Hsh_formula}
\end{eqnarray}
which is the general formula of the superheating field for a homogeneous dirty superconductor, 
valid for arbitrary $\Gamma$ and $T$.

\begin{figure}[tb]
   \begin{center}
   \includegraphics[width=0.494\linewidth]{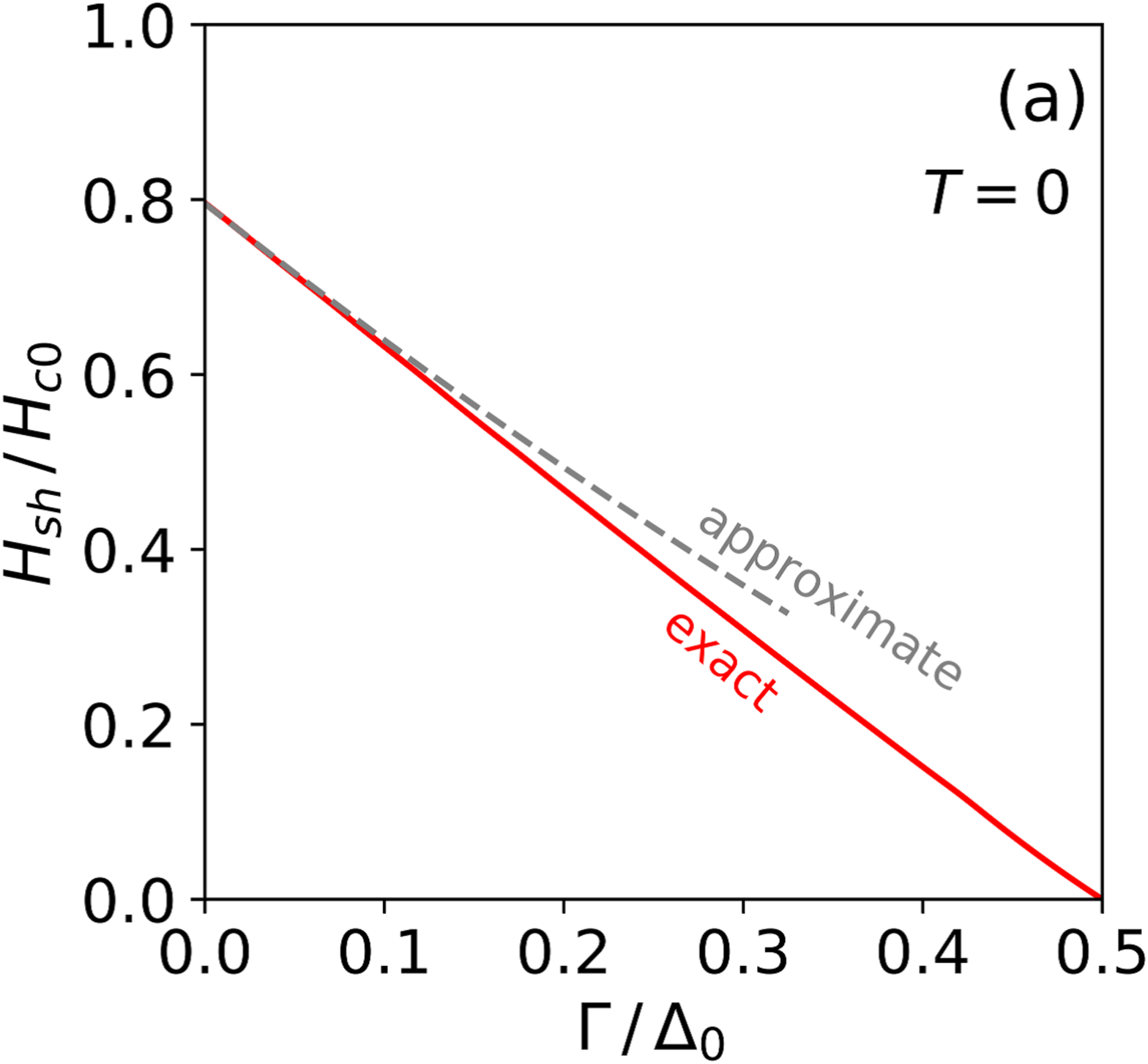}
   \includegraphics[width=0.494\linewidth]{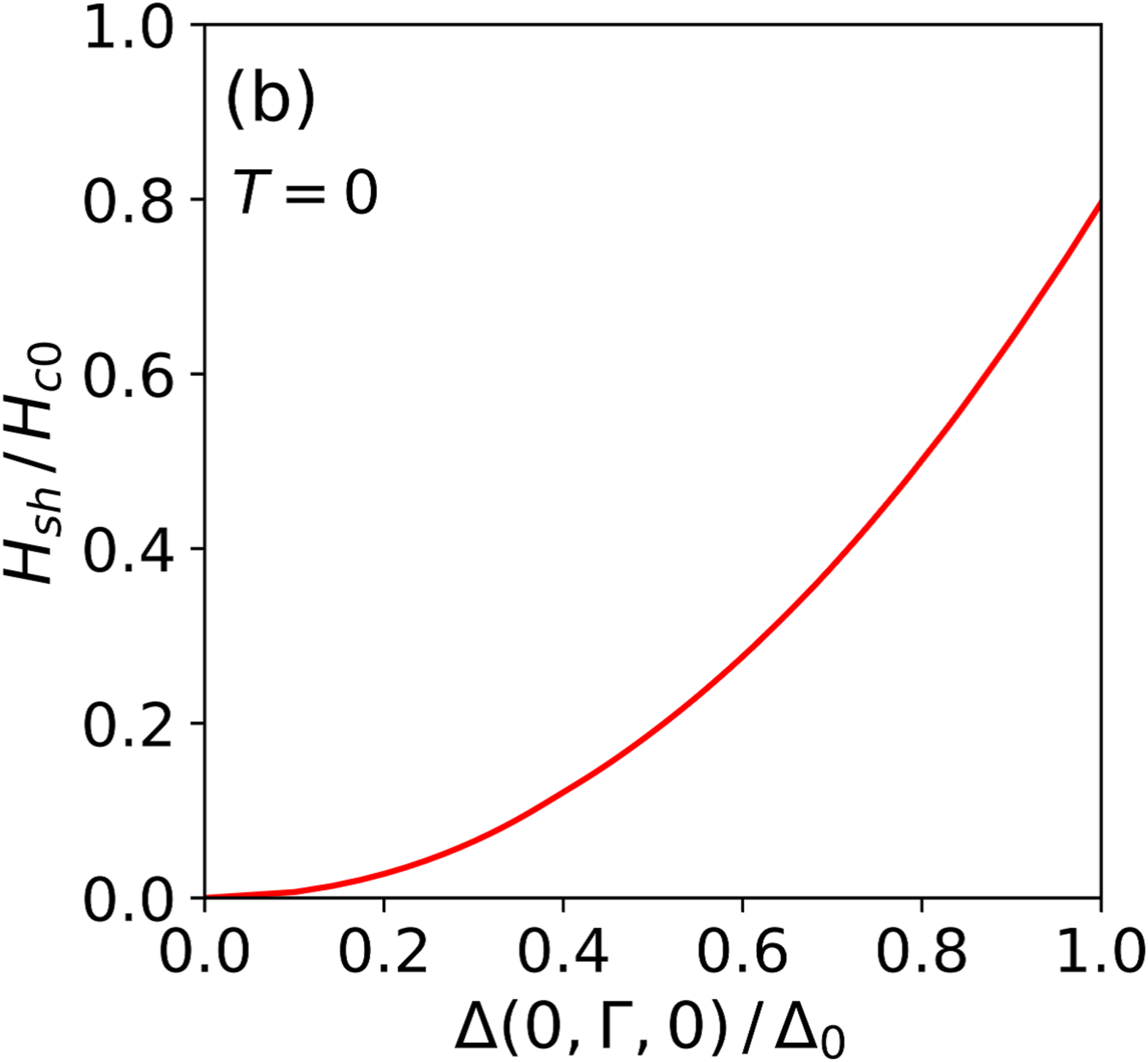}
   \end{center}\vspace{0 cm}
   \caption{
Superheating field $H_{sh}(\Gamma, T)$ at $T=0$. 
(a) $H_{sh}$ as a function of $\Gamma$. 
The dashed curve is calculated from the approximate formula given by Eqs.~(\ref{Hsh_Gamma_0})-(\ref{Hsh_Gamma_0_zd}). 
(b) $H_{sh}$ as a function of $\Delta$ in the zero-current state.   
   }\label{fig15}
\end{figure}

For $T = 0$, we can evaluate Eq.~(\ref{Hsh_formula}) by using Eqs.~(\ref{self-consistency_Tzero_1}) and (\ref{superfluid_density_Tzero_1}) [see also Figs.~\ref{fig5} (b) and \ref{fig7} (b)].
Shown as the solid red curve in Fig.~\ref{fig15} (a) is $H_{sh}(\Gamma,T)|_{T=0}$ as a function of $\Gamma$. 
As $\Gamma$ increases, $H_{sh}$ decreases and vanishes at $\Gamma=1/2$. 
Shown in Fig.~\ref{fig15} (b) is $H_{sh}(\Gamma,0)$ as a function of $\Delta(0,\Gamma,0)/\Delta(0,0,0)$.

For $T=0$ and $\Gamma \ll 1$ (such that $\Gamma \ll \Delta-s$, 
Eq.~(\ref{Hsh_formula}) reduces to a formula (see Appendix~\ref{a5}), 
\begin{eqnarray}
H_{sh} (\Gamma, 0) = H_{c0} \sqrt{ I(\Gamma) - \frac{2}{3}s_d^2(\Gamma) -2\Gamma s_d(\Gamma)} , \label{Hsh_Gamma_0}
\end{eqnarray}
where
\begin{eqnarray}
&& I (\Gamma)=  1 - \biggl( 1- \frac{\pi z_d}{2} \biggr)e^{-\frac{\pi z_d}{2}} - 4\Gamma \bigl( 1- e^{-\frac{\pi z_d }{4}} \bigr) , \label{Hsh_Gamma_0_I} \\
&& z_d(\Gamma) = \zeta_d \frac{1 - \frac{\pi}{4}\zeta_d }{1-\frac{\pi}{4}\zeta_d + \frac{\Gamma}{\Delta_d} } , \label{Hsh_Gamma_0_zd} 
\end{eqnarray}
$\zeta_d = s_d/\Delta_d$ and $s_d$ are given by Eqs.~(\ref{Delta_d}) and (\ref{sd}). 
For the ideal dirty BCS superconductor ($\Gamma=0$), we have $z_d = \zeta_{d0}$ and
\begin{eqnarray}
H_{sh} (0,0)
&=& H_{c0} \sqrt{ 1 - \biggl( 1 -\frac{\pi \zeta_{d0}}{2} \biggr) e^{-\frac{\pi\zeta_{d0}}{2}} -\frac{2}{3}s_{d0}^2 } \nonumber \\
&=& 0.79 H_{c0} . 
\end{eqnarray}
This is slightly smaller than the clean-limit value, $H_{sh}^{\rm clean}(0,0)=0.84 H_{c0}$~\cite{1966_Galaiko, 2008_Catelani}, 
and is consistent with the previous study~\cite{2012_Lin_Gurevich}, 
in which $H_{sh}$ takes the maximum at the mean free path (mfp) $= 5.32 \xi_0$ and decreases with mfp.

For $T\simeq T_c$, we use the GL results. 
Substituting Eqs.~(\ref{GL_Delta}) and (\ref{GL_ns}) into Eq.~(\ref{Hsh_formula}) and using the depairing value of the $s$ parameter in the GL regime, $s_d(\Gamma,T)=s_m/3 = (4T_c/3\pi)(1-T/T_c)$, we obtain 
\begin{eqnarray}
H_{sh}  (\Gamma,T)|_{T\simeq T_c}
= \frac{\sqrt{5}}{3} H_c(\Gamma, T) 
= 0.745 H_c(\Gamma, T) .
\label{GL_Hsh}
\end{eqnarray}
The coefficient is independent of $\Gamma$ and coincident with the well-known GL result obtained for $\Gamma=0$~\cite{1968_Kramer, 2011_Transtrum}.

\begin{figure}[tb]
   \begin{center}
   \includegraphics[width=0.494\linewidth]{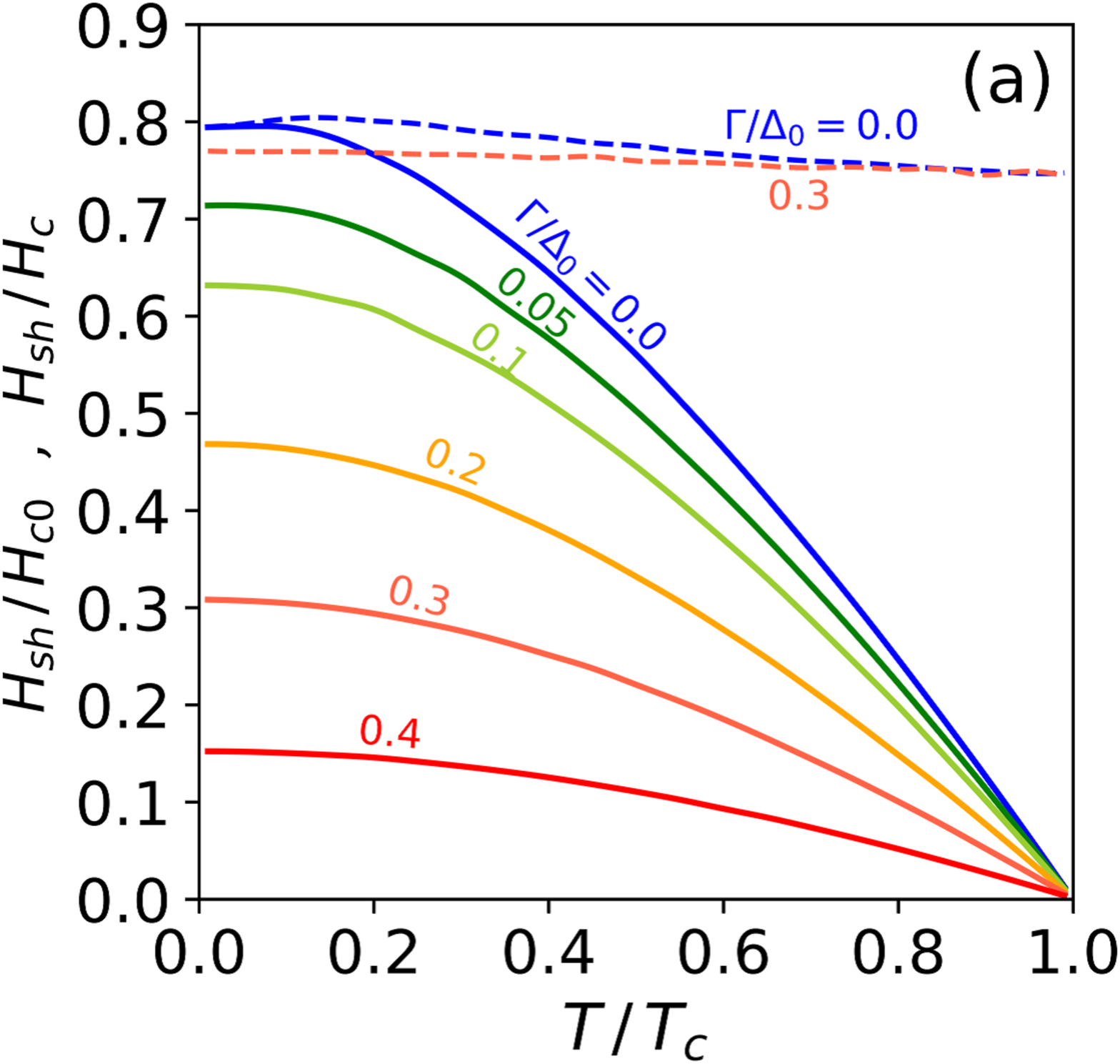}
   \includegraphics[width=0.494\linewidth]{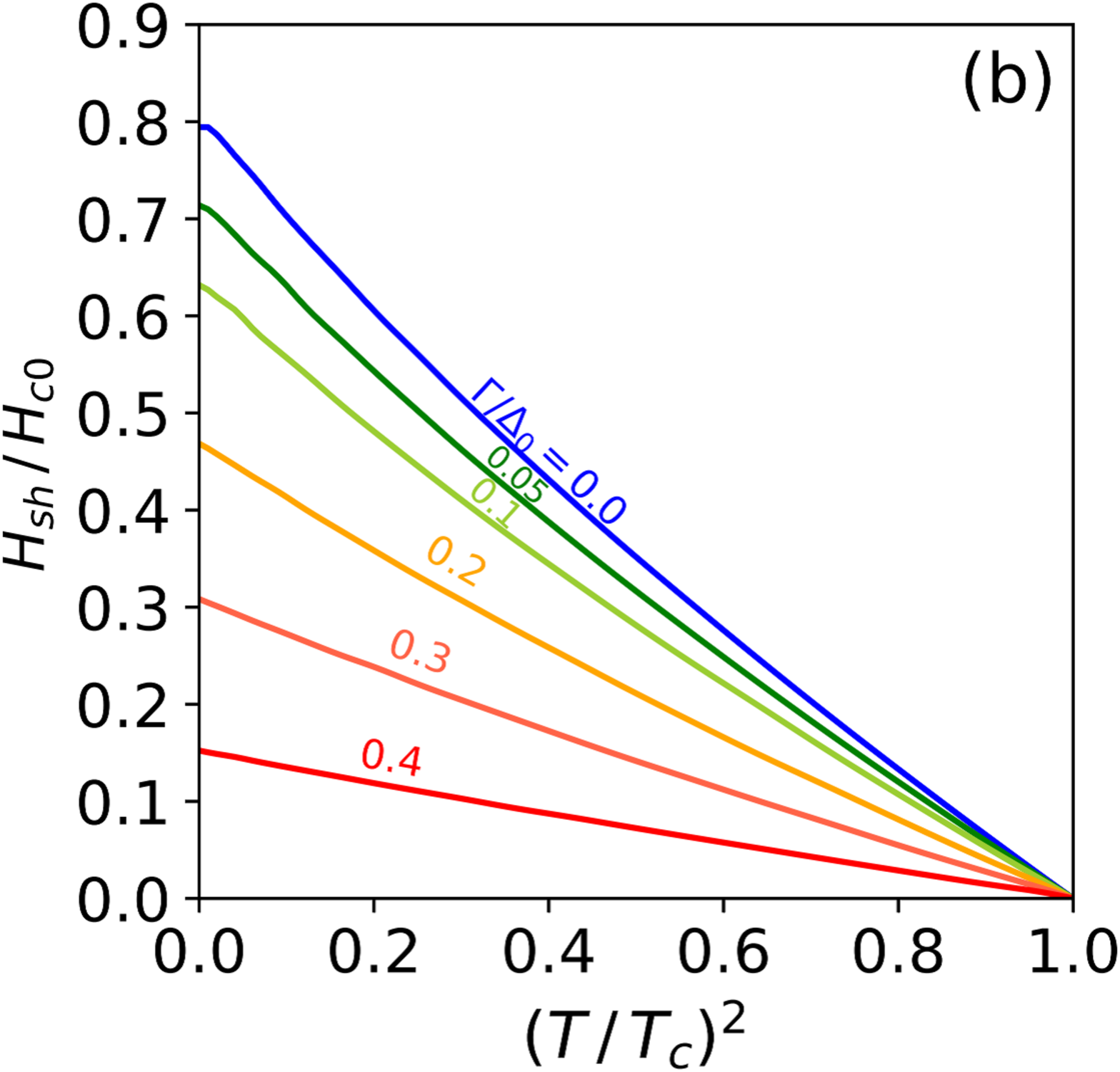}
   \end{center}\vspace{0 cm}
   \caption{
Superheating field $H_{sh}(\Gamma, T)$ at finite temperatures. 
(a) $H_{sh}(\Gamma, T)/H_{c0}$ (solid curves) and $H_{sh}(\Gamma, T)/H_{c}(\Gamma,T)$ (dashed curves) as functions of $T/T_c$ for different $\Gamma$. 
(b) $H_{sh}$ as functions of $(T/T_c)^2$. 
   }\label{fig16}
\end{figure}

For $0 < T < T_c$, we use the numerical solutions of $n_s(s,\Gamma,T)$ and $s_d(\Gamma,T)$ [see Fig.~\ref{fig6} (b)]. 
Shown as the solid curves in Fig.~\ref{fig16} (a) is $H_{sh}(\Gamma, T)$ as functions of $T/T_c$. 
We find $H_{sh}$ is a monotonically decreasing function of $\Gamma$ and $T$. 
Shown as the dashed curves in Fig.~\ref{fig16} (a) is $H_{sh}(\Gamma, T)$ normalized by $H_c(\Gamma,T)$. 
The curves merge at $T \to T_c$ and reproduce the GL coefficient $\sqrt{5}/3$. 
It is often useful to plot $H_{sh}$ as functions of $(T/T_c)^2$ (see, e.g., Ref.~\cite{2015_Posen_PRL}). 
Shown in Fig.~\ref{fig16} (b) is $H_{sh}$ as functions of $(T/T_c)^2$. 
The slope of $H_{sh}$ is steeper than $[1-(T/T_c)^2]$ decreases as $\Gamma$ increases.

\section{Discussions}

In Sec.~\ref{sec_solutions}, we have investigated a disordered superconductor with a finite Dynes $\Gamma$ parameter by solving the Usadel equation. 
We have calculated $\Delta(s, \Gamma, T)$, $n_s(s, \Gamma, T)$, $\lambda(s, \Gamma, T)$, and $j_s(s, \Gamma, T)$ for all $T$, all $\Gamma$, and all superfulid flow parameter $s \propto q^2$. 
Besides, we have derived the formulas of $\Delta$, $n_s$, $\lambda$, and $j_s$ at $T=0$, 
taking the effects of $\Gamma$ into account [Eqs.~(\ref{self-consistency_Tzero_1})-(\ref{js_Tzero_3})]. 
The formulas for $T \simeq T_c$ have also been obtained, which have the similar forms as the usual GL results except that $T_c$ depends on $\Gamma$. 
Using these results, we have investigated a narrow thin-film in Sec.~\ref{sec_film} and a semi-infinite superconductor in Sec.~\ref{sec_bulk}. 
In the following, we summarize the results and discuss their implications.

\subsection{Depairing current density}

In Sec.~\ref{sec_film_jd}, we have calculated the depairing current density $j_d(\Gamma, T)$ for all $T$ and all $\Gamma$ (see Fig.~\ref{fig8}). 
Also, we have derived the analytical formulas for $j_d$ valid for $T=0$ and $\Gamma \ll 1$ [see Eqs.~(\ref{jd_approx})-(\ref{alpha})]. 
The formulas for $T \simeq T_c$ has the similar form as the well-known GL depairing current except that $T_c$ depends on $\Gamma$ [see Eqs.~(\ref{jd_GL_1})-(\ref{jd0_GL})].

Our results show that $j_d$ is given by the Kupriyanov-Lukichev-Maki (KLM) theory for $\Gamma=0$ and decreases as $\Gamma$ increases. 
Hence, we can expect that real materials, which usually have $\Gamma>0$, 
exhibit smaller $j_d$ than the ideal KLM value. 
The previous measurements do not contradict this expectation, 
but the correlation between $j_d$ and $\Gamma$ is still unclear. 
Simultaneous measurements of $j_d$ and $\Gamma$ can provide with a deeper insight into observed values of $j_d$. %

While materials mechanisms behind $\Gamma$ are not well-understood, 
it would be possible to engineer $\Gamma$ by combining tunnel measurements and various materials processing. 
Finding a better materials processing method which can reduce $\Gamma$, 
we can ameliorate $j_d$.

In SNSPD, the detection efficiency (DE) depends on the bias current $j_{\rm bias}/j_d$. 
Then, an increase or decrease of $j_d$ via a $\Gamma$ engineering (see Fig.~\ref{fig7}) would result in a shift of the necessary dc bias. 
For instance, when $\Gamma=0.1$, we have more than $10\%$ degradation of $T_c$ and $30\%$ degradation of $j_d$ compared with the ideal BCS superconductor, 
which would reduce the necessary dc bias by $30\%$.  
Simultaneous measurements of $\Gamma$, $j_d$ and DE before and after materials treatments (e.g., ion irradiation~\cite{2019_Zhang_You}) can test the theory.

\subsection{Kinetic inductance}

In Sec.~\ref{sec_film_Lk_Zero}, we have derived the zero-current kinetic inductance formula, Eq.~(\ref{Lk_0_Gamma_T}). 
By using this formula, we have calculated $L_k(0,\Gamma,T)$ for all $T$ and all $\Gamma$. 
Our results show that $\Gamma$ affects the $T$ dependence of $L_k(0,\Gamma,T)/L_k(0,\Gamma,0)$ as shown in Fig.~\ref{fig9} (b).  
Simultaneous measurements of $\Gamma$ and the zero-current kinetic inductance for different $T$ can confirm the theoretical prediction.

In Sec.~\ref{sec_film_Lk_Fast} and Sec.~\ref{sec_film_Lk_Slow}, 
we have numerically calculated the current dependent nonlinear kinetic inductance in the fast- and the slow-measurement regimes for all $T$, all $\Gamma$, and all current up to $j_d$ (see Figs.~\ref{fig11} and \ref{fig13}).  
In the fast measurement regime, the current-induced increase of $L_k(j_{\rm bias},\Gamma, T)/L_k(0,\Gamma,T)$ is at most $\sim 1.5$ even at $j_d$. 
The coefficient of the quadratic expansion at $T=0$ is $C_{\rm fm}=0.136$ [see Eqs.~(\ref{C_fast_measure_1})]. 
On the other hand, in the slow measurement regime, $L_k$ diverges at $j_d$, 
and the coefficient of the quadratic expansion at $T=0$ is given by $C_{\rm sm}=4C_{\rm fm}=0.544$ [see Eqs.~(\ref{C_slow_measure_1})]. 
The difference between the fast- and the slow-measurement results would be detectable in experiments. 
The effects of $\Gamma$ on $C_{\rm fm}$ and $C_{\rm sm}$ are not significant.  
To test the theory, measurements of the current-dependent nonlinear kinetic inductance should be combined with a measurement of $j_d$ and tunneling spectroscopy to extract $\Gamma$.

Also, our theory suggests that it would be possible to tune $L_k$ by engineering $\Gamma$ as well as by the dc or ac current and controlling $T$.  
While $L_k$ always increases with the bias current and $\Gamma$, 
the dissipative conductivity $\sigma_1$ can be smaller than that of the ideal BCS superconductor by tuning the dc bias and $\Gamma$ (see Fig.~7 in Ref.~\cite{2020_Kubo_jd}). 
Then, we can simultaneously increase $L_k$ and reduce $\sigma_1$ (e.g., $\Gamma=0.05$ and $j_s \ll j_d$ lead to $10\%$ increase of $L_k$ and $20\%$ reduction of $\sigma_1$~\cite{2017_Gurevich_SUST, 2017_Gurevich_Kubo, 2020_Kubo_jd}). 
These results might be useful for developing superconducting circuit elements (e.g., superinductor~\cite{2019_Niepce}).

\subsection{Superheating field}

In Sec.~\ref{sec_bulk}, we have derived the general formula of the superheating field for a disordered superconductor $H_{sh}$, 
taking the effects of $\Gamma$ into account [see Eq.~(\ref{Hsh_formula})].
Using this formula, we have calculated $H_{sh}(\Gamma,T)$ for all $T$ and all $\Gamma$. 
A simple analytical formula for $T=0$ and $\Gamma \ll 1$ is also derived,  
which is given by Eq.~(\ref{Hsh_Gamma_0}). 
For the ideal dirty BCS superconductor with $\Gamma=0$, we have obtained $H_{sh}(0,0) =0.79 H_{c0}$ at $T=0$, 
which is slightly smaller than $H_{sh}^{\rm clean}=0.84 H_{c0}$ for a clean limit superconductor with a large $\lambda/\xi$~\cite{1966_Galaiko, 2008_Catelani} and consistent with the previous study~\cite{2012_Lin_Gurevich} in which $H_{sh}$  decreases with mfp when ${\rm mfp} < 5.32\xi_0$.

According to our results, 
we can expect that the maximum operating field of an SRF cavity made from a homogeneous dirty BCS superconductor with $\Gamma=0$ is given by $H_{sh}(0,0) = 0.79 H_{c0}$.
Taking dirty Nb materials for example, 
$\mu_0 H_{c0}=200\,{\rm mT}$ yields $\mu_0 H_{sh}(0,0)=160\,{\rm mT}$, 
which translates into the accelerating field $E_{acc}=37\,{\rm MV/m}$ for the Tesla-shape SRF cavity.  
This can be tested by measuring quench fields of impurity-doped dirty Nb cavities with ${\rm mfp} \ll \xi$. 
It should be noted that $H_{sh}$ can increase as impurities decreases~\cite{2012_Lin_Gurevich}, 
then materials with ${\rm mfp} \sim \xi$~\cite{2013_Grassellino, 2013_Dhakal, 2017_Maniscalco, 2018_Yang, 2019_Gonnella} can have slightly higher $H_{sh}$ than $0.79 H_{c0}$.

Strong local-heating (and resultant quenches) of SRF cavities are often attributed to geometrical defects on the surface~\cite{2008_Iwashita, 2011_Ge, 2013_Yamamoto, 2019_Wenskat, 2020_Pudasaini, 1999_Knobloch, 2015_Kubo_PTEP, 2015_Kubo_PTEP_mag, 2016_Xu}. 
Our theory suggest there can be another source of local heating. 
Since real materials have a finite $\Gamma$~\cite{2015_Becker, 2019_Groll} and $H_{sh}(\Gamma,0) < H_{sh}(0,0)$, 
an area with a large $\Gamma$ on the inner surface can be a hot spot even when $H_0 \ll H_{sh}(0,0)$. 
For instance, $H_{sh}$ at an area with $\Gamma=0.3$ on the surface of ${\rm Nb_3 Sn}$ can be estimated as $\mu_0 H_{sh}=160\,{\rm mT}$ at $T=0$ from Fig.~\ref{fig15}, 
which is much smaller than the ideal value $\mu_0 H_{sh}(0,0)=430\,{\rm mT}$. 
Here $\mu_0 H_{c0} = 540\,{\rm mT}$ is used. 
Taking another example, 
$H_{sh}$ at an area with $\Gamma=0.15$ on the surface of disordered Nb materials can be estimated as $\mu_0 H_{sh}=110\,{\rm mT}$ at $T=0$,
which is smaller than the ideal dirty Nb by $30\%$. 
Such an area with $\Gamma>0$ can be a source of local heating and may cause quenches. 
Simultaneous measurements of an onset field of local heating and $\Gamma$ at the hot spot can test the theory. 
Materials processing that can reduce $\Gamma$ would improve the accelerating field of SRF cavities.

The multilayer structure~\cite{2006_Gurevich, 2014_Kubo, 2015_Gurevich, 2017_Kubo_SUST, 2019_Kubo_Gurevich} or a superconductor including inhomogeneous impurities in the vicinity of the surface~\cite{2017_Kubo_SUST, 2019_Sauls, 2019_Kubo_Gurevich} can suppress the surface current and enhance $H_{sh}$. 
The formula of $H_{sh}$ given by Eq.~(\ref{Hsh_formula}) is not applicable to these structure because $j_s(x)$ takes the maximum at the inside ($x_0 > 0$). 
Study on effects of $\Gamma$ on $H_{sh}$ in these structures would be useful for comparison of the theory and experiments~\cite{2013_Antoine, 2016_Tan, 2017_Anne-Marie, 2019_Antoine, 2019_Kubo_JJAP, 2020_Ito, 2019_R_Ito_SRF, 2019_H_Ito_SRF, 2019_Keckert_SRF, 2019_Katayama_SRF, 2019_Oseroff_SRF, 2019_Thoeng_SRF,2019_Turner_SRF, 2019_Senevirathne_SRF}.


\begin{acknowledgments}
I would like to express the deepest appreciation to Alex Gurevich for his hospitality during my visit to Old Dominion University. 
This work was supported by Japan Society for the Promotion of Science (JSPS) KAKENHI Grants No. JP17H04839, No. JP17KK0100, and JP19H04395. 
\end{acknowledgments}

\appendix
\section{Derivations of Eqs.~(\ref{Delta_0_Gamma_0}), (\ref{Hc_Gamma_0}), and (\ref{ns_0_Gamma_0})} \label{a1}

The $\theta$ parameter of the Matsubara Green's function for the zero-current state is given by 
\begin{eqnarray}
u = \cot\theta = \frac{\omega_n + \Gamma}{\Delta} .
\end{eqnarray}
Then the self-consistency equation at $T = 0$ can be written as 
\begin{eqnarray}
0 &=& \lim_{M \to \infty} \biggl[ \int_{\Gamma/\Delta}^{(M+\Gamma)/\Delta} \!\!\!\! \frac{du}{\sqrt{1+u^2}} -\int_{0}^{M}\!\!\! \frac{d\omega}{\sqrt{\omega^2 + 1}} \biggr] \nonumber \\
&=& \lim_{M \to \infty} \biggl(\sinh^{-1} \frac{M+\Gamma}{\Delta} -\sinh^{-1}M - \sinh^{-1} \frac{\Gamma}{\Delta} \biggr) \nonumber \\
&=& -\ln \Delta - \sinh^{-1} \frac{\Gamma}{\Delta} , 
\end{eqnarray}
resulting in Eq.~(\ref{Delta_0_Gamma_0}). 
Here, the Matsubara sum $2\pi T \sum_{\omega_n}$ is replaced with integration $\lim_{M \to \infty} \int_{0}^{M}d\omega$, and $d\omega = \Delta du$ and $\sinh^{-1} M = \ln M \, (M\to \infty)$ are used.

The thermodynamic critical field can be calculated from Eqs.~(\ref{Hc}) and (\ref{Omega}). 
At $T =0$, we find
\begin{eqnarray}
\Omega(0, \Gamma, 0) 
&=& -N_0 \Delta^2 \lim_{M \to \infty} \int_{\Gamma/\Delta}^{(M+\Gamma)/\Delta} du \biggl[ 
\frac{1}{\sqrt{1+u^2}} \nonumber \\
&&+  \frac{2u^2}{\sqrt{1+u^2}} -2u  \biggr] \nonumber \\
&=& -N_0 \Delta^2 \lim_{M \to \infty} \biggl[ -u^2 + u\sqrt{1+u^2} \biggr]_{\Gamma/\Delta}^{(M+\Gamma)/\Delta} \nonumber \\
&=& -\frac{1}{2}N_0 \Delta^2 \biggl( 1 + \frac{2\Gamma^2}{\Delta^2} - \frac{2\Gamma}{\Delta} \sqrt{1+\frac{\Gamma^2}{\Delta^2}} \biggr) ,
\end{eqnarray}
resulting in Eq.~(\ref{Hc_Gamma_0}).

The superfluid density can calculated from Eq.~(\ref{superfluid_density}). 
At $T = 0$, we have 
\begin{eqnarray}
&&\frac{n_s(0,\Gamma,0)}{n_{s0}} = \frac{\lambda_0^2}{\lambda^2(0,\Gamma,0)} 
= \frac{2}{\pi} \int_{0}^{\infty}\!\!\! \frac{d\omega}{1+u^2} \nonumber \\
&&= \frac{2\Delta}{\pi} \int_{\Gamma/\Delta}^{\infty} \frac{du}{1+u^2} 
= \Delta \biggl( 1 - \frac{2}{\pi} \tan^{-1}\frac{\Gamma}{\Delta} \biggr) , 
\end{eqnarray}
namely, Eq.~(\ref{ns_0_Gamma_0}).

\section{Derivations of Eqs.~(\ref{self-consistency_Tzero_1}) and (\ref{superfluid_density_Tzero_1})} \label{a2}

Generalizing the procedures in Appendix~\ref{a1}, we can derive Eqs.~(\ref{self-consistency_Tzero_1} ) and (\ref{superfluid_density_Tzero_1}). 
The Mastubara Green's function $u=\cot\theta$ for the current carrying state satisfies
\begin{eqnarray}
\biggl( 1-\frac{\zeta}{\sqrt{1+u^2}} \biggr) u = \frac{\omega_n + \Gamma}{\Delta}
\end{eqnarray}
where $\zeta=s/\Delta$. 
The self-consistency equation at $T = 0$ is given by
\begin{eqnarray}
0 &=& \int_{0}^{\infty} \!\!\! d\omega \biggl( \frac{1}{\Delta \sqrt{1+u^2}}  - \frac{1}{\sqrt{\omega^2 +1}} \biggr) \nonumber \\
&=& \int_{u_0}^{\infty} \!\!\! du \biggl( 1- \frac{\zeta}{(1+u^2)^{3/2}} \biggr) 
\Biggl( \frac{1}{\sqrt{1+u^2}} - \nonumber \\
&& \frac{1}{ \sqrt{\Delta^{-2} + [( 1- \zeta/\sqrt{1+u^2} ) u - \Gamma/\Delta]^2 } }  \Biggr)  \nonumber \\
&=& -\ln\Delta - \sinh^{-1}u_0 -\frac{\zeta}{2} \biggl( \frac{\pi}{2} -\tan^{-1}u_0 - \frac{u_0}{1+u_0^2}  \biggr) \nonumber \\
\end{eqnarray}
resulting in Eq.~(\ref{self-consistency_Tzero_1}). 
Here $u_0(s,\Gamma)$ is defined by $(1-\zeta/\sqrt{1+u_0^2}) u_0 = \Gamma/\Delta$.

The superfluid density can calculated from Eq.~(\ref{superfluid_density}). 
At $T = 0$, we find 
\begin{eqnarray}
&&\frac{n_s(s,\Gamma,0)}{n_{s0}} = \frac{\lambda_0^2}{\lambda^2(s,\Gamma,0)} 
= \frac{2}{\pi} \int_{0}^{\infty}\!\!\! \frac{d\omega}{1+u^2} \nonumber \\
&&= \frac{2\Delta}{\pi} \int_{u_0}^{\infty} \!\!du \biggl( \frac{1}{1+u^2} - \frac{\zeta}{(1+u^2)^{\frac{3}{2}}} + \frac{\zeta u^2}{ (1+u^2)^{\frac{5}{2}} } \biggr) \nonumber \\
&&=\Delta \biggl[ 1- \frac{2}{\pi}\tan^{-1} u_0 - \frac{4\zeta}{3\pi} \biggl\{ 1 - \frac{u_0 (3+2u_0^2)}{2(1+u_0^2)^{\frac{3}{2}}} \biggr\} \biggr]
.
\end{eqnarray}
This is Eq.~(\ref{superfluid_density_Tzero_1}).

\section{Derivations of Eqs.~(\ref{jd_approx})-(\ref{alpha})} \label{a3}

For $\Gamma \ll \Delta(s, \Gamma, 0)-s$, we can calculate $j_s$ from Eqs.~(\ref{self-consistency_Tzero_2})-(\ref{js_Tzero_2}), 
which takes the maximum when $\partial j_s /\partial s =0$: 
\begin{eqnarray}
&&\biggl( \frac{\pi}{4} - \zeta_d -\frac{\Gamma}{2\Delta_d} \biggr) 
\biggl( 1 - \frac{\pi}{4} \zeta_d - \frac{\Gamma}{\Delta_d} \biggr) = \frac{\pi^2}{8} \zeta_d , \label{a3:sd_Deltad_1} \\
&& \Delta_d =  \exp \biggl[-\frac{\pi \zeta_d}{4} - \frac{\Gamma}{\Delta_d} \biggr] \label{a3:sd_Deltad_2} ,\\
&& \zeta_d = s_d/ \Delta_d . \label{a3:sd_Deltad_3}
\end{eqnarray}
Here 
$\partial \Delta /\partial s = -(\pi/4)[1-(\pi/4)(s/\Delta)-\Gamma/\Delta]^{-1}$ is used. 
Then the maximum value of $j_s$ is given by 
\begin{eqnarray}
j_d(\Gamma, T)|_{T\to 0} 
= \sqrt{\pi s_d} \biggl[ \Delta_d - \frac{4s_d}{3\pi} -\frac{2\Gamma}{\pi} \biggr] \frac{H_{c0}}{\lambda_0} , 
\end{eqnarray}
which is the depairing current density.

For $\Gamma=0$, Eq.~(\ref{a3:sd_Deltad_1}) is a simple quadratic equation for $\zeta_d$. 
Then the solution is given by
\begin{eqnarray}
&&\zeta_{d0}= \frac{2}{\pi} + \frac{3\pi}{8} - \sqrt{\biggl( \frac{2}{\pi} + \frac{3\pi}{8} \biggr)^2 -1}  = 0.300.  
\\
&& \Delta_{d0} = \exp \biggl[ -\frac{\pi}{4} \zeta_{d0} \biggr] = 0.790, \\
&& s_{d0} = \Delta_{d0} \zeta_{d0} = 0.237 ,
\end{eqnarray}
resulting in $j_d (0, 0) = 0.595H_{c0}/\lambda_0$, 
the well-known result obtained by Maki~\cite{1963_Maki_I, 1963_Maki_II} and Kupriyanov and Lukichev~\cite{1980_Kupriyanov}.

For $0 \le \Gamma \ll 1$, we can solve Eq.~(\ref{a3:sd_Deltad_1}) by expanding $\zeta_d$ about $\zeta_{d0}$. 
Substituting 
\begin{eqnarray}
\zeta_{d}= \zeta_{d0} - \alpha \frac{\Gamma}{\Delta_{d0}}  , \label{zeta_d_a3}
\end{eqnarray}
into Eq.~(\ref{a3:sd_Deltad_1}), we find 
\begin{eqnarray}
\alpha = \frac{1+ \pi/2 -2 (1+8/\pi) \zeta_{d0}}{2+ 3\pi^2/8 - \pi \zeta_{d0}} = 0.365,
\end{eqnarray}
Then Eqs.~(\ref{a3:sd_Deltad_2}) and (\ref{a3:sd_Deltad_3}) result in
\begin{eqnarray}
&& \Delta_{d} = e^{-\frac{\pi}{4} \zeta_{d} -\frac{\Gamma}{\Delta_d}} = \Delta_{d0} \exp \biggl[ \biggl( \frac{\pi \alpha}{4} -1 \biggr)\frac{\Gamma}{\Delta_{d0}} \biggr] , \label{Delta_d_a3} \\
&& s_{d} = \Delta_{d} \zeta_{d} = (s_{d0} -\alpha \Gamma) \exp \biggl[ \biggl( \frac{\pi \alpha}{4} -1 \biggr)\frac{\Gamma}{\Delta_{d0}} \biggr]  . \label{sd_a3}
\end{eqnarray}
%

\section{Derivations of Eqs.~(\ref{C_fast_measure_1}) and (\ref{C_slow_measure_1})} \label{a4}

\subsection{Fast measurement}
For $s\ll 1$ and $\Gamma \ll 1$, the superfluid density is given by Eq.~(\ref{superfluid_density_Tzero_3}). 
Then, Eq.~(\ref{Lk_s_Gamma_T_fast}) yields
\begin{eqnarray}
\frac{L_k (s_{\rm bias}, \Gamma, 0)}{L_{k0}} = 1 + \biggl( 1+ \frac{2}{\pi} \biggr) \Gamma + \biggl(\frac{\pi}{4}+\frac{4}{3\pi} \biggr) s_{\rm bias} \label{appendix_Lk_1} ,
\end{eqnarray}
then
\begin{eqnarray}
\frac{L_k (s_{\rm bias}, \Gamma, 0)}{L_{k}(0,\Gamma,0)} = 1 + \biggl(\frac{\pi}{4}+\frac{4}{3\pi} \biggr) s_{\rm bias} . \label{appendix_Lk_2}
\end{eqnarray}
Here the current-momentum relation, Eq.~(\ref{js_Tzero_3}), can be written as
\begin{eqnarray}
s_{\rm bias} 
&=& \frac{1}{\pi} \biggl( \frac{j_{\rm bias}}{j_{s0}} \biggr)^2
= \frac{1}{\pi} \biggl( \frac{j_d}{j_{s0}} \biggr)^2 \biggl( \frac{j_{\rm bias}}{j_d} \biggr)^2 \nonumber \\
&=& s_d \biggl[ \Delta_d(\Gamma) - \frac{4 s_d(\Gamma)}{3\pi} -\frac{2\Gamma}{\pi} \biggr]^2  \biggl( \frac{j_{\rm bias}}{j_d} \biggr)^2 \nonumber \\
&\simeq & s_{d0} \biggl( \Delta_{d0} - \frac{4 s_{d0}}{3\pi} \biggr)^2  \biggl( \frac{j_{\rm bias}}{j_d} \biggr)^2 . \label{appendix_sbias}
\end{eqnarray}
Substituting Eq.~(\ref{appendix_sbias}) into Eq.~(\ref{appendix_Lk_2}), 
we find
\begin{eqnarray}
&&\frac{L_k (s_{\rm bias}, \Gamma, 0)}{L_{k}(0,\Gamma,0)}  = 1+ C_{\rm fm} \biggl( \frac{j_{\rm bias}}{j_d} \biggr)^2 , \\
&&C_{\rm fm} = \frac{3\pi^2+16}{12\pi} s_{d0} \biggl( \Delta_{d0} - \frac{4 s_{d0}}{3\pi}  \biggr)^2 = 0.136 .
\end{eqnarray}
%

\subsection{Slow measurement}

Substituting the superfluid density, Eq.~(\ref{superfluid_density_Tzero_3}), for $s\ll 1$ and $\Gamma \ll 1$ into Eq.~(\ref{Lk_s_Gamma_T_slow_0}), 
we obtain
\begin{eqnarray}
\frac{L_k (s_{\rm bias}, \Gamma, 0)}{L_{k0}} = 1 + \biggl( 1+ \frac{2}{\pi} \biggr) \Gamma + 3\biggl(\frac{\pi}{4}+\frac{4}{3\pi} \biggr) s \label{appendix_Lk__slow_1} ,
\end{eqnarray}
then
\begin{eqnarray}
\frac{L_k (s_{\rm bias}, \Gamma, 0)}{L_{k}(0,\Gamma,0)} = 1 + 3\biggl(\frac{\pi}{4}+\frac{4}{3\pi} \biggr) s \label{appendix_Lk_slow_2}
\end{eqnarray}
Using Eq.~(\ref{appendix_sbias}), we find 
\begin{eqnarray}
&&\frac{L_k (s, \Gamma, 0)}{L_{k}(0,\Gamma,0)}  = 1+ C_{\rm sm} \biggl( \frac{j_s}{j_d} \biggr)^2 , \\
&&C_{\rm sm} = \frac{3\pi^2+16}{4\pi} s_{d0} \biggl( \Delta_{d0} - \frac{4 s_{d0}}{3\pi}  \biggr)^2 = 0.544 .
\end{eqnarray}
%

\section{Derivations of Eq.~(\ref{Hsh_Gamma_0})} \label{a5}

To evaluate Eq.~(\ref{Hsh_formula}) for small $\Gamma$ regions, 
we rewrite Eqs.~(\ref{self-consistency_Tzero_2}) and (\ref{superfluid_density_Tzero_2}) in more convenient forms. 
Let us expand $\Delta(s,\Gamma,0)$ around $\Delta (s,0,0)$ and write 
\begin{eqnarray}
\Delta(s,\Gamma,0)=\Delta (s,0,0) - \beta \Gamma. \label{Delta_a5}
\end{eqnarray}
Substituting Eq.~(\ref{Delta_a5}) into Eq.~(\ref{self-consistency_Tzero_2}), we find
\begin{eqnarray}
\beta = \biggl( 1-\frac{\pi s}{4\Delta(s,0,0)} \biggr)^{-1} .
\end{eqnarray}
Then Eq.~(\ref{superfluid_density_Tzero_2}) yields
\begin{eqnarray}
\frac{n_s(s,\Gamma,0)}{n_{s0}} 
= \Delta (s,0,0) - \frac{\Gamma}{1-\frac{\pi s}{4\Delta(s,0,0)}} - \frac{4s}{3\pi} - \frac{2\Gamma}{\pi} 
\end{eqnarray}
Now Eq.~(\ref{Hsh_formula}) reduces to
\begin{eqnarray}
\biggl( \frac{H_{sh} (\Gamma, 0)}{H_{c0}} \biggr)^2
&=&  \pi  \int_0^{s_d} \biggl[ \Delta (s,0,0) - \frac{\Gamma}{1-\frac{\pi s}{4\Delta(s,0,0)}} \biggr] ds \nonumber \\
&&- \frac{2}{3}s_d^2 -2\Gamma s_d . 
\end{eqnarray}
Here $s_d$ is given by Eq.~(\ref{sd_a3}). 
To perform the integration, we change the variable from $s$ to $z = s/\Delta(s,0,0)$.  
Using $d \Delta(s,0,0)/ds = -(\pi/4)[ 1-\pi s/ 4\Delta(s,0,0)]^{-1}$ and $\Delta(s,0,0) = e^{-\pi z/4}$, 
we find 
$ds = dz/(dz/ds) = e^{-\pi z/4}  ( 1 - \pi z/4) dz$.
Then the integration becomes
\begin{eqnarray}
&&\int_0^{s_d} \biggl[ \Delta (s,0,0) - \frac{\Gamma}{1-\frac{\pi s}{4\Delta(s,0,0)}} \biggr] ds \nonumber \\
&& = \int_0^{z_d} \biggl[ e^{-\frac{\pi z}{2}} \biggl( 1 - \frac{\pi z}{4}\biggr) - \Gamma e^{-\frac{\pi z}{4}} \biggr]  dz \nonumber \\
&& = \frac{1}{\pi} \biggl[ 1 - \biggl( 1- \frac{\pi z_d}{2} \biggr)e^{-\frac{\pi z_d}{2}} - 4\Gamma (1- e^{-\frac{\pi z_d}{4}}) \biggr]
\end{eqnarray}
where $z_d = s_d/ \Delta(s_d,0,0)$. 
To obtain $z_d$, we expand $\Delta(s_d,0,0)$ around $\Delta_d = \Delta(s_d, \Gamma, 0)$: 
\begin{eqnarray}
\Delta(s_d,0,0) = \Delta_d + \eta \Gamma . \label{Delta_sd_0_0_a5} 
\end{eqnarray}
Here $\Delta_d$ is given by Eq.~(\ref{Delta_d_a3}). 
Substituting Eq.~(\ref{Delta_sd_0_0_a5}) into $\Delta(s_d,0,0)=\exp [- \pi s_d/\Delta(s_d,0,0)]$ and using $\Delta_d = \exp[-\pi s_d/\Delta_d -\Gamma/\Delta_d]$, 
we find
\begin{eqnarray}
\eta = \biggl( 1-\frac{\pi \zeta_d}{4} \biggr)^{-1}  , \label{eta_a5} 
\end{eqnarray}
where $\zeta_d = s_d/\Delta_d$ is given by Eq.~(\ref{zeta_d_a3}). 
Then we find
\begin{eqnarray}
z_d = \frac{s_d}{\Delta_d + \eta \Gamma} = \zeta_d \frac{1 - \frac{\pi}{4}\zeta_d }{1-\frac{\pi}{4}\zeta_d + \frac{\Gamma}{\Delta_d} } .
\end{eqnarray}

\end{document}